\documentclass[pdflatex,sn-mathphys-num]{sn-jnl}

\usepackage{graphicx}%
\usepackage{adjustbox}%
\usepackage{multirow}%
\usepackage{amsmath,amssymb,amsfonts}%
\usepackage{amsthm}%
\usepackage{mathrsfs}%
\usepackage[title]{appendix}%
\usepackage{xcolor}%
\usepackage{textcomp}%
\usepackage{manyfoot}%
\usepackage{booktabs}%
\usepackage{amsmath}
\usepackage{stmaryrd}
\usepackage{import}
\usepackage{mathtools}
\usepackage{relsize}
\usepackage[autostyle=true]{csquotes}
\usepackage{verbatim}
\usepackage{tikz}
\usepackage{tikz-cd}
\usepackage{bbm}
\usepackage{microtype}
\usepackage[british]{babel}
\usepackage{enumitem}
\usepackage{wrapfig}
\usepackage{stackengine}
\usepackage{algorithm}
\usepackage[nameinlink, capitalize]{cleveref}

\usepackage{amsopn}
\DeclareMathOperator{\diag}{diag}

\theoremstyle{thmstyleone}%
\newtheorem{theorem}{Theorem}
\theoremstyle{thmstyletwo}%
\newtheorem{example}{Example}%
\newtheorem{remark}{Remark}%
\newtheorem{lemma}{Lemma}%

\theoremstyle{thmstylethree}%
\newtheorem{definition}{Definition}%

\newcommand{\Z}{\mathbb{Z}}
\newcommand{\R}{\mathbb{R}}

\DeclareMathOperator{\Ima}{Im}

\newcommand{\encircle}[1]{\stackMath\mathbin{\stackinset{c}{0ex}{c}{0ex}{\bigcirc}{#1}}}

\DeclareMathOperator{\divg}{div}
\newcommand{\cc}{\mathcal{C}}

\newcommand{\celltri}{\raisebox{-1pt}{\includegraphics[height=9pt]{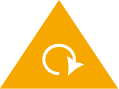}}}
\newcommand{\cellsq}{\raisebox{-1pt}{\includegraphics[height=9pt]{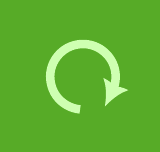}}}

\newcommand{\CW}{\textsc{cw}}
\newcommand{\CC}{\IfSmallCapsTF{\textsmaller{CC}}{\textsc{cc}}}

\makeatletter
\newcommand*{\IfSmallCapsTF}{%
	\ifx\f@series\my@test@sc
	\expandafter\@firstoftwo
	\else
	\expandafter\@secondoftwo
	\fi
}
\newcommand*{\my@test@sc}{b}
\makeatother

\begin{document}

\title{Don't be Afraid of Cell Complexes!}
\subtitle{An Introduction from an Applied Perspective}


\author[1]{\fnm{Josef} \sur{Hoppe}}
\equalcont{These authors contributed equally to this work.}

\author[1]{\fnm{Vincent P.} \sur{Grande}}
\equalcont{These authors contributed equally to this work.}

\author[1]{\fnm{Michael T.} \sur{Schaub}}

\affil[1]{\orgdiv{Faculty of Computer Science \& Department of Biology}, \orgname{RWTH Aachen University}, \orgaddress{\street{Ahornstr. 55}, \city{Aachen}, \postcode{52074}, \country{Germany}}\\
\texttt{hoppe@cs.rwth-aachen.de}, \texttt{grande@cs.rwth-aachen.de}}


\abstract{Cell complexes (\CC s) are a higher-order network model deeply rooted in \emph{algebraic topology} that has gained interest in \emph{signal processing} and \emph{network science} recently.
The processing of signals supported on \CC{}s can be described in terms of easily-accessible algebraic or combinatorial notions.
However, the commonly presented definition of \CC{}s is grounded in abstract concepts from topology and remains disconnected from the signal processing methods developed for \CC{}s.
In this paper, we aim to bridge this gap by providing a simplified definition of \CC s that is accessible to a wider audience and can be used in practical applications.
Specifically, we first introduce a simplified notion of abstract regular cell complexes (\textsc{arcc}s).
These \textsc{arcc}s only rely on notions from algebra and are equivalent to regular cell complexes for most practical applications.
Second, using this new definition we provide an accessible introduction to (abstract) cell complexes from a perspective of network science and signal processing, including an introduction to central methods and recent developments.
Furthermore, as many practical applications work with \CC s of dimension 2 and below, we provide an even simpler definition for this case that significantly simplifies understanding and working with \CC s in practice.}




\maketitle
\begingroup
\renewcommand\thefootnote{}\footnote{Illustrative Examples: \url{https://github.com/josefhoppe/cell-complex-playground}}
\addtocounter{footnote}{-1}%
\endgroup

\section{Introduction}

\begin{figure}[t]
    \centering
    \includegraphics[width=.8\linewidth]{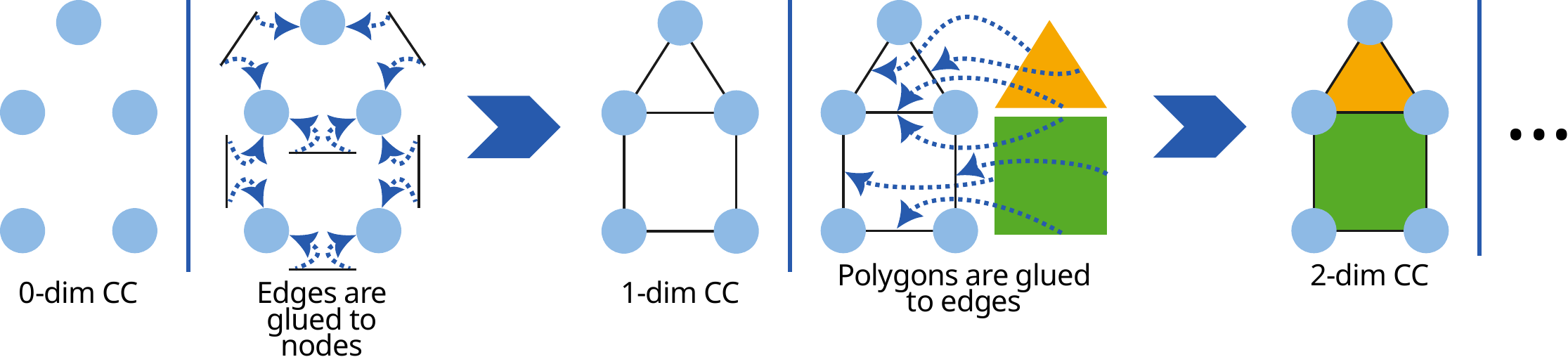}
    \caption{\CC{}s are obtained by starting with a set of vertices ($0$-cells) and iteratively glueing $k$-cells to $(k-1)$-cells (dotted blue arrows). Left: Initial set of vertices. Center left: Edges ($1$-cells) are glued to the vertices ($0$-cells). Center right: Polygons ($2$-cells) are glued to the edges ($1$-cells).}\label{fig:glueing}
\end{figure}
 
\newcommand{\zero}{{\color{gray} 0}}

Graphs have been a staple in research to abstract, model and analyse many real-world systems such as social networks, transportation networks, molecules, and brain connectivity~\cite{newman2018networks,strogatz2001exploring,mulder2018network,derrible2011applications}.
More recently, researchers have started to explore the use of higher-order networks, to model complex, non-pairwise relationships between nodes~\cite{battiston2020networks,bick2023higher,barbarossa2020topological,schaub2021signal}.
\emph{Cell complexes} (\CC s), which include~\emph{simplicial complexes} and \emph{cubical complexes} as special cases, are one such higher-order network abstraction that has gained particular interest due to their deep connections to algebraic topology~\cite{hatcher2002algebraic} and topological data analysis~\cite{wasserman2018topological,carlsson2020topological,chazal2021introduction}.
A thorough introduction to algebraic topology from a perspective of pure mathematics can be found in one of the classic  textbooks \cite{hatcher2002algebraic, Bredon:1993, tomDieck:2008}.

In algebraic topology, \emph{cell complexes} (or \CW{}\footnote{\CW{} denotes the technical condition of being \textbf{c}losure-finite and having the \textbf{w}eak topology.} complexes) represent objects of flexible shape which are built out of basic ball-shaped building blocks (\emph{cells}) of arbitrary dimension.
Cells of different dimensions are rigidly related.
For example, an area is enclosed by lines, which in turn are enclosed by points.
This rigid structure describes the underlying topology.
The right balance between flexibility and structure allows cell complexes to model the spaces underlying a vast range of applications.
In particular, they have been used for signal processing of signals defined on the edges of graphs and complexes~\cite{schaub2018flow, schaub2021signal,roddenberry2022signal,sardellitti2024topological} and to expand the expressivity of neural networks \cite{bodnar2021weisfeiler,hajij2022topological}.
Notably, the above applications only exploit the underlying topology of \CC{}s through their algebraic representations via \emph{boundary matrices}.
The definitions of \CC{}s provided in the majority of the current literature in signal processing, network science, or machine learning, however, commonly use more complicated notions from point-set topology, which obfuscates the connection to the algebraic structure used and makes cell complexes less accessible to a broader audience.

\paragraph{Contribution and outline}
This paper has three main contributions.
First, we give an intuitive introduction to cell complexes and their algebraic structure.
Second, we give formal definitions for \emph{abstract} regular cell complexes, i.e., cell complexes without explicitly specifying the underlying topological space.
In an effort to make \CC{}s useful for a broader audience, our definition aims to only introduce concepts relevant to signal processing and network science.
In particular, abstract regular \CC{} are equivalent to ordinary regular \CC{}s for dimension up to $2$ and in arbitrary dimensions, all ordinary regular \CC{}s are abstract regular \CC{}s and both are equivalent for the computational applications of this paper (\cref{thm:equivalence}).
Finally, we introduce important concepts and methods that are relevant for applications in signal processing and network science; where necessary, we also generalize methods from simplicial complexes to cell complexes.

To this end, we first give an informal, intuitive introduction to the central ideas underpinning the construction of cell complexes in \cref{sec:intuition}.
\Cref{sec:2dim-def} contains a simple definition for $2$-dimensional (abstract) cell complexes, which are the most commonly-used cell complexes in applications.
We will then continue to give a definition of abstract regular cell complexes in arbitrary dimensions in \cref{sec:def}, and provide a theorem that all regular cell complexes are abstract regular cell complexes, and prove the converse for dimensions up to $2$.
\Cref{sec:examples} provides an overview of common methods to obtain cell complexes.
With that in place, we discuss the topology and geometry (\cref{sec:topology}) of abstract cell complexes, and weights (\cref{sec:weights}) on abstract cell complexes.
This forms a basis for our introduction to different applications of cell complexes in \cref{sec:applications}, some of which we generalize from their original definitions on simplicial complexes or put into our unified framework.
Finally, in \cref{sec:outlook} we discuss limitations and give an outlook on current challenges and desirable future developments.
We also provide jupyter notebooks with examples for central concepts in a companion repository: \url{https://github.com/josefhoppe/cell-complex-playground}
\paragraph{Related work}

Higher-order network models beyond pairwise interactions have received significant attention over the last years \cite{bick2023higher}.
However, most of the work focusses on hypergraphs or simplicial complexes.
An overview over the field of topological or geometric deep learning can be found in \cite{hajij2022topological} and \cite{bronstein2017geometric}, where higher-order network models are combined with insights from topology and geometry and common deep learning techniques.
The adjacent field of topological data analysis tries to extract robust structure and information from complex data sets by the means of algebraic topology and persistent homology; it is surveyed in \cite{munch2017user}, \cite{carlsson2021topological}, and \cite{wasserman2018topological}.
Signal Processing on cell complexes, similar to traditional signal processing of time-varying signals and of signal processing of edge signals on graphs, has been explored in \cite{roddenberry2022signal} and in \cite{sardellitti2021topological}.
Neural network architectures based on cell complexes have been proposed in \cite{hajij2020cell} and \cite{bodnar2021weisfeiler}.
Hansen and Ghrist \cite{hansen2019toward} have studied applications of sheaves, a construct that supports flexible data types and relations from algebraic geometry, on cell complexes.
In \cite{hoppe2023representing}, the authors study sparse flow representation using cell complexes.
Adaptive learning on cell complexes is considered in \cite{marinucci2024topological}.
We give a more detailed overview over applications of cell complexes in \cref{sec:applications}.

\section{Cell Complexes: Intuition and Motivation}
\label{sec:intuition}

Intuitively, graphs may be thought of geometrically as a collection of points (or \emph{vertices}) connected by lines (or \emph{edges}).
Cell complexes extend graphs to higher dimensions by allowing the use of building blocks (\enquote{cells}) of arbitrary dimension in addition to the $0$-dimensional vertices and $1$-dimensional edges of ordinary graphs.
Most prominently, $2$-dimensional cells are polygons that are attached to a cycle of edges ($1$-cells) of the graph, as illustrated in \cref{fig:glueing}.
In general, a $k$-dimensional cell or $k$-cell is represented by a  $k$-dimensional object (a $k$-disc or equivalently $k$-ball).
The intuition behind this is that cell complexes model real-world structures with \emph{geometric} or \emph{topological} properties,
e.g., in physical networks where $2$-cells correspond to areas, or in flow systems where we can measure the flow \emph{around} a part of the system represented by $2$-cells.
In this regard, cell complexes inherently differ from other network models like hypergraphs which are more flexible but do not enforce an underlying geometry in the network.

We will use cell complexes in practice to \emph{model} complex data and facilitate geometry- and topology-informed computations.
Thus, we need an abstract computational model of cell complexes.
In particular, we are interested in the combinatorial and algebraic structure on cells enforced by the geometry and \emph{not} in the point-set topology itself, which we will abstract away.
For abstract cell complexes, we will thus consider the combinatorial structure of cells induced by the geometry, but not the infinite continuous euclidean space itself.

Cells in a cell complex are related via the \emph{boundary} relation (or \emph{boundary map}). A $k$-cell is enclosed by multiple $(k-1)$-cells which together form its \emph{boundary}.
This boundary relation is the central combinatorial structure of the abstract cell complex.
Like on graphs, discrete geometry, global structure and global phenomena emerge from this local relation.

\begin{wrapfigure}{r}{.5\linewidth}
	\includegraphics[width=\linewidth]{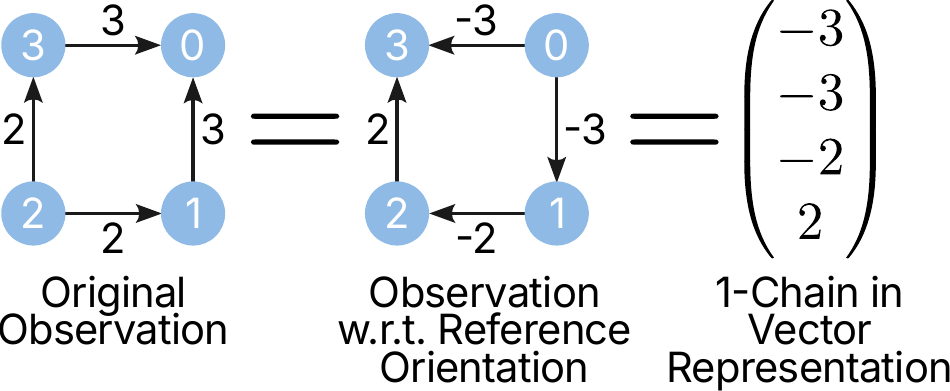}
	\caption{Flows on the edges of an oriented graph (or $1$-dimensional \CC).
		Left: Flow values with arrows indicating the direction of each flow.
		Center: The same flow, represented with respect to a reference orientation of the edges induced by an arbitrary ordering of the vertices.
		Right: The $1$-chain corresponding to the displayed flow in the center in vector notation.
	}\label{fig:chains}
\end{wrapfigure}

Data on spaces with geometric or topological structure usually has a direction:
For instance, we are not only interested in the absolute value of an edge flow, but also in its direction.
To represent such data, cells are thus endowed with a reference orientation which is arbitrarily chosen at construction.
For edges, the orientation is often also represented by an arrow.
The orientation is not a property of the underlying space, but rather a means of defining a reference direction for the data.
Hence, changing the orientation of a cell only changes the sign of the associated data.
\Cref{fig:chains} illustrates the bookkeeping of flows with respect to the reference orientation of edges.
For polygons, an orientation can be represented by a cyclic orientation of the boundary.
In embedded orientable settings, this can equivalently be viewed as choosing a normal direction, or an \enquote{upper} side, of the polygon.
The normal direction specifies which cyclic orientation is viewed as clockwise.
The boundary of a volume consists of polygons oriented so that the chosen normal direction points outwards.

Boundaries also express \emph{how} a cell relates to cells it is attached to.
The boundary of a polygon can either contain an edge in accordance with or opposite to its reference orientation.

\section{Cell Complexes of Dimension at most 2}\label{sec:2dim-def}

\begin{figure}[t!]
	\centering
	\begin{minipage}[b]{.15\linewidth}
		\centering
		\includegraphics[width=.9\linewidth]{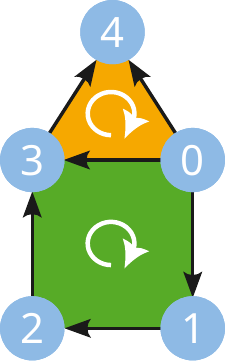}
		Cell Complex
	\end{minipage}
	\hfill
	\begin{minipage}[b]{.57\linewidth}
		\centering
		\begin{equation*}
			\bordermatrix{ &\hspace{-.5pt}0\hspace{-1pt}\shortrightarrow\hspace{-1pt}1 & 0\hspace{-1pt}\shortrightarrow\hspace{-1pt}3 & 0\hspace{-1pt}\shortrightarrow\hspace{-1pt}4 & 1\hspace{-1pt}\shortrightarrow\hspace{-1pt}2 & 2\hspace{-1pt}\shortrightarrow\hspace{-1pt}3 & 3\hspace{-1pt}\shortrightarrow\hspace{-1pt}4\hspace{-.5pt}\cr
				\encircle{0} & \mathbf{-1} & \mathbf{-1} & \mathbf{-1} & \zero & \zero & \zero \cr
				\encircle 1 & \mathbf{1} & \zero & \zero & \mathbf{-1} & \zero & \zero \cr
				\encircle 2 & \zero & \zero & \zero & \mathbf{1} & \mathbf{-1} & \zero \cr
				\encircle 3 & \zero & \mathbf{1} & \zero & \zero & \mathbf{1} & \mathbf{-1} \cr
				\encircle 4 & \zero & \zero & \mathbf{1} & \zero & \zero & \mathbf{1}}
		\end{equation*}
		$B_1$
	\end{minipage}
	\begin{minipage}[b]{.26\linewidth}
		\centering
		\begin{equation*}
			\bordermatrix{ & \text{\includegraphics[height=12pt]{figures/cc-illustration-orange.pdf}} & \text{\includegraphics[height=12pt]{figures/cc-illustration-green.pdf}} \cr
				0\hspace{-1pt}\rightarrow\hspace{-1pt}1 & \zero & \mathbf{1} \cr
				0\hspace{-1pt}\rightarrow\hspace{-1pt}3 & \mathbf{1} & \mathbf{-1} \cr
				0\hspace{-1pt}\rightarrow\hspace{-1pt}4 & \mathbf{-1} & \zero \cr
				1\hspace{-1pt}\rightarrow\hspace{-1pt}2 & \zero & \mathbf{1} \cr
				2\hspace{-1pt}\rightarrow\hspace{-1pt}3 & \zero & \mathbf{1} \cr
				3\hspace{-1pt}\rightarrow\hspace{-1pt}4 & \mathbf{1} & \zero }
		\end{equation*}
		$B_2$
	\end{minipage}
	\caption{The $2$-dimensional toy cell complex introduced in \cref{fig:glueing}. Arrows indicate the reference orientation of $1$- and $2$-cells (edges and polygons). The boundary matrices $B_1$ and $B_2$ show which $(k-1)$-cells are in the boundary of which $k$-cells. The sign of the entry indicates whether the $(k-1)$-cell is in the boundary in its reference orientation ($1$) or opposite to its reference orientation ($-1$). The boundary matrices are annotated with the reference orientations of the cells.}\label{fig:1}
\end{figure}

The vast majority of applications in signal processing and network science use $2$-dimensional cell complexes.
Signals that are observed on edges can be processed using polygons, but so far specific volume or hypervolume data \cite{barbarossa2020topological,schaub2021signal} has not been used.
So far, structural inference work \cite{hoppe2023representing,battiloro2024latent,canbolat2025online}, lifting procedures \cite{bodnar2021weisfeiler}, and generative models \cite{hoppe2024random} are limited to $2$-dimensional cells.
Higher-dimensional cells are difficult to handle due to the increased combinatorial complexity; except in special cases like cubical complexes.
While future work may tackle this successfully, the current state of the art is focused on $2$-dimensional cell complexes.

By limiting the maximum dimension of cells to $2$, we can give a significantly more concise definition of cell complexes that still supports all concepts that are required for most state-of-the-art methods and applications.
Thus, the subsequent chapters can be fully understood with the definition given at the beginning of this section; the reader may skip the general definition in the next section.
In particular, we provide an elementary definition for \emph{regular} cell complexes, a subset of all cell complexes, without the need for any advanced concepts from topology.
If it is clear from the context, we will use \emph{cell complexes} to refer to abstract regular cell complexes from now on.

We will now first introduce a notion of graphs as abstract regular cell complexes of dimension $1$, which we then later use to define cell complexes of dimension $2$.
We do not attempt to replace regular CW complexes as topological objects in algebraic topology.
Instead, we try to isolate the finite boundary-matrix structure that is sufficient for the chain complexes, Hodge Laplacians, homology groups, and signal-processing constructions used in the applied literature and throughout this paper.

\begin{definition}[Abstract regular cell complexes of dimension $1$]
	\label{def:1dCellComplexes}
	An abstract regular cell complex $\cc$ of dimension $1$ consists of an ordered set $\cc_0=\{c_0^1,\dots,c_0^{n_0}\}$, called $0$-\emph{cells}, a non-empty ordered set $\cc_1=\{c_1^1,\dots,c_1^{n_1}\}$, called $1$-\emph{cells}, and a matrix $B_1\in \{0,\pm 1\}^{|\cc_0|\times |\cc_1|}$, called \emph{boundary matrix of dimension $1$} or \emph{first boundary matrix}, such that every column of $B_1$ contains exactly one positive and one negative entry.
\end{definition}
\Cref{fig:1} shows an example of a  \CC{} and its first boundary matrix $B_1$, also known as the \emph{signed incidence matrix}.
In the example, we have five $0$-cells (nodes), $~\encircle{0}~,~\encircle{1}~, \dots, ~\encircle{4}~\in\cc_0$ and six $1$-cells (edges) in $\cc_1$.
The first boundary matrix is then given by
\begin{equation*}
	B_1 = \bordermatrix{ &\hspace{-.5pt}0\hspace{-1pt}\shortrightarrow\hspace{-1pt}1 & 0\hspace{-1pt}\shortrightarrow\hspace{-1pt}3 & 0\hspace{-1pt}\shortrightarrow\hspace{-1pt}4 & 1\hspace{-1pt}\shortrightarrow\hspace{-1pt}2 & 2\hspace{-1pt}\shortrightarrow\hspace{-1pt}3 & 3\hspace{-1pt}\shortrightarrow\hspace{-1pt}4\hspace{-.5pt}\cr
		\encircle{0} & \mathbf{-1} & \mathbf{-1} & \mathbf{-1} & \zero & \zero & \zero \cr
		\encircle 1 & \mathbf{1} & \zero & \zero & \mathbf{-1} & \zero & \zero \cr
		\encircle 2 & \zero & \zero & \zero & \mathbf{1} & \mathbf{-1} & \zero \cr
		\encircle 3 & \zero & \mathbf{1} & \zero & \zero & \mathbf{1} & \mathbf{-1} \cr
		\encircle 4 & \zero & \zero & \mathbf{1} & \zero & \zero & \mathbf{1}}.
\end{equation*}
We can easily check that the condition of \cref{def:1dCellComplexes} is fulfilled and every column of $B_1$ contains exactly one $1$ and one $-1$ entry.
In the corresponding cell complex on the left of \cref{fig:1}, we have an oriented edge for every column of $B_1$ pointing from the $0$-cell (node) corresponding to the $-1$ entry to the $0$-cell corresponding to the $+1$ entry.
We can identify the above construction with oriented multigraphs without self-loops.
We call a closed traversal in such an oriented multigraph a \emph{simple oriented loop}, or simple oriented cycle, if it uses each $1$-cell at most once, visits no $0$-cell twice except for the common start and endpoint, and fixes a cyclic orientation of the traversal.
\begin{definition}[Abstract regular cell complexes of dimension $2$]
	\label{def:2dCellComplexes} 
	An abstract regular cell complex $\cc$ of dimension $2$ consists of an abstract regular cell complex of dimension $1$ together with a non-empty ordered set $\cc_2=\{c_2^1,\dots,c_2^{n_2}\}$, called $2$-cells, and a matrix $B_2\in \{0,\pm 1\}^{|\cc_1|\times |\cc_2|}$, called \emph{second boundary matrix} or \emph{boundary matrix of dimension $2$} such that each column of $B_2$ corresponds to a simple oriented loop in $\cc_1$ where the sign of the non-zero entries corresponds to the orientation of the edges.
\end{definition}
Equivalently, a 2‑dimensional cell complex can be viewed as an oriented (multi)graph with additional structure.
Each 2‑cell is attached along such a simple oriented loop.
The non-zero signs in the corresponding column of $B_2$ record whether the reference-oriented edges agree with this chosen cyclic orientation.
In the case of a multigraph, such a loop may also be formed by two distinct parallel $1$-cells with opposite signs in the boundary column.

\begin{remark}[On the two conditions for candidate $2$-cell boundaries]
Assume that we have a $1$-skeleton of a cell complex and we want to attach a $2$-cell to it.
Adding a $2$-cell means adding a column to the (initially-empty) boundary matrix $B_2$.
However, not every potential column would define a valid $2$-cell, and thus we cannot add arbitrary columns.
Given a candidate column $b$ for $B_2$, there are two separate ways in which it may fail to define the boundary of a valid $2$-cell, corresponding \emph{(1)} to the condition of being a superposition of \emph{oriented cycles}, or \emph{(2)} the condition of being precisely one \emph{simple oriented loop} on top of this.
First, a valid column $b$ must satisfy the boundary-of-boundary condition $B_1b=0$, corresponding to being a superposition of cycles.
For instance, in the $1$-skeleton of \cref{fig:1}, the candidate column
\[
b_{\mathrm{path}}=
\bordermatrix{ & b_{\mathrm{path}}\cr
	0\hspace{-1pt}\rightarrow\hspace{-1pt}1 & \mathbf{1}\cr
	0\hspace{-1pt}\rightarrow\hspace{-1pt}3 & \zero\cr
	0\hspace{-1pt}\rightarrow\hspace{-1pt}4 & \zero\cr
	1\hspace{-1pt}\rightarrow\hspace{-1pt}2 & \mathbf{1}\cr
	2\hspace{-1pt}\rightarrow\hspace{-1pt}3 & \zero\cr
	3\hspace{-1pt}\rightarrow\hspace{-1pt}4 & \zero}
\]
already fails this necessary condition, since $B_1b_{\mathrm{path}}=-\encircle{0}+\encircle{2}\ne 0$.
Thus the selected edges form an oriented path, not a cycle.
For practical applications, this would define an ill-defined $2$-cell where no flow can wrap around the cell.

Second, even satisfying $B_1b=0$ is not enough.
If a $1$-skeleton contains two disjoint oriented cycles $\gamma$ and $\gamma'$, then $B_1(\gamma+\gamma')=0$, but the support of $\gamma+\gamma'$ is disconnected.
Such a column satisfies the algebraic boundary-of-boundary condition, but it is not the boundary of a single $2$-cell in the sense of \cref{def:2dCellComplexes}.
The reason for this is clear: In practice, we want the cells to correspond to connected \enquote{single} entities, where in dimension $2$, a single circular flow can wrap around the $2$-cell.
This is why the definition asks for a simple oriented loop rather than only for a vector in $\ker B_1$, which would suffice if we only cared about producing a chain complex.
\end{remark}

\begin{figure}[t]
	\centering
	\includegraphics[width=\linewidth]{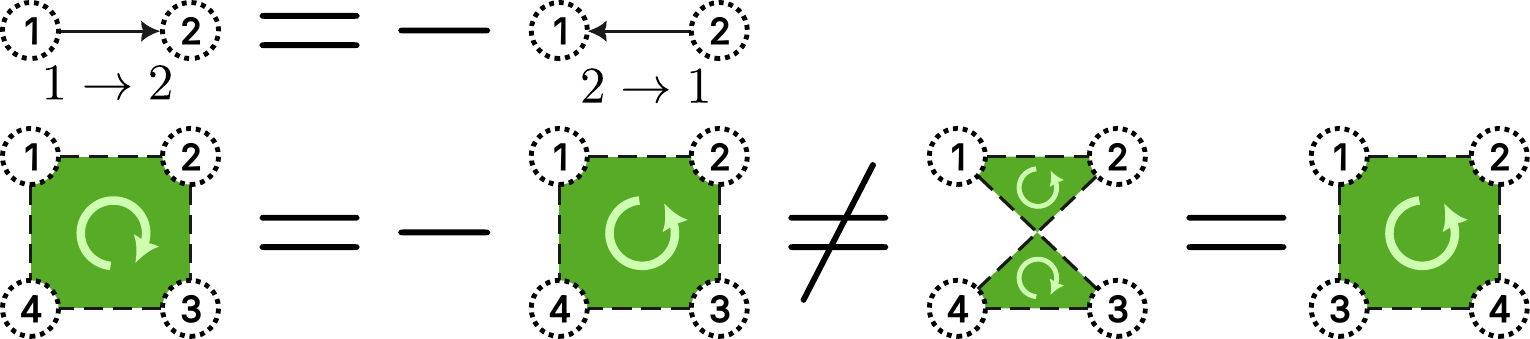}
	
	\caption{\textbf{Different cells with the same vertex set.} \textbf{Top:} A \emph{$1$-cell} can be identified by the $0$-cells it is glued to, and reversing the orientation reverses the \enquote{start} and \enquote{end} $0$-cells. \textbf{Bottom:} \emph{$2$-cell}s are attached to edges, and their orientation may be viewed as designating an \enquote{upper} side of the $2$-cell or, as portrayed here, via the orientation of its boundary. Note that $2$-cells are not uniquely determined by the vertices that define their edges: the center right $2$-cell is completely different from the two on the left as its boundary includes the edges $(2,4)$ and $(1,3)$, but not $(2,3)$ and $(1,4)$ like the cell on the left. The rightmost illustration pictures the same $2$-cell with a different $2$-dimensional embedding of the nodes.}\label{fig:vertices-cells}
\end{figure}

\Cref{fig:1} shows an example of how $2$-cells may be attached to the $1$-dimensional abstract \CC{} introduced earlier via the second boundary matrix
\[
	B_2 = 
	\bordermatrix{ & \text{\includegraphics[height=12pt]{figures/cc-illustration-orange.pdf}} & \text{\includegraphics[height=12pt]{figures/cc-illustration-green.pdf}} \cr
		0\hspace{-1pt}\rightarrow\hspace{-1pt}1 & \zero & \mathbf{1} \cr
		0\hspace{-1pt}\rightarrow\hspace{-1pt}3 & \mathbf{1} & \mathbf{-1} \cr
		0\hspace{-1pt}\rightarrow\hspace{-1pt}4 & \mathbf{-1} & \zero \cr
		1\hspace{-1pt}\rightarrow\hspace{-1pt}2 & \zero & \mathbf{1} \cr
		2\hspace{-1pt}\rightarrow\hspace{-1pt}3 & \zero & \mathbf{1} \cr
		3\hspace{-1pt}\rightarrow\hspace{-1pt}4 & \mathbf{1} & \zero }.
\]
We can easily check that traversing the oriented $1$-cell from $~\encircle{0}~$ to $~\encircle{3}~$, then traversing the oriented $1$-cell from $~\encircle{3}~$ to $~\encircle{4}~$ according to the specified orientation and then traversing the $1$-cell from $~\encircle{0}~$ to $~\encircle{4}~$ in opposite direction forms a simple oriented loop, thus fulfilling \cref{def:2dCellComplexes}.
We say a $1$-cell $c_1^i$ and an attached $2$-cell $c_2^j$ have \emph{matching orientation} if the corresponding entry of the boundary matrix $B_2^{i,j}$ is $1$, and \emph{opposite orientation} if it is~$-1$.
\begin{definition}[Boundary of a cell]
\label{def:boundaryofcell}
The \emph{boundary} $\partial c_k^j$ of a $k$-cell $c_k^j$ is the set of $(k-1)$-cells $c_{k-1}^i$ such that the corresponding entry $B_k^{i,j}$  of the boundary matrix is non-zero.
\end{definition}
The boundary of a cell is therefore not, in general, determined only by the set of incident $0$-cells; \cref{fig:vertices-cells} illustrates this distinction and the effect of reversing orientations.
Equivalently, $\partial c_k^j$ is the support of the oriented boundary chain $B_kc_k^j=\sum_i B_k^{i,j}c_{k-1}^i$; the signs in this chain carry the orientation information.

\begin{remark}[Orientation]
	Having a built-in notion of \emph{orientation} allows for a natural representation of physical data, like flows.
	Given a $k$-cell $c_k^i$, we can view the \emph{orientation} of $c$ as the collection of signs of the non-zero entries $\pm1$ in the associated column $B_k^{-,i}$.
	However, out of all possible assignments of signs, only two can be realized by \CC{}s.
	If we flip the reference orientation of a cell $c_k^i$, thereby creating the oppositely oriented cell $\hat{c}_k^i$, this changes the sign of the corresponding column of $B_k$, i.e.\ $B_k^{-,i} =-\hat{B}_k^{-,i}$.
	If $k<n$, the corresponding row of $B_{k+1}$ has to be changed in the same manner.
\end{remark}
When looking at the first column of the first boundary matrix $B_1$ of the example of \cref{fig:1}, there is a single $1$ at the row corresponding to $~\encircle{1}~$ and a single $-1$ at the row corresponding to $~\encircle{0}~$, corresponding to a $1$-cell oriented from $~\encircle{0}~$ to $~\encircle{1}~$:
\begin{equation*}
\bordermatrix{ &\hspace{-.5pt}0\hspace{-1pt}\shortrightarrow\hspace{-1pt}1\cr
		\encircle{0} & \mathbf{-1}  \cr
		\encircle 1 & \mathbf{1}  \cr
		\encircle 2 & \zero \cr
		\encircle 3 & \zero \cr
		\encircle 4 & \zero }.
\end{equation*}
Analogously, the column of $B_2$ corresponding to the orange $2$-cell is given by

\[\bordermatrix{ & \text{\includegraphics[height=15pt]{figures/cc-illustration-orange.pdf}}  \cr
	0\hspace{-1pt}\rightarrow\hspace{-1pt}1 & \zero  \cr
	0\hspace{-1pt}\rightarrow\hspace{-1pt}3 & \mathbf{1} \cr
	0\hspace{-1pt}\rightarrow\hspace{-1pt}4 & \mathbf{-1}\cr
	1\hspace{-1pt}\rightarrow\hspace{-1pt}2 & \zero  \cr
	2\hspace{-1pt}\rightarrow\hspace{-1pt}3 & \zero \cr
	3\hspace{-1pt}\rightarrow\hspace{-1pt}4 & \mathbf{1} }.
\]
This indicates that this cell is oriented according to the orientation of the $1$-cells $0\hspace{-1pt}\rightarrow\hspace{-1pt}3 $ and $3\hspace{-1pt}\rightarrow\hspace{-1pt}4$ but with opposite orientation of the $1$-cell $0\hspace{-1pt}\rightarrow\hspace{-1pt}4$.
Changing the signs of this column would result in an equivalent $2$-cell with opposite orientation:
\[\bordermatrix{ & \text{\reflectbox{\includegraphics[height=15pt]{figures/cc-illustration-orange.pdf}}}  \cr
	0\hspace{-1pt}\rightarrow\hspace{-1pt}1 & \zero  \cr
	0\hspace{-1pt}\rightarrow\hspace{-1pt}3 & \mathbf{-1} \cr
	0\hspace{-1pt}\rightarrow\hspace{-1pt}4 & \mathbf{1}\cr
	1\hspace{-1pt}\rightarrow\hspace{-1pt}2 & \zero  \cr
	2\hspace{-1pt}\rightarrow\hspace{-1pt}3 & \zero \cr
	3\hspace{-1pt}\rightarrow\hspace{-1pt}4 & \mathbf{-1} }.
\]

We will now introduce vector spaces associated to a cell complex that represent oriented data on the \CC.

\begin{definition}[Signal/Chain space of a cell complex]
	\label{def:SignalSpace}
Given a two-dimensional abstract regular cell complex $\cc$ with cells $\cc_0$, $\cc_1$, and $\cc_2$ and boundary matrices $B_1$ and $B_2$, a \emph{signal on the $k$-cells} $s$ assigns a real number to every $k$-cell in $\cc_k$.
We denote by $C_k$ the $k$-th signal space (also called chain space) of all possible signals on $k$-cells.
We can then identify $C_k$ with $\R^{|\cc_k|}$ with formal basis of the $k$-cells $\cc_k$, where $\alpha c_k^i$ denotes a signal of strength $\alpha\in \R$ on cell $c_k^i\in \cc_k$ and $0$ on all other cells.
For $k>0$, the boundary matrix $B_k$ becomes a linear map between the space of signals on $k$-cells and on $(k-1)$ cells by sending $s_k\in C_k$ to $B_ks_k\in C_{k-1}$.
Thus we obtain an associated sequence of signal/chain spaces
\[
\begin{tikzcd}
C_0&\ar[l,"B_1"]C_1&\ar[l,"B_2"]C_2
\end{tikzcd}
\]
\end{definition}
With this identification, the set-valued boundary from \cref{def:boundaryofcell} is the support of the oriented chain boundary. In particular, for a basis element $c_k^j\in C_k$, the boundary map $B_k\colon C_k\to C_{k-1}$ is given by
\[
	B_k c_k^j=\sum_i B_k^{i,j}c_{k-1}^i .
\]
 
\begin{figure}[t]
	\centering
	\includegraphics[width=\linewidth]{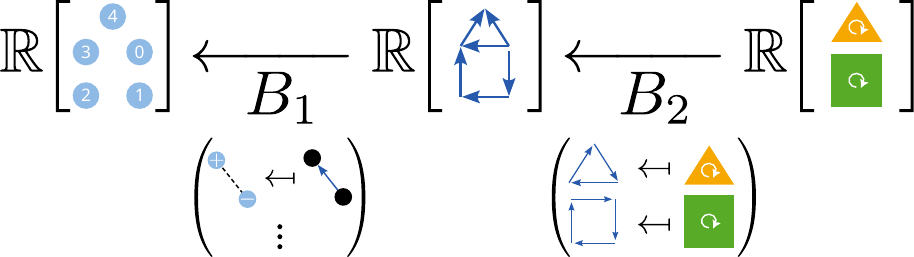}
	\caption{The chain complex of the toy cell complex from \cref{fig:1}. The boundary matrices are the same, but illustrated as maps on the basis elements of the chain spaces.
	}
	\label{fig:chaincomplex}
\end{figure}

\Cref{fig:chaincomplex} illustrates the chain complex of the toy cell complex (and its boundary matrices) from \cref{fig:1}, where $C_0\cong\R^{|\cc_0|}\cong \R^5$ because of the five $0$-cells, $C_1\cong\R^{|\cc_1|}\cong \R^6$ because of the six $1$-cells, and $C_2\cong\R^{|\cc_2|}\cong \R^2$ because of the two $2$-cells.

The above sequence satisfies the conditions of a so-called \emph{chain complex}, namely that $B_1\circ B_2 =0$.
The following lemma verifies this condition.
\begin{lemma}
    For a $2$-dimensional abstract regular cell complex $\cc$ with boundary matrices $B_1$ and $B_2$, every signal $s_2\in C_2$ on the $2$-cells vanishes after applying $B_2$ and $B_1$
    \[
    B_1B_2s_2 = 0.
    \]
\end{lemma}
\begin{proof}
    We only need to show that $B_1 \circ B_2 = 0$:
    Since the boundary maps are linear, we show the claim on the basis elements $\cc_2$ of the signal/chain space $C_2$.
    Let $c \in \cc_2$ be a $2$-cell. From \cref{def:2dCellComplexes}, applying the second boundary matrix to the basis element in $C_2$ corresponding to the $2$-cell $c$, $B_2 c\in C_1$, represents an oriented cycle of edges. This means that for each $0$-cell that is in the boundary of one of these edges, there is also another edge in the support of $B_2c$ with opposite orientation with respect to the $0$-cell. In other words, applying $B_1$ adds one positive and one negative value to obtain $0$.
\end{proof}
We will verify the above claim for the boundary matrices $B_1$ and $B_2$ of our example cell complex in \cref{fig:1}:
\begin{align*}
B_1B_2&= \begin{pmatrix}
	 \mathbf{-1} & \mathbf{-1} & \mathbf{-1} & \zero & \zero & \zero \\
	 \mathbf{1} & \zero & \zero & \mathbf{-1} & \zero & \zero \\
	\zero & \zero & \zero & \mathbf{1} & \mathbf{-1} & \zero \\
	\zero & \mathbf{1} & \zero & \zero & \mathbf{1} & \mathbf{-1} \\
	 \zero & \zero & \mathbf{1} & \zero & \zero & \mathbf{1}
	\end{pmatrix}
	\cdot 
	\begin{pmatrix}
	\zero & \mathbf{1} \\
	\mathbf{1} & \mathbf{-1} \\
	 \mathbf{-1} & \zero \\
	\zero & \mathbf{1} \\
	 \zero & \mathbf{1} \\
	 \mathbf{1} & \zero 
	\end{pmatrix}\\
	&=\begin{pmatrix}
	\zero +(\mathbf{-1})+\mathbf{1}+\zero+\zero+\zero &(\mathbf{-1})+\mathbf{1}+\zero+\zero+\zero +\zero\\
	\zero +\zero+\zero+\zero +\zero +\zero&\mathbf{1}+\zero+\zero +(\mathbf{-1})+\zero+\zero\\
	\zero +\zero+\zero+\zero + \zero +\zero & \zero + \zero + \zero + \mathbf{1}+(\mathbf{-1})+\zero\\
	\zero + \mathbf{1}+\zero + \zero + \zero + (\mathbf{-1}) &\zero +  (\mathbf{-1})+\zero + \zero + \mathbf{1}+\zero \\
	\zero + \zero + (\mathbf{-1})+\zero + \zero + \mathbf{1} & \zero+\zero + \zero + \zero + \zero + \zero
	\end{pmatrix}\\
	& = \begin{pmatrix}
	\zero&\zero\\
	\zero&\zero\\
	\zero&\zero\\
	\zero&\zero\\
	\zero&\zero
	\end{pmatrix}.
\end{align*}
\begin{remark}[Intuition behind boundary operators on data]
	We will now briefly discuss the physical meaning behind applying boundary operators on data:
	We will start with the first boundary operator $B_1$ and a signal supported on the oriented edges $s\in C_1$.
	Applying $B_1$ to $s$ will yield a signal on the vertices $B_1s\in C_0$.
	In particular, $B_1$ assigns every $0$-cell the sum of the signals on the incident $1$-cells, multiplied by $\pm 1$ depending on the orientation.
	Let us take the cell complex of \cref{fig:1} and start with an edge signal/flow of $1$ on every edge in the direction of the reference orientation, i.e.\ $s=(1,1,1,1,1,1)^\top$.
	Now we can compute 
	\begin{align*}
	B_1s&= \bordermatrix{ &\hspace{-.5pt}0\hspace{-1pt}\shortrightarrow\hspace{-1pt}1 & 0\hspace{-1pt}\shortrightarrow\hspace{-1pt}3 & 0\hspace{-1pt}\shortrightarrow\hspace{-1pt}4 & 1\hspace{-1pt}\shortrightarrow\hspace{-1pt}2 & 2\hspace{-1pt}\shortrightarrow\hspace{-1pt}3 & 3\hspace{-1pt}\shortrightarrow\hspace{-1pt}4\hspace{-.5pt}\cr
		\encircle{0} & \mathbf{-1} & \mathbf{-1} & \mathbf{-1} & \zero & \zero & \zero \cr
		\encircle 1 & \mathbf{1} & \zero & \zero & \mathbf{-1} & \zero & \zero \cr
		\encircle 2 & \zero & \zero & \zero & \mathbf{1} & \mathbf{-1} & \zero \cr
		\encircle 3 & \zero & \mathbf{1} & \zero & \zero & \mathbf{1} & \mathbf{-1} \cr
		\encircle 4 & \zero & \zero & \mathbf{1} & \zero & \zero & \mathbf{1}}
	\begin{pmatrix}
		1\\
		1\\
		1\\
		1\\
		1\\
		1
	\end{pmatrix}\\
	&=\begin{pmatrix}
	-1-1-1\\
	1-1\\
	1-1\\
	1+1-1\\
	1+1
	\end{pmatrix}
	=
	\begin{pmatrix}
		-3\\
		0\\
		0\\
		1\\
		2
	\end{pmatrix}
\end{align*}
	
	This corresponds to the net-inflow at every node, with a flow of $3$ originating at node $\encircle{0}$.
	We can easily verify that this sums to $0$.
	Analogously, we can consider a signal $t$ of strength $1$ in direction of the reference orientation on both $2$-cells \celltri{} and \cellsq{} of \cref{fig:1}.
	The boundary matrix $B_2$ transforms this signal into the superposition of edge flows of strength $1$ along the boundary of the $2$-cells, which we can verify computationally:
	\[
	B_2t=\bordermatrix{ & \text{\includegraphics[height=12pt]{figures/cc-illustration-orange.pdf}} & \text{\includegraphics[height=12pt]{figures/cc-illustration-green.pdf}} \cr
		0\hspace{-1pt}\rightarrow\hspace{-1pt}1 & \zero & \mathbf{1} \cr
		0\hspace{-1pt}\rightarrow\hspace{-1pt}3 & \mathbf{1} & \mathbf{-1} \cr
		0\hspace{-1pt}\rightarrow\hspace{-1pt}4 & \mathbf{-1} & \zero \cr
		1\hspace{-1pt}\rightarrow\hspace{-1pt}2 & \zero & \mathbf{1} \cr
		2\hspace{-1pt}\rightarrow\hspace{-1pt}3 & \zero & \mathbf{1} \cr
		3\hspace{-1pt}\rightarrow\hspace{-1pt}4 & \mathbf{1} & \zero }
\begin{pmatrix}
	1\\
	1
\end{pmatrix}=\begin{pmatrix}
1\\
0\\
-1\\
1\\
1\\
1
\end{pmatrix}.
	\]
	Because of the opposing orientations of the $2$-cells at edge $0\shortrightarrow 3$, the flow cancels out and the edge gets assigned a flow of $0$, whereas the resulting flow of $B_2t$ corresponds to a circular flow around the boundary of the union of \celltri{} and \cellsq.
	
\end{remark}
\begin{remark}[Intuition behind adjoints $B_k^*$ of boundary operators on data.]
	\label{rmk:coboundary}
	The boundary matrices return data in $1$ dimension below the input dimension, i.e., it takes a signal supported on $k$-cells and returns a signal on $(k-1)$-cells.
	Now, for a standard inner product on the signal spaces $C_k$, the adjoint or transpose of the boundary matrices $B_k^\top$ send data in the opposite direction.
	We will call the adjoints of the boundary matrices the \emph{coboundary matrices} and sometimes write them as $B^k=B_k^*$.
	Given any signal on $k$-cells $s_k$ and any signal on $(k-1)$-cells $s_{k-1}$, the adjoint $B_k^*$ of $B_k$ is defined by the following relationship:
	\[
	\langle B_k s_k,s_{k-1}\rangle_{C_{k-1}}=	\langle s_k,B_k^* s_{k-1}\rangle_{C_{k}}.
	\]
	More explicitly, for a signal $s_0$ on $0$-cells, $B_1^* s_0$ is a signal on the $1$-cells that assigns every $1$-cell a flow corresponding to the difference of $s_0$ on the two $0$-cells (nodes).
	Thus, $B_1^* s_0$ is the \emph{gradient flow} corresponding to the \emph{node potential} given by $s_0$.
	In our case, the adjoint $B_k^*$ and the transpose $B_k^\top$ coincide.
	
	Now, given a signal $s_1$ on $1$-cells (edges), $B_2^\top s_1$ is a signal on $2$-cells that assigns every $2$-cell the sum of edge flows along its boundary, respecting orientation in the form of signs.
	If $s_1$ contains a strong circular flow around a $2$-cell $c_2^i$, the respective value of $c_2^i$ in $B_2^\top s_1$ will be high.
	In contrast, if the flow of $s_1$ is zero around $c_2^i$, or if the flow divides along two sides of $c_2^i$, then the respective value of $c_2^i$ in $B_2^\top s_1$ will be small.
\end{remark}
We will now introduce the cell complex analogue of simple graphs.
This subclass appears often in applications and allows for a very easy notation akin to graphs and simplicial complexes:
\begin{definition}[Simple \CC{}s]
    A cell complex $\cc$ is called \emph{simple} if all $k$-cells for $k>0$ are uniquely determined by their image under the boundary maps up to change of orientation:
    \[\forall c^i_k,c_k^j\in \cc_k :B_k c^i_k = \pm B_k c^j_k \Rightarrow c^i_k = c^j_k.\]
\end{definition}
Because the cells in simple \CC{}s are uniquely determined by their boundaries, we can reference them using their boundaries instead of having to use non-descriptive names like $c_k^i$.
The cell complex from our running example of \cref{fig:1} is a simple \CC{}, as no two $1$-cells or $2$-cells have the same oriented boundary up to sign.
This would change if we would add another $1$-cell with corresponding column in $B_1$ of $(-1,1,\zero,\zero,\zero)^\top$, as then there would be two $1$-cells between $0$-cells $~\encircle{0}~$ and $~\encircle{1}~$.
This would even be the case for an addition of a cell with corresponding column $(1,-1,\zero,\zero,\zero)^\top$ because of the $\pm$ in the definition.
The following remark introduces some notation for simple \CC{}s.
\begin{remark}[Notation for simple \CC{}s]
		\label{rem:CanonicalNotation}
    In a simple \CC{}, a $1$-cell $c_1^i$ with boundary $B_1 c_1^i = c_0^2- c_0^1$ can be identified with a tuple $(c_0^1,c_0^2)$ or \enquote{$c_0^1 \shortrightarrow c_0^2$}.
    Similarly, a $2$-cell $c_2$ with a simple cycle $ c_0^1 \shortrightarrow c_0^2 \shortrightarrow \dots \shortrightarrow c_0^{n} \shortrightarrow c_0^1$ as boundary can be represented as a tuple $c_2=(c_0^1, c_0^2, \dots, c_0^{n})$.
    With this notation, any cyclic permutation of $(c_0^1, c_0^2, \dots, c_0^{n})$ refer to the same $2$-cell $c_2$.
   When identifying cells with their corresponding basis elements in the chain spaces $\R^{|\cc_k|}$,
    we can write $(c_0^1,c_0^2)=-(c_0^2,c_0^1)$ for edges or $(c_0^1, c_0^2, \dots, c_0^{n})=-(c_0^n,c_0^{n-1},\dots,c_0^1)$ for $2$-cells to express the relations between different orientations of the same cell, implicitly working in the chain space $C_k$.
    There is just one point we need to be careful about: Because we define cell complexes using explicit boundary matrices $B_k$, there always needs to be \emph{one} reference orientation $(c_0^1,c_0^2)$ or $(c_0^1,c_0^2,\dots,c_0^n)$ for every cell that represents the orientations imposed by $B_k$.
\end{remark}
Because the \CC{} of \cref{fig:1} is a simple \CC{}, we can use the above naming scheme:
When simply referring to the $0$-cells as integers $0,1,\dots , 4$, the $1$-cell $0\shortrightarrow 1$ then is called $(0,1)$.
Furthermore, we can refer to the orange $2$-cell \celltri{} as $\text{\celltri}=(0,3,4)$.
Equivalently, both the names $(3,4,0)$ and $(4,0,3)$ refer to the same $2$-cell, whereas $(0,1,2,3)$ refers to the green $2$-cell \cellsq.

For simple \CC{}s, the ordering on the $0$-cells imposed by \cref{def:1dCellComplexes} lets us choose a canonical reference orientation for $1$-cells and $2$-cells.
\begin{remark}[Canonical orientation of $1$- and $2$-cells in simple \CC{}s]
	\label{rem:CanonicalOrientation}
    A $1$-cell $c_1=(c_0^i,c_0^j)$ is in canonical orientation if $i < j$.
    A $2$-cell $c_2=(c_0^{i_1}, \dots, c_0^{i_n})$ is in canonical orientation if the first $0$-cell is the smallest ($i_1 = \min_j i_j$) and the direction of traversal continues with the smaller neighbor of the smallest node, i.e., $i_2 < i_n$.
    We note that when considering the $2$-cell $(c_0^1,c_0^3,c_0^2,c_0^4)$, there is no orientation of the cell in a purely ascending order, i.e.\ $(c_0^1,c_0^2,c_0^3,c_0^4)$, as nodes $c_0^2$ and $c_0^1$ are not neighbours.
\end{remark}
We can easily verify that the  $1$-cells
$(0,1)$, $(0,3)$, $(0,4)$, $(1,2)$, $(2,3)$, $(3,4)$
and the $2$-cells $(0,3,4)$ and $(0,1,2,3)$ 
 of \cref{fig:1} are in canonical orientation.
 
 In the following \cref{sec:def}, we will give an extension of the above definition to arbitrary dimensions, again directly checking all conditions on the boundary matrices, as the primary interface for cell complex applications.
 We invite the application-oriented reader to skip the technical details of the definition and the following section, as the theory already developed in the current section will suffice for the vast majority of applications.
After the theory-heavy section, we will give some examples of how cell complexes can be used to model real-world data in \cref{sec:examples}.

\section{Abstract Regular Cell Complexes of Arbitrary Dimension}
\label{sec:def}

In this section, we will give a purely combinatorial and algebraic definition of what we call abstract regular cell complexes.
In particular, the definition will only depend on the computational objects we will work with in applications, namely the boundary matrices (and derived operators).
Thus, the definition will not rely on any continuous or geometric structure, and can be explicitly verified for any given set of cells and boundary matrices.
This will come at a cost: while all finite regular cell complexes will fulfil our definition of abstract regular cell complexes, there are some abstract regular cell complexes that do not have a topological realization as regular cell complexes in higher dimensions.
This is unavoidable, as a topological space $X$ being a regular cell complex depends on the boundaries of the cells of $X$ being homeomorphic to spheres, which is a condition that cannot be verified computationally in higher dimensions \cite[Section 10, a theorem of S.P.\ Novikov]{volodin1974problem}.
However, for all applications we have in mind the definition we will introduce will be sufficient.
In \cref{sec:historical-definitions}, we will give an overview of different historical definitions that led to the definition of regular \CW{} complexes, which we will discuss in \cref{sec:topdefinition}.

In 2D, being a \enquote{valid boundary} can be checked graph-theoretically simply by checking whether a sequence of edges forms a simple cycle.
In higher dimensions, this condition of specifying \enquote{valid boundaries} becomes more complicated, as the boundary of a $k$-cell of \CW{} complexes is a $(k-1)$-dimensional sphere.
In our definition, we will instead use the homology as a computable algebraic surrogate:

\begin{definition}[Abstract regular cell complex]
	\label{def:abstractcellcomplex}
		Let $n\in\Z_{\geq 0}$ be a non-negative integer and $\cc=(\cc_*,B_*)$ a pair of a finite sequence of ordered non-empty sets $\cc_k = \{c_k^1,c_k^2, \dots\}$ for $0\le k\le n$ called \emph{cells} and matrices $B_k\in\{0,\pm 1\}^{|\cc_{k-1}|\times |\cc_{k}|}$ for $1\le k\le n$ called \emph{boundary matrices}. Given a cell $c_k^i$, we call the set of cells $c_{k-1}^j$ associated to the non-zero entries $B_k^{j,i}$ of the $i$-th column of $B_k$ the \emph{boundary} $\partial c_k^i$ of $c_k^i$ and write $\partial S =\bigcup_{c_k^i\in S}\partial c_k^i$ for a set of cells $S$.
	We now call $\cc$ an \emph{abstract regular cell complex} if 
	\begin{enumerate}
			\item If $n\geq 1$, all columns of $B_1$ have exactly one positive and one negative entry. \label{cond:B1columns}
			\item For every cell $c^i_k\in\cc_k$ of positive dimension $1\leq k\leq n$ and sets of cells $\hat{\cc}_{k-r}=\partial^{r}\{c^i_k\}$ for $0\le r\le k$,
		the associated pair $(\hat{\cc}_*,\hat{B}_*)$ with boundary matrices $\hat{B}_l\colon \Z^{|\hat{\cc}_l|}\to\Z^{|\hat{\cc}_{l-1}|}$ restricted from $B_l$ fulfills
		\label{cond:individualcells}
		\begin{enumerate}
			\item $\ker \hat{B}_k =0$ and $\ker \hat{B}_{l-1} = \Ima \hat{B}_l$ for all $2\le l\le k$,\label{cond:trivialhomology}
			\item $\Z^{|\hat{\cc}_0|}/\Ima \hat{B}_1\cong \Z$.\label{cond:connected}
		\end{enumerate}
	\end{enumerate}
\end{definition}

\begin{table}[t]
	\centering
	\footnotesize
	\setlength{\tabcolsep}{3pt}
	\begin{tabular}{p{.34\linewidth}p{.25\linewidth}p{.32\linewidth}}
		\toprule
		Intuition for the generated subcomplex & Homology statement & Algebraic condition \\
		\midrule
		It is connected. & $H_0(\hat{\cc})\cong \Z$ & $\Z^{|\hat{\cc}_0|}/\Ima \hat{B}_1\cong \Z$ \\
		It has no holes below the top dimension. & $H_{l-1}(\hat{\cc})=0$ for $2\le l\le k$ & $\ker \hat{B}_{l-1}=\Ima \hat{B}_l$ for $2\le l\le k$ \\
		The top cell is not a cycle by itself. & $H_k(\hat{\cc})=0$ & $\ker \hat{B}_k=0$ \\
		\bottomrule
	\end{tabular}
	\caption{\textbf{Local homology conditions for one cell.}
	For a cell $c_k^i$, the subcomplex $\hat{\cc}$ generated by $c_k^i$ and its iterated boundaries is required to have the homology of a point, as the closure of a topological cell would.
	In homology conditions, we can effectively combine the bottom two rows; as we chose with concrete boundary matrices and there is no $\hat{B}_{k+1}$, we have to write the top-dimensional condition separately.
	}
	\label{tab:local-homology-condition}
\end{table}
We will give more details on how to explicitly check these conditions in \cref{sec:checking-arcc}.
In dimension $n=0$, we will only have a set of $0$-cells $\cc_0$ and no boundary matrices.
\begin{remark}[On the definition of abstract cell complexes]
	We will now try to give an intuition behind the above definition.
	The integer $n$ denotes the dimension, i.e. the highest dimension of cells of the \CC{} $\cc$.
	The set $\cc_k$ is the set of abstract $k$-cells which we assume is ordered and finite.
	The order on $\cc_k$ gives a canonical association between entries $B_k^{i,j}$ of the boundary matrix $B_k$ and pairs of cells $(c_{k-1}^i,c_k^j)\in\cc_{k-1}\times \cc_k$.
	The entry $B_k^{i,j}$ then encodes whether $c_{k-1}^i$ is not contained in the boundary of $c_k^j$ (i.e.\ $B_k^{i,j}=0$), or is contained in the boundary with matching orientation (i.e.\ $B_k^{i,j}=1$) or opposite orientation (i.e.\ $B_k^{i,j}=-1$).
	We can view this definition as building a cell complex iteratively:
	We start with $0$-cells, which are just abstract points without any conditions.
	In the next step we will attach $1$-cells $c_1^j$ according to the entries $B_1^{-,j}$, which assigns a $\pm1$ to the cells on the boundary of $c^j_1$.
	This then continues for arbitrary $k$-cells which are attached to $(k-1)$-cells.
		
		However, the definition does not allow for arbitrary attachments.
		In the usual definition of topological regular cell complexes, one attaches $k$-disks, i.e.\ $D^k=\{x\in\R^k : |x|\le1\}$.
		Condition~\ref{cond:individualcells} checks the homological consequences of a closed cell being a disk.
		For every cell $c_k^i$, we consider the minimal cell complex $\hat{\cc}$ containing $c_k^i$, i.e.\ $c_k^i$, its boundary, the boundary of its boundary, etc.
	To this, we can associate a cellular chain complex
	\[
	\begin{tikzcd}
		\Z ^{|\hat{\cc}_0|}&\ar[l, "\hat{B}_1"']\Z^{|\hat{\cc}_1|} &\ar[l, "\hat{B}_2"']\dots &\ar[l, "\hat{B}_{k-1}"']\Z^{|\hat{\cc}_{k-1}|} & \ar[l,"\hat{B}_k"'] \Z&\ar[l] 0.
	\end{tikzcd}
	\]
	The algebraic requirements in Condition \ref{cond:individualcells} say that this chain complex has the homology of a point, as summarized in \cref{tab:local-homology-condition}.
	Thus, Condition \ref{cond:connected}, $\Z^{|\hat{\cc}_0|}/\Ima \hat{B}_1\cong \Z$, asks for the generated subcomplex to be connected, while Condition \ref{cond:trivialhomology} rules out positive-dimensional homology inside it.
	Because it is not computationally possible to algorithmically determine whether a given simplicial complex is an $n$-sphere for general~$n$ \cite[Section 10, a theorem of S.P.\ Novikov]{volodin1974problem}, these homology conditions cannot be equivalent to the homeomorphism condition checked by \CW{} complexes.
	However, for the computational aspects in this paper, they are sufficient and much better behaved.
	Finally, condition \ref{cond:B1columns} is a bookkeeping condition: There is no canonical orientation for $k$-cells for $k\ge 1$ and we can thus flip their orientation by multiplying the associated entries by $-1$ in $B_k$ and $B_{k+1}$.
	However, the \enquote{orientation} of $0$-cells corresponds to a scalar and we can enforce a consistent orientation by requiring every edge to have one head (entry $+1$) and one tail (entry $-1$). (Condition \ref{cond:individualcells} already ensures all $1$-cells to connect precisely two $0$-cells.)
	
\end{remark}

\begin{definition}[Signal/Chain space of a cell complex]
	Given an abstract regular cell complex $(\cc_*,B_*)$ of dimension $n$, we have an associated sequence of \emph{signal/chain spaces} $C_k=\R^{|\cc_k|}$:
	\[
	\begin{tikzcd}
		C_0&\ar[l,"B_1"]C_1&\ar[l,"B_2"]\dots &\ar[l,"B_{n-1}"]C_{n-1}&\ar[l,"B_{n}"]C_n
	\end{tikzcd}
	\]
\end{definition}
The $k$-th signal space $C_k$ encodes real-valued signals on the $k$-cells of $\cc$.
I.e., we can think of $C_0$ as node signals, elements of $C_1$ represent oriented signals on the edges like flows, and so on.
We will identify $\cc_k$ with the canonical basis of $\R^{|\cc_k|}$ and thus think of $c_k^i=e_i\in \R^{|\cc_k|}=C_k$ being the $i$-th basis vector.
The matrices $B_k$ are then boundary maps $B_k\colon C_k=\R^{|\cc_k|}\to C_{k-1}=\R^{|\cc_{k-1}|}$ between the signal spaces.
When looking at the combined chain space $\R^{|\cc|}=\R^{|\cc_0|}\oplus \dots\oplus \R^{|\cc_n|}$, we can view the boundary maps $B_k$, extended by zero to the other summands, as the combined boundary matrix $\mathbf{B}$.
The above sequence satisfies the conditions of being something called a \emph{chain complex}, namely that $B_{k-1}\circ B_k =0$ for $2\le k\le n$, or equivalently $\mathbf{B}^2=0$.
The following lemma verifies this condition.
\begin{lemma}
	The above definition of chains on \CC{}s together with the boundary maps forms a chain complex $(C_*, B_*)$.
\end{lemma}
\begin{proof}
	We will identify a cell $c_k^i$ with the associated basis element in $\R^{|\cc_k|}$.
	It suffices to show that $B_{k-1}B_kc_k^i=0$ for arbitrary cells $c_k^i$ for $2\le k\le n$.
	Because of the definition, we can work in the smallest subcomplex $\hat{\cc}$ containing $c_k^i$ with associated chain spaces
	\[
	\begin{tikzcd}
		\dots&\ar[l,"\hat{B}_{k-2}"]\hat{C}_{k-2}&\ar[l,"\hat{B}_{k-1}"]\hat{C}_{k-1}&\ar[l,"\hat{B}_k"]\hat{C}_k= \langle c_k^i\rangle\cong \R
	\end{tikzcd}
	\]
	Because of condition \ref{cond:trivialhomology} of \cref{def:abstractcellcomplex}, we have that
	\[
	\ker \hat{B}_{k-1}=\Ima \hat{B}_k
	\]
	and thus $0=\hat{B}_{k-1} \hat{B}_kc_k^i=B_{k-1} B_kc_k^i$ which concludes the proof.
\end{proof}
We now compare abstract regular cell complexes with regular \CW{} complexes.
Every finite regular \CW{} complex induces an abstract regular cell complex with the same cellular boundary matrices.
Conversely, in dimensions at most $2$, every abstract regular cell complex can be realized by a regular \CW{} complex with the same boundary matrices.
The obstruction to such a converse statement in higher dimensions is discussed after the theorem.
\begin{theorem}
	\label{thm:equivalence}
	For every finite regular \CW{} complex $X$ there is an abstract regular cell complex $\cc_X$ with the same boundary matrices. For every abstract regular cell complex $\cc$ of dimension up to $2$ there is a regular \CW{} complex $X_\cc$ with the same boundary matrices.
\end{theorem}
\begin{proof}
	We first assume that we are given a regular \CW{} complex $X$ with cells $\cc^X_k$.
	We can choose an arbitrary orientation and ordering of cells to obtain boundary matrices $B_k^X$.
	We now construct our abstract regular cell complex by setting $\cc_k=\cc^X_k$ and use the same ordering as the boundary $B_k^X$.
	Furthermore, we set $B_k=B_k^X$.
	We claim that $\cc=(\cc_*,B_*)$ is an abstract regular cell complex.
	In regular cell complexes, all attaching maps are homeomorphisms onto their image. Thus, their mapping degrees are $\pm 1$ and hence the entries of the boundary matrices are in $\{0,\pm 1\}$.
	Condition \ref{cond:B1columns} of \cref{def:abstractcellcomplex} means that every $1$-cell is connected to exactly one $0$-cell with degree $1$ and one $0$-cell with degree $-1$, which holds by convention.
	Thus condition \ref{cond:individualcells} remains.
	Because the closure of every regular $k$-cell, together with its iterated boundary faces, is homeomorphic to a closed disk, its associated homology is trivial except in degree $0$ where it is $\Z$.
	This is equivalent to the second condition and we conclude the first part of the proof.
	
	Now we assume we are given an abstract regular cell complex $\cc=(\cc_*,B_*)$ of dimension up to $2$ and we want to construct a regular \CW{} with the same boundary matrices.
		For the first two dimensions, this is straightforward: For every $0$-cell in $\cc_0$ we add a point to the $0$-skeleton of $X$, and for every $1$-cell $c_1^i$ of $\cc_1$ we attach an interval to the two $0$-cells specified by the column $B^{-,i}_1$ of the boundary matrix.
		Now we consider a $2$-cell $c_2^i\in\cc_2$.
		We have that $B_2c_2^i=\sum \pm c_1^j$ and $B_1B_2c_2^i=B_1\sum \pm c_1^j =0$.
		Thus the selected oriented $1$-cells define an integral $1$-cycle in $X_1$.
		By the graph-theoretic argument spelled out in \cref{rmk:twoways}, using connectedness and trivial homology of the subcomplex generated by $c_2^i$, this cycle is supported on a single simple cycle.
		We can then attach a cell $\bar{c}_2^i$ to $X_1$ by an attaching map to this simple cycle.
	Doing this for all cells $c_2^i$ gives us a $X$ with the desired cells and boundary matrices.
\end{proof}
The above theorem means two things:
\begin{enumerate}
	\item In the setting of most cell complex applications, we will be in dimension $2$ for which the above theorem shows an equivalence of definitions. In this case, we can even use the simpler \cref{def:2dCellComplexes}.
	\item If our cell-complex construction is geometry-based, the resulting combinatorial objects in our data representation will fulfill the conditions of \cref{def:abstractcellcomplex} and thus be abstract regular cell complexes.
\end{enumerate}
We will conclude the chapter with two remarks.
In \cref{rmk:twoways}, we will shine more light onto the ways in which we do not (and partially cannot) capture the full extent of the topological definition.
Finally, in \cref{rmk:NonregularAbstractCellComplexes}, we will discuss why non-regular cell complexes do not make a good computational model, and why definition attempts in terms of their boundary matrices are not fruitful.
\begin{remark}[Obstructions beyond dimension two]
\label{rmk:twoways}
	There are three separate obstructions to extending the converse direction in \cref{thm:equivalence} to a concise boundary-matrix characterization in all dimensions.
	The first obstruction is about local structure of candidate cells: starting with $3$-cells, homology does not detect whether the candidate boundary has the local manifold structure of a sphere.
	The second obstruction is more involved: starting with $4$-cells, even a manifold boundary with the homology of a sphere need not be a sphere.
	The third obstruction is computational: for $d\ge5$, exact sphere recognition for finite $d$-dimensional simplicial complexes is undecidable \cite[Section 10, a theorem of S.P.\ Novikov]{volodin1974problem}.

	These obstructions are absent for $2$-cells.
	If $b=\hat{B}_2c_2$ is the boundary of a $2$-cell in the subcomplex generated by $c_2$, then $\hat{B}_1b=0$ says that $b$ is an integral $1$-cycle supported on the selected $1$-cells.
	The connectedness condition in \cref{cond:connected} says that the selected graph is connected, and \cref{cond:trivialhomology} says that $\ker \hat{B}_1=\Ima \hat{B}_2=\Z b$.
	Thus the selected graph has a one-dimensional cycle space, generated by a cycle using every selected edge.
	If there were a tree branch, it would contain a bridge, but every element of $\ker \hat{B}_1$ has coefficient $0$ on a bridge.
	Since all selected edges occur in $b$ with coefficient $\pm1$, this cannot happen.
	Hence the boundary is a simple cycle, i.e. a cell decomposition of $S^1$.

	For a $3$-cell, the boundary is two-dimensional and this argument no longer works.
	The conditions of \cref{cond:individualcells} ensure that the closure of the candidate cell has the homology of a point.
	This forces the candidate boundary to have the homology of $S^2$.
	However, this does not force the boundary to be a surface.
	Several $2$-cells may meet along the same $1$-cell, which is not a manifold.
	Thus the missing information is local manifold information, not the chain-complex condition $B_{k-1}B_k=0$ or the homology of the generated chain complex.
	We give an example of this in \cref{subsec:nonmanifold-3cell-boundary}.

	For a $4$-cell, there is an additional global obstruction.
	Even if a candidate boundary is known to be a manifold, i.e. behaves as desired locally, having the homology of $S^3$ does not imply being homeomorphic to $S^3$.
	The Poincaré homology sphere is the standard example: it is a closed $3$-manifold with the same integral homology as $S^3$, but with non-trivial fundamental group.
	It cannot be the boundary of a regular $4$-cell, whose boundary has to be homeomorphic to $S^3$.
	Hence, even after local manifoldness has been checked, the exact condition is no longer a homology computation, but a sphere recognition problem.
	The Novikov obstruction shows that this problem cannot be solved in high dimensions.
	If we had an algorithmic boundary-matrix characterization equivalent to regular \CW{} complexes in all dimensions, we could apply it to the incidence data of an arbitrary finite $d$-dimensional simplicial complex $K$ together with one formal $(d+1)$-cell whose intended boundary is $K$.
	If $K$ were homeomorphic to $S^d$, this top cell could be regular; otherwise it could not be.
	Such a test would therefore decide whether $K$ is homeomorphic to $S^d$.
	For $d\ge5$, this would decide the sphere recognition problem, contradicting Novikov's undecidability result.

	The definition in \cref{def:abstractcellcomplex} therefore takes an, in the eyes of the authors, useful middle ground: it keeps the data representation used in applications, is directly checkable, and enforces the homological conditions needed for the chain complex, the Hodge Laplacians, and the homology groups used later.
\end{remark}
\begin{remark}[Non-regular abstract cell complexes]
	\label{rmk:NonregularAbstractCellComplexes}
	We note that the above definition is for abstract \emph{regular} cell complexes, i.e. an algebraic version of \emph{regular} cell complexes.
	This is not a problem for the goal of this paper, as both applications in signal processing as well as in network science call for \emph{regular} cell complexes.
	We do not know an analogous abstract yet concise definition of (non-regular) cell complexes.
	The main obstacle to this is that while the chain complex of regular cell complexes determines combinatorially which cells attach to which other cells, this is not the case for general (non-regular) cell complexes.
	For example, the standard non-regular \textsc{cw} representation of a $2$-torus contains one $0$-cell, two $1$-cells, and one $2$-cell and is, on the level of chain complexes, indistinguishable from the wedge of a $2$-sphere with two $1$-spheres.
	However, as every cell complex is homotopy equivalent to some regular cell complex (with potentially more cells), this does not present a problem in practice.
\end{remark}

\section{On How to Find a Cell Complex in the Wild} \label{sec:examples}
In many cases, we do not get access to data that is already in the form of cell complexes.
However, a closer look often reveals underlying topological and geometrical structures which are exploitable and representable by cell complexes.
In this section, we will introduce the most important techniques by revealing the hidden cell complex structure in different data sets and modalities.
We hope that this enables the reader to employ a topological and geometric perspective on data analysis for many downstream applications.
\Cref{tab:cc-wild-methods} gives an overview of the different methods discussed in the following.
\begin{table}[t]
	\centering
	\footnotesize
	\begin{tabular}{p{.25\linewidth}p{.35\linewidth}p{.3\linewidth}}
		\toprule
		Construction principle & Methods & Natural input \\
		\midrule
		\hyperref[subsec:cc-wild-special-cases]{Established special cases} & Simplicial complexes, cubical complexes & Existing complex data \\
		\hyperref[subsec:cc-wild-metric-proximity]{Geometric information} & Vietoris--Rips, Delaunay, and alpha complexes & Point clouds (with a scale parameter) \\
		\hyperref[subsec:cc-wild-product-structure]{Product structure} & Products of graphs, simplicial complexes, or \CC{}s & Separable coordinates, space--time domains \\
		\hyperref[subsec:cc-wild-planar-structure]{Planar embeddings and parcellations} & Window lifting, spherical closure, and dual complexes & Embedded planar graphs, tessellations, and physical partitions \\
		\hyperref[subsec:cc-wild-graph-cycle-structure]{Graph-cycle structure} & Chordless cycles, cycle bases, 	spanning-tree liftings & Arbitrary graphs without trusted geometry \\
		\hyperref[subsec:cc-wild-signal-task-structure]{Signal/task structure} & Sparse-flow cell inference, online topology identification, differentiable cell-complex modules & Edge flows, trajectories, supervised tasks \\
		\hyperref[subsec:cc-wild-null-generative-structure]{Random \CC{} models} & Random liftings and generative models & Synthetic data and existing graphs \\
		\bottomrule
	\end{tabular}
	\caption{\textbf{Ways to obtain cell complexes from data.} 
	We will give an overview of different methods to obtain cell complexes from data in this section, which can roughly be divided into the above categories.
	}\label{tab:cc-wild-methods}
\end{table}

\subsection{Established special cases}\label{subsec:cc-wild-special-cases}

\paragraph{Simplicial complexes are cell complexes}
Cell complexes, and in particular \textsc{arcc}s, generalise simplicial complexes.
In short, simplicial complexes only permit the simplest possible $k$-cells, that is edges for $1$-cells, triangles for $2$-cells, tetrahedra for $3$-cells and so on.
Thus, all techniques discussed in this paper are applicable to simplicial complexes as well.
This means that \CC{}s are able to model more complex data, whereas simplicial complexes enjoy a simple definition and combinatorial structure.
\begin{definition}[Simplicial Complex]
	A \emph{finite simplicial complex} $\mathcal{S}=(V,S)$ is a pair of a finite set $V$ called vertices and a set $S$ of non-empty finite subsets of $V$ called \emph{simplices} such that
	\begin{enumerate}
		\item $S$ is closed under taking non-empty subsets and
		\item the union over all simplices recovers $V$, i.e.\ we have $\bigcup_{\sigma \in S}\sigma =V$, or in other words, $\forall v \in V: \{v\} \in S$.
	\end{enumerate}
\end{definition}
We call the set of subsets of $V$ in $S$ with $k+1$ elements the $k$-\emph{simplices} $S_k$.
Given an ordering of $V$, we can order the $k$-simplices for all $k$ lexicographically.
Writing an oriented $k$-simplex as $[v_0,\dots,v_k]$ with $v_0<\dots<v_k$, we define the corresponding column of $B_k$ by
\[
	B_k[v_0,\dots,v_k]=\sum_{r=0}^k(-1)^r[v_0,\dots,\widehat{v_r},\dots,v_k],
\]
where $[v_0,\dots,\widehat{v_r},\dots,v_k]$ denotes the $(k-1)$-face obtained by removing $v_r$.
This gives boundary matrices with entries in $\{0,\pm1\}$ and $B_kB_{k+1}=0$.
A $1$-simplex (edge) $\{u,v\}$ is a $1$-cell $(u,v)$, and a $2$-simplex (triangle) $\{u,v,w\}$ is a $2$-cell $(u,v,w)$.
This also holds for arbitrary dimensions (with our general definition from \cref{sec:def}).
An important example of simplicial complexes are triangular meshes known from computer geometry.
Cubical complexes form another important special case; since they arise naturally as products of path graphs, we return to them in \cref{subsec:cc-wild-product-structure}.

\subsection{Geometric information}\label{subsec:cc-wild-metric-proximity}
In this subsection, we will discuss how to obtain simplicial complexes, which are \CC{}s, from point clouds.
By using later refinements, we can then obtain more general \CC{}s from point clouds.
\paragraph{Vietoris--Rips complexes}
The Vietoris--Rips complex is a standard way to build a cell complex, in particular a simplicial complex, on a given point cloud and is widely used in topological data analysis.
Essentially, given a distance $\varepsilon$, we connect points that are within $\varepsilon$ of each other and fill in all cliques with simplices.
We note that this construction works for any data with pairwise distances.
While VR complexes are easy to compute and very general, they are sensitive to the choice of $\varepsilon$ and can be very large, as every clique in the graph of pairwise connections is filled with a simplex, even if we only consider simplices/cells up to a given small dimension $k$.
\begin{definition}
Given a real number $\varepsilon$ and a point cloud $X\subset \R^n$, the associated Vietoris--Rips complex $VR_\varepsilon$ is the simplicial complex with vertex set $X$ and simplex set
\[
S_\varepsilon=\{\emptyset\not =\sigma \subset X:  \lVert x_1- x_2\rVert_2\leq \varepsilon \quad \forall x_1,x_2\in \sigma\}.
\]
\label{def:VRcomplex}
\end{definition}

\paragraph{Delaunay triangulations and alpha complexes}
While Vietoris--Rips complexes only use pairwise distances, alpha complexes use the ambient Euclidean geometry of the point cloud more explicitly \cite{edelsbrunner1994three}.
They use this information to parcellate the ambient space into Delaunay simplices, similar to Delaunay triangulations in 2D.
The advantage is that this produces much smaller complexes with fewer cells.
In particular, where the degree of local cell density of VR complexes scales polynomially with the local point density, for Delaunay-based complexes this mainly depends on the ambient dimension.
On the other hand, Delaunay-based complexes are computationally infeasible even in moderate ambient dimensions, and the existence of individual cells is very sensitive to the exact positions of points, which can be an interpretability problem for certain tasks on noisy data.
Let $X\subset \R^d$ be a finite point cloud.
For every point $x\in X$, its \emph{Voronoi cell} is
\[
	V_x=\{p\in \R^d:\lVert p-x\rVert_2\leq \lVert p-y\rVert_2\text{ for all }y\in X\}.
\]
The Voronoi cells decompose $\R^d$ into regions of points that are closest to the same data point.
The corresponding \emph{Delaunay complex} $\operatorname{Del}(X)$ is the simplicial complex whose simplices are those non-empty finite subsets $\sigma\subseteq X$ with a common Voronoi intersection,
\[
	\bigcap_{x\in \sigma}V_x\neq\emptyset.
\]
Equivalently, one can think of a Delaunay simplex as a small group of points that can all lie on the boundary of a ball whose interior contains no data point.
For an edge, this means that there is an empty ball touching both endpoints; for a triangle in the plane, there is an empty circle through its three vertices.
This is essentially a condition guaranteeing that no foreign point would be contained in the simplex, and that we pick a \enquote{nice} and well-behaved parcellation.
If $X$ affinely spans $\R^d$ and no $d+2$ of its points lie on a common sphere, then the Delaunay complex has dimension $d$: its maximal simplices are triangles in the plane, tetrahedra in $\R^3$, and $d$-simplices in $\R^d$.
They will parcellate the convex hull of the point cloud.

The \emph{alpha complex} at scale $\alpha\geq 0$ is a scale-dependent subcomplex of the Delaunay complex.
\begin{definition}[Alpha complex]
Let $B_\alpha(x)$ be the closed ball of radius $\alpha\geq 0$ around $x$ and consider the clipped Voronoi cells
\[
	V_x^\alpha=V_x\cap B_\alpha(x).
\]
The alpha complex $\operatorname{Alpha}_\alpha(X)$ is the simplicial complex with vertex set $X$ and simplex set
\[
	\{\emptyset\neq\sigma\subseteq X:\bigcap_{x\in\sigma}V_x^\alpha\neq\emptyset\}.
\]
\end{definition}
Equivalently, the alpha complex keeps exactly those Delaunay simplices whose common Voronoi witness can be chosen within distance $\alpha$ of all vertices.
Since every $V_x^\alpha$ is convex, the nerve theorem identifies the homotopy type of $\operatorname{Alpha}_\alpha(X)$ with the union of balls $\bigcup_{x\in X}B_\alpha(x)$.
The Voronoi diagram is a decomposition of the underlying space into polyhedral regions: its $d$-dimensional regions are the Voronoi cells $V_x$, and their shared faces provide the lower-dimensional cells.
Choosing orientations of all faces yields signed incidence matrices $B_k$.
They are the boundary matrices of a finite regular cell complex and hence define an \textsc{arcc}.
The Delaunay complex records the dual incidence structure of this diagram.

\subsection{Product structure}\label{subsec:cc-wild-product-structure}

\paragraph{Products of graphs and simplicial and cell complexes}
Cell complexes naturally arise as the product of easier structures:
Let us consider two very simple graphs $G_1$ and $G_2$ with
\[G_{1/2}=(V_{1/2},E_{1/2}) = (\{v_{1/2},u_{1/2}\},\{(v_{1/2},u_{1/2})\})\]
both corresponding to a single edge with nodes.
For example, each graph may correspond to a different direction in physical space.
The vertex set of their cartesian graph product $G_1\square G_2$ now consists of pairs of vertices $(u,v)\in V_1\times V_2$.
The edges set of $G_1\square G_2$ is the union
\[E_{G_1\square G_2}=V_1\times E_2\cup V_2\times E_1,\]
\begin{wrapfigure}{r}{.44\linewidth}
	\centering
	\includegraphics[width=\linewidth]{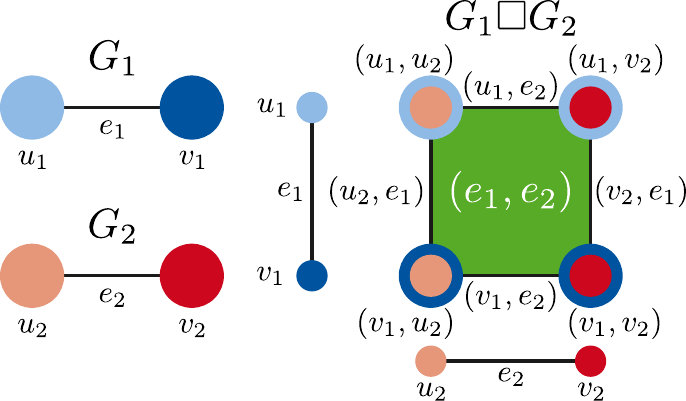}
	\vspace*{-8pt}
	\caption{Two graphs and their product cell complex with four $0$-cells, four $1$-cells and one $2$-cell.}\label{fig:product-complex}
\end{wrapfigure}
where the edge $(v_1,(u_2,v_2))$ connects the vertices $(v_1,u_2)$ and $(v_1,v_2)$ in $G_1\square G_2$.
After having assigned $0$-dimensional objects to pairs of vertices and $1$-dimensional objects to mixed pairs of one vertex and one edge, this construction would naturally continue to assign $2$-dimensional objects to pairs of edges of different graphs.
In the case of the graph product, we are bound by the language of graph theory not to do this.
However, the world of cell complexes allows for the generalisation of graph products by assigning a $2$-cell to every pair of edges.
\Cref{fig:product-complex} shows this example product.

More generally, we can construct a product of two given cell complexes, using the following definition which we will explain below and which preserves abstract regularity by \cref{thm:products-preserve-arcc}.
\begin{definition}[Products of cell complexes]
	\label{def:prodcellcomplexes}
We let $\cc=(\cc_*,B_*)$ and $\cc'=(\cc'_*, B'_*)$ be two cell complexes.
Their product $\cc\times\cc'$ is given by the cells
\[
(\cc\times\cc')_{\hat{k}}=\mathop{\dot\bigcup}_{k+k'=\hat{k}}\cc_k\times \cc'_{k'}.
\]
This means that for every pair of cells $c_k\in \cc_k$ and $c'_{k'}\in \cc'_{k'}$ we will get a new cell $c_k\times c'_{k'} = (c_k,c'_{k'})\in (\cc\times\cc')_{k+k'}$ in the product in dimension $k+k'$.

Note that boundary of cell $c_k\times c'_{k'}$ consists of cells in dimension $k+k'-1$ and can be obtained from the boundaries of $c_k$ and $c'_{k'}$.
Specifically, the boundary matrices are defined via their action on the basis elements:
\[
B^{\cc\times\cc'}_{k+k'}\left(c_k\times c'_{k'}\right)=(B^\cc_kc_k)\times c'_{k'}+(-1)^{k}c_k\times (B^{\cc'}_{k'} c'_{k'}).
	\]
Here $\dot\bigcup$ denotes a disjoint union of cells, and the products involving chains are extended linearly in both components.
Thus the displayed formula determines the boundary map on all chains.
\end{definition}
If we had applied the boundary relation in both factors simultaneously we would have obtained a set $\partial c_k\times \partial' c'_{k'}$ of cells in dimension $k+k'-2$, which clearly would not form a valid boundary.
Instead, we need to apply the boundary to each of the factors individually and take the union to obtain the set of cells in the boundary
\[
\partial (c_k\times c'_{k'}) = \partial c_k\times \{c'_{k'}\}\cup\{c_k\}\times \partial' c'_{k'},
\]
leading to the above definition of the boundary matrices.

In the case of our simple example in \cref{fig:product-complex} where both $\cc$ and $\cc'$ corresponded to a set of a single edge $e_{1/2}$ with associated vertices $u_{1/2}$, $v_{1/2}$, the product $\cc\times \cc'$ is a square consisting of
\begin{enumerate}
	\item Four $0$-cells $u_1\times u_2$, $u_1\times v_2$, $v_1\times u_2$, and $v_1\times v_2$ representing the four vertices of a square.
	\item Four $1$-cells $u_1\times e_2$, $v_1\times e_2$, $e_1\times u_2$, and $e_1\times v_2$ representing the four edges of a square.
	\item A single $2$-cell $e_1\times e_2$ corresponding to the area of the square.
\end{enumerate}
This aligns with our intuition of what a product of two lines should be.

Let us now explicitly compute the associated boundary matrices.
Let us first consider $B_1^{\cc\times\cc'}$.
With vertex order $u_i,v_i$, we orient both factor edges from $v_i$ to $u_i$.
The only boundary matrices of $\cc$ and $\cc'$ are $B_1=B_1'=\begin{pmatrix}
		1 &
		-1
\end{pmatrix}^\top$.
Using the tensor-product ordering of rows and columns, and ordering the $1$-cells as $u_1\times e_2$, $v_1\times e_2$, $e_1\times u_2$, and $e_1\times v_2$, we can now write down an explicit boundary matrix as follows:
\[
B^{\cc\times\cc'}_1=\left[I_{|\cc_0|}\otimes B'_1\quad B_1\otimes I_{|\cc'_0|}\right]=\begin{pmatrix}
	1&0&1&0\\
	-1&0&0&1\\
	0&1&-1&0\\
	0&-1&0&-1
\end{pmatrix}.
\]
With respect to this order of $1$-cells, the remaining boundary matrix in the example is $B_2^{\cc\times\cc'}=\begin{pmatrix}1&-1&-1&1\end{pmatrix}^\top$.
In the general case, the boundary matrices are more complicated because they combine the boundaries of both factors.

Based on the definition of the set of $k$-cells in $\cc\times\cc'$, we can write down an explicit description of the $k$-th chain space based on the chain spaces $C_i$ of $\cc$ and $C_j'$ of $\cc'$:
\[
C_k^{\cc\times\cc'} =\bigoplus_{i+j=k}C_i\otimes C'_j.
\]
In \cref{fig:prodChainComplex}, we give an overview of the interplay between the boundary matrices and the chain spaces of a product cell complex in the general case.

\begin{figure}[h!]
\centering
\begin{adjustbox}{max width=\linewidth}
\begin{tikzcd}
	\text{Dim }m+n& & & C_m\otimes C'_n \ar[dl, "B_m\otimes I_{|\cc'_n|}"]\ar[dr, "(-1)^mI_{|\cc_m|}\otimes B'_n"]& & & \\
	\text{Dim }m+n-1\text{\hspace{-6pt}}&& C_{m-1}\otimes C'_n \ar[dl, "B_{m-1}\otimes I_{|\cc'_n|}"]\ar[dr, "(-1)^{m-1}I_{|\cc_{m-1}|}\otimes B'_n"]& &C_m\otimes C'_{n-1}\ar[dl, "B_{m}\otimes I_{|\cc'_{n-1}|}"]\ar[dr, "(-1)^{m}I_{|\cc_{m}|}\otimes B'_{n-1}"] & &\\
	\text{Dim }m+n-2\text{\hspace{-6pt}}&\dots& &\dots& &\dots\\
	\vdots&\vdots & &\vdots & &\vdots & &\\
	\text{Dim }2&\dots\ar[dr, "I_{|\cc_0|}\otimes B'_2"] & &\dots\ar[dl, "B_1\otimes I_{|\cc'_1|}"]\ar[dr, "-I_{|\cc_1|}\otimes B'_1"]& & \dots\ar[dl, "B_2\otimes I_{|\cc'_0|}"] &\\
	\text{Dim }1& & C_0\otimes C'_1 \ar[dr, "I_{|\cc_0|}\otimes B'_1"]& & C_1\otimes C'_0 \ar[dl, "B_1\otimes I_{|\cc'_0|}"]& \\
	\text{Dim }0&& & C_0\otimes C'_0& & & \\
	\end{tikzcd}
\end{adjustbox}
\caption{The product chain complex and its boundary maps. Each chain space $C_k^{\cc\times\cc'}$ is the direct sum of tensor products $C_i\otimes C_j'$ with $i+j=k$. Each boundary matrix $B_k^{\cc\times\cc'}$ combines the maps represented by the arrows from dimension $k$ to dimension $k-1$ in the diagram.}
\label{fig:prodChainComplex}
\end{figure}

\begin{figure}[t]
	\centering
	\vspace{-6pt}
	\includegraphics[width=.25\linewidth]{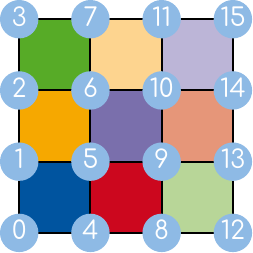}
	\caption{The $2$-dimensional cubical complex $C_{4,4}$ ($2$-cells drawn as coloured squares). Note that the vertices ($0$-cells, numbered) and edges ($1$-cells) are part of the structure as well.}\label{fig:squarelattice}
\end{figure}

\paragraph{The square lattice and cubical complexes}
Another natural variant of \CC{}s are finite cubical complexes, which are \CC{}s on the $k$-dimensional square lattice.
The one-dimensional cubical complex is just the path graph $P_n$, with $n$ vertices and edges $(i,i+1)$ between consecutive vertices.
Formally, the $k$-dimensional cubical complex of size $n_1 \times n_2 \times ... \times n_k$ can be defined as the cell complex product of path graphs:

\begin{equation}
	C_{n_1,n_2,...,n_k} := P_{n_1} \times P_{n_2} \times ... \times P_{n_k}.
\end{equation}

\Cref{fig:squarelattice} shows $C_{4,4}$ as an example.
The $k$-dimensional cubical complex is also a $k$-dimensional \CC{}; its $k$-cells correspond to the $k$-dimensional cubes in the square lattice.

\subsection{Planar embeddings and parcellations}\label{subsec:cc-wild-planar-structure}

When a graph comes with a meaningful planar embedding, its enclosed regions provide natural candidates for $2$-cells.
Conversely, spatial parcellations already define such regions directly and may also be represented through their dual complexes.

\begin{figure*}[t!]
    \centering
    \includegraphics[width=.5\linewidth]{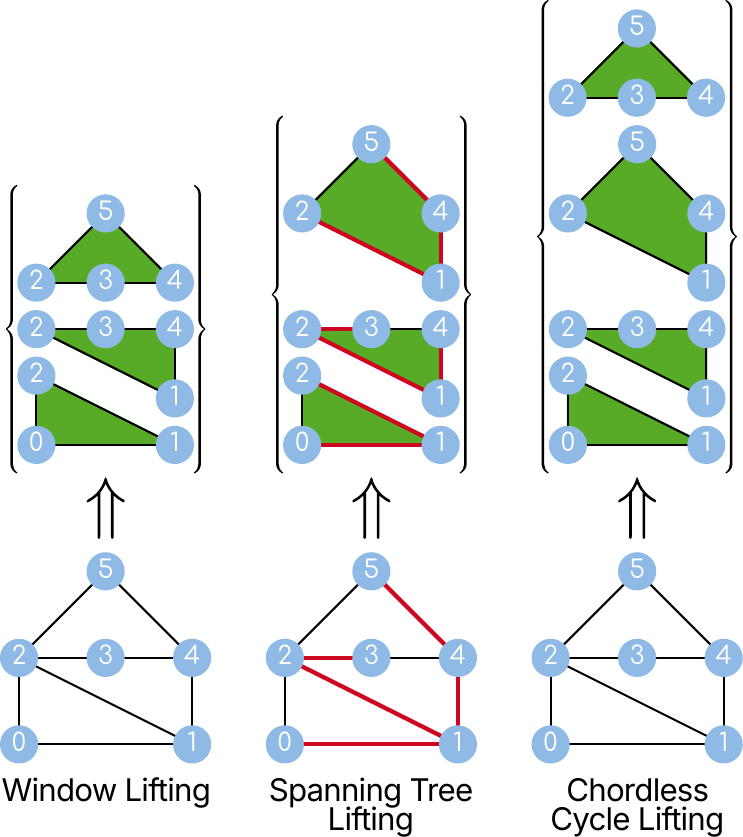}
    \caption{Procedures to lift existing graphs to $2$-dimensional \CC{}s. For graphs with a natural planar embedding, the \emph{window lifting} provides a straightforward way to add higher-order structure based on real-world geometry. The \emph{spanning tree lifting} works on arbitrary graphs, but cells tend to have considerable overlap. Finally, the \emph{lifting of all chordless cycles} in a graph is, in general, larger than the two others and creates $2$-dimensional holes (homology classes). In the worst case, the chordless cycle lifting results in exponentially many cycles.}\label{fig:liftings}
\end{figure*}

\paragraph{Window lifting from planar embeddings}
For graphs that have a natural planar embedding, for example road networks, we can take all inner windows as cells, spanning the entire graph with $2$-cells.
\Cref{fig:liftings} \emph{left} shows a small example for the window lifting: Each of the three windows in the planar embedding is lifted to a $2$-cell.
Note that this lifting is dependent on the specific embedding: If the positions of nodes 3 and 5 were swapped, the second $2$-cell shown would be $(1,2,5,4)$ instead.
Thus, the lifting is limited in expressiveness for planar graphs without a natural embedding.

\paragraph{Spherical closure and duality}

\begin{figure}[t]
	\includegraphics[width=\linewidth]{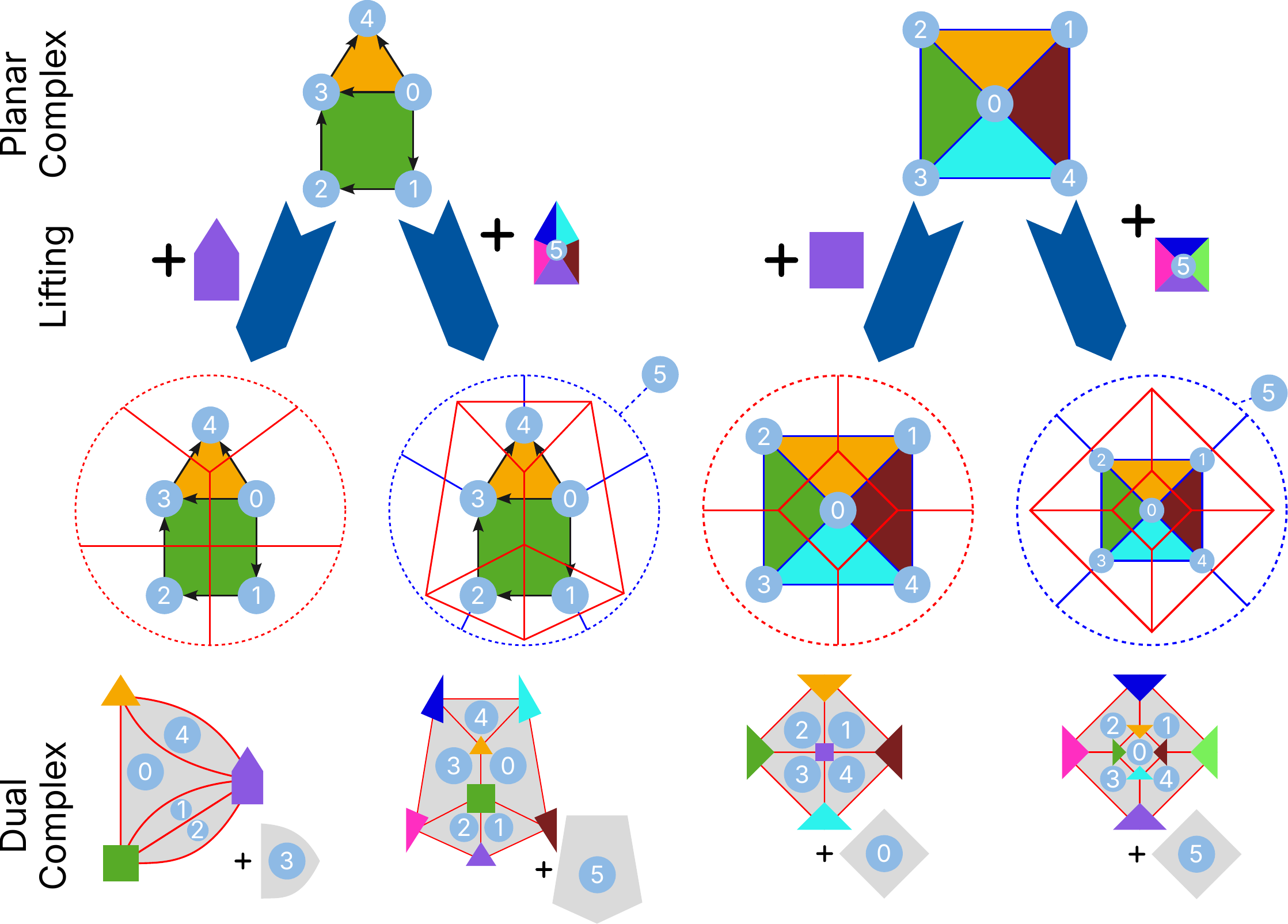}
	\caption{\textbf{Planar complexes, their lifting to spherical complexes, and their dual complexes.}
	Blue lines denote the original complex; red lines denote the dual complex. The big blue and red circles represent a single node that makes the complex spherical (and is dual to the outer window).
	\textbf{Left:} The outer window lifting of our toy example from \cref{fig:1} adds one cell spanning the five outer edges. The dual complex consequently has one \enquote{outer} node represented by the red circle.
	The dual complex is not a simple \CC{} as there are multiedges and $2$-cells enclosed by two edges sharing endpoints.
	The visualization on the bottom is an unfolded variant where all outer points are the same point and the triangles are adjacent to the cells living between multiedges.
	\textbf{Middle Left:} The outer point lifting of our toy example adds five triangles that all share a point, spanning a tent above the complex.
	In the dual complex, the outer point is turned into an outer window.
	\textbf{Middle Right:} The second toy complex consists of a square containing four triangles.
	Its outer window lifting results in a pyramid, which has another pyramid as its dual complex.
	The tip of the original pyramid becomes the base of the dual pyramid and vice versa.
	\textbf{Right:} The outer point lifting of the second complex results in an octahedron.
	The dual of an octahedron, as is well-known, is a cube, which is also the result in our definition of dual complexes.
	}\label{fig:dual-complex}
\end{figure}

Planar \CC{}s also have dual complexes, where nodes are replaced by polygons and vice versa.
For a straightforward construction, we require that the complex is connected both on the level of nodes and on the level of polygons, and that each node adjacent to the outer window is only adjacent to two edges in the outer window.
To arrive at a dual complex, we require a spherical complex (i.e., a cell complex homeomorphic to the $2$-sphere).
A planar complex can be made spherical by either adding a single cell representing the outer window or by adding an additional node connected to all nodes in the outer window, and filling the additional triangles.
Given a spherical complex $\cc$ with boundaries $B_k$, we first adjust the reference orientations of all $2$-cells s.t. $B_2 \mathbbm{1}=0$.
We then obtain the dual complex $\cc'$ with $\cc_0'=\cc_2, \cc_1'=\cc_1, \cc_2'=\cc_0$ and boundaries $B_1'=B_2^\top, B_2'=B_1^\top$.
\Cref{fig:dual-complex} shows two examples of both constructions to take the dual complex.
It is also possible to take the dual directly from the planar complex, but that results in edges that are attached to just one node and cells with open boundaries.
After removing incomplete cells, the two examples from \cref{fig:dual-complex} would turn into a single edge and a square, respectively.
However, if applied twice, the result will not be the original complex, but a subcomplex.

A well-known special case of this is the duality between Delaunay triangulation and Voronoi diagram, already introduced in \cref{subsec:cc-wild-metric-proximity}.
Similarly, with a suitable modification of the outer windows, a division of the plane into hexagons is dual to a triangulation of the plane.
The dual of a square lattice is also a square lattice.
This shows an interesting advantage of cell complexes over simplicial complexes: The dual of a (planar / spherical) simplicial complex is only a simplicial complex if the degree of nodes is at most three, whereas cell complexes are closed under taking the dual complex.
This theory extends to arbitrary parcellations of compact subsets of $\R^d$ by cell complexes.

\subsection{Graph-cycle structure}\label{subsec:cc-wild-graph-cycle-structure}

\paragraph{Filling cycles to lift arbitrary graphs}
Filling all simple cycles is not generally feasible as, in the worst case, the number of possible simple cycles on a graph may be exponential in the number of nodes.
One way to limit this is by lifting only \emph{chordless} cycles, i.e., cycles which do not have any two nodes connected by a non-cycle edge.
Theoretically, chordless cycles have the same scaling problem: Any graph can be made chordless by splitting each edge into two edges with a node in between.
However, in practice, the number of chordless cycles is often much smaller than the number of simple cycles.
Thus, chordless cycles have been used and proven to be useful to capture relevant structural information \cite{bodnar2021weisfeiler}.
Alternatively, one can add a complete cycle basis, e.g.\ obtained from a spanning tree \cite{syslo1979cycle}.
One possible spanning tree lifting is shown in \cref{fig:liftings} \emph{centre}; however, the obtained cycles depend heavily on the chosen spanning tree.

\subsection{Signal/task structure}\label{subsec:cc-wild-signal-task-structure}
When applying methods that utilize cell structure to graphs without a natural lifting, the best performance can be achieved by selecting cells specifically for the task at hand.
This section focuses on the different task-based cell inference approaches \cite{battiloro2024latent,canbolat2025online,hoppe2023representing,liu2026topologicalkalman}; we will discuss the tasks addressed in these publications in \cref{sec:applications}.

In general, given a cell complex, we assume that we have a way to evaluate the performance of the given task on this complex in a cost-like function.
Inferring higher-order cells based on a signal processing task is then a discrete optimization problem over the space of all cell complexes with a specific skeleton.
The number of possible cell complexes (combinations of cells) is exponential in the number of added $2$-cells.
Consequently, finding the optimal cell complex can be NP-hard \cite{hoppe2023representing}.
Furthermore, the number of possible cells (simple cycles) is, in the worst case, exponential in the number of nodes.
Finally, the signal processing task (and therefore the cost function) is often somewhat expensive to perform, e.g. (approximately) solving systems of linear equations.
Thus, if this is evaluated for a large number of cells that could be added, it will again limit scalability in practice.

All three challenges have been addressed by existing approaches, trading the quality of the solution for an improved runtime.
First, instead of considering all combinations of cells, scalability can be improved by adding cells greedily \cite{hoppe2023representing,liu2026topologicalkalman}.
Second, in every step, instead of evaluating the cost function directly on every cell candidate, a heuristic can be used to pre-select candidates.
This way, the more expensive cost function only has to be evaluated on a fixed number \cite{hoppe2023representing} or until a cell meeting a configured threshold of improvement has been found \cite{liu2026topologicalkalman}.
Finally, if this heuristic is sufficiently simple to calculate, one can use spanning trees to find a set of already good candidates and quickly evaluate the heuristic using the spanning tree structure \cite{hoppe2023representing}.

\subsection{Random and generative structure}\label{subsec:cc-wild-null-generative-structure}
Random models are useful both as baselines for assessing observed or inferred structures and as generators of controlled synthetic data.
They typically preserve selected properties exactly or approximately while randomizing the remaining structure.
In this subsection, we discuss these two uses for cell complexes and describe several corresponding construction principles.

Given a method that infers $2$-cells to optimize the cost function of a specific task, it is important to examine how good the inferred structure is.
This is where the random model can serve as a baseline.
If the method provides a meaningful inference of structure, it should significantly outperform a random lifting.
For the baseline random model, we fix the graph skeleton and define some properties of the $2$-cells to fix as well.
For example, given a \CC{} with $2$-cells $c_2^1,\dots,c_2^n$, a null model on the same $1$-skeleton could randomly select $2$-cells $\hat c_2^1,\dots,\hat c_2^n$ while preserving the multiset of boundary sizes, i.e.\ $|\partial c_2^i|=|\partial\hat c_2^i|$ after a suitable reordering.
Hoppe and Schaub introduced an algorithm to sample random $2$-cells on graphs\footnote{\url{https://pypi.org/project/py-raccoon/}}~\cite{hoppe2024random}.
By sampling multiple such complexes, we can estimate the distribution of arbitrary metrics under these constraints, such as Betti numbers, eigenvalues of cell complex operators, or the performance of a given method.

Given just one cell complex, we may ask if it has different structural properties (e.g.\ Betti numbers, biggest eigenvalue, ...) compared to cell complexes that have certain characteristics.
For this, we could fix the skeleton and use the same model as before.
However, depending on the application, this may be too restrictive.
Instead, we can also sample the $1$-skeleton from a graph random model and then add randomly selected $2$-cells \cite{hoppe2024random}.
Alternatively, we could apply a top-down approach and start by sampling $2$-cells (adding their boundaries as well) and then completing the complex by adding the missing edges.

Finally, random cell complexes can be used for synthetic experiments.
To investigate scalability, we only require a random model that can produce progressively larger instances of cell complexes.
To evaluate the ability of a method to recover certain structure, it is often useful to generate synthetic cell complexes that are biased towards having a certain structure.
This would be possible by modifying the aforementioned top-down or bottom-up approaches to include additional structure or biases.
Geometric cell complexes can also be generated by starting from a Delaunay triangulation and forming $2$-cells by randomly combining adjacent triangles \cite{hoppe2023representing}.
This construction has been used in an ad hoc manner, but its statistical properties have not yet been studied systematically.

\section{Topology and Geometry on Cell Complexes}\label{sec:topology}

One of the key strengths of modelling data as and on cell complexes is the close connection between \CC{}s and topology and geometry.
In this section, we will highlight some of the most important connections.

\subsection{Homology}
Homology is one of the key concepts of algebraic topology.
Intuitively, the $k$-th homology of a topological space or a cell complex encodes the $k$-dimensional loops/holes of the space.
This is most natural for $k=1$, where $1$-homology just encodes the ordinary loops of the complex.
For $k=0$, homology measures the connected components of the cell complex.
\Cref{fig:toy-homology} shows a modified version of our toy cell complex with a single $1$-homology.
\begin{definition}[Homology of a cell complex]
	Given a cell complex $\cc$ with chain spaces $C_i$ and boundary maps $B_i$, we denote by the $k$-th (real-valued) homology of $\cc$ the $\R$-vector space $H_k(\cc)$ given by
	\[
	H_k(\cc)=\ker B_k/\Ima B_{k+1}.
	\]
	More generally, if we consider chain spaces $C_i^\mathcal{R}=\mathcal{R}^{|\cc_i|}$ with coefficients in a commutative ring $\mathcal{R}$ with unit and view $B^\mathcal{R}_i$ as a map $B^\mathcal{R}_i\colon C_{i}^\mathcal{R}\to C_{i-1}^\mathcal{R}$, we define the $k$-th homology in $\mathcal{R}$-coefficients as the $\mathcal{R}$-module $H_k(\cc;\mathcal{R})$ given by
	\[
	H_k(\cc;\mathcal{R})=\ker B_k^\mathcal{R}/\Ima B^\mathcal{R}_{k+1}.
	\]
	Here, we use the convention $B_0=0$ and $B_{n+1}=0$ at the endpoints.
	\label{def:homology}
\end{definition}
\begin{example}[Homology classes of the toy cell complex]
\begin{figure}[t]
	\centering
	\includegraphics[width=.5\linewidth]{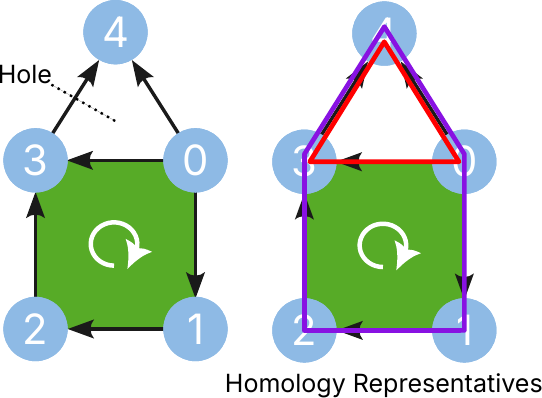}
	\caption{\textbf{Homology representatives. Left:} Modified toy example after removal of triangular $2$-cell resulting in a hole.
	\textbf{Right:}
	Two (red and purple) equivalent non-trivial $1$-homology representatives for the hole.
	}
	\label{fig:toy-homology}
\end{figure}
Let us now consider the cell complex $\cc$ of \cref{fig:1} after removing the orange $2$-cell $(0,3,4)$.
Then, the edges $0\rightarrow 3$, $3\rightarrow 4$ and the reverse of $0\rightarrow 4$ form a cycle around the triangular hole in $\cc$.
We can represent this cycle as a chain $c=(0\rightarrow 3)+(3\rightarrow 4)-(0\rightarrow 4)$ in $C_1$.
The cycle representatives of $H_1(\cc)$ lie in the kernel of $B_1$, and a quick computation reveals that $B_1c$ is $0$,
\begin{align*}
B_1((0\rightarrow 3)+(3\rightarrow 4)-(0\rightarrow 4)) &= B_1(0\rightarrow 3)+B_1(3\rightarrow 4)-B_1(0\rightarrow 4)\\
&=(~\encircle{3}~-~\encircle{0}~)+(~\encircle{4}~-~\encircle{3}~)-(~\encircle{4}~-~\encircle 0~)\\
&=0.
\end{align*}
Hence $c$ lies in the kernel of $B_1$ and thus represents a homology class $[c]$ corresponding to the hole in $\cc$.
This means that the cycle enters and leaves every vertex of $\cc$ an equal number of times.
However, two elements $c$ and $c'$ that go around the same hole should represent the same homology class as well.
By quotienting out the image of $B_2$ from the kernel of $B_1$, the homology $H_1(\cc)$ achieves exactly that.
Assume that $c'$ corresponds to a cycle corresponding to the edges $0\rightarrow 1$, $1\rightarrow 2$, $2\rightarrow 3$, $3\rightarrow 4$ and the reverse of $0\rightarrow 4$, i.e.
\[
c'=(0\rightarrow 1)+(1\rightarrow 2)+(2\rightarrow 3)+(3\rightarrow 4)-(0\rightarrow 4)\in C_1.
\]
Again, $c'$ lies in the kernel of $B_1$ and is thus a homology representative, i.e. $B_1c'=0$.
Intuitively, $c$ and $c'$ represent the same hole.
We now verify that this is true in homology, i.e. $[c]=[c']\in H_1(\cc)$.
In order to do this, we need to show that $c'-c\in \Ima B_2$:
\begin{align*}
c'-c=&(0\rightarrow 1)+(1\rightarrow 2)+(2\rightarrow 3)+(3\rightarrow 4)\\&
-(0\rightarrow 4)-\left((0\rightarrow 3)+(3\rightarrow 4)-(0\rightarrow 4)\right)\\
&= (0\rightarrow 1)+(1\rightarrow 2)+(2\rightarrow 3) -(0\rightarrow 3)\\
&=B_2(\text{\cellsq})
\end{align*}
where $\text{\cellsq}\in C_2$ denotes the basis vector corresponding to the green $2$-cell of $\cc_2$.
Thus, $c$ and $c'$ only differ by the boundary of a $2$-cell and they indeed represent the same hole.
In summary, enforcing homology representatives to \emph{lie in $\ker B_1$} ensures that they \emph{represent} a collection of \emph{cycles}, whereas \emph{quotienting by} $\Ima B_2$ ensures that we \emph{identify} two homology classes which \emph{show the same behaviour} with respect to the holes.
\end{example}
In applications, we mostly do not need the additional algebraic vector space or module structure.
Thus, we consider the \emph{Betti numbers}, where the $i$-th Betti number simply counts the number of $i$-dimensional loops or topological features and is defined as:
\begin{definition}[Betti number of a cell complex]
	Given a cell complex $\cc$ with homology $H_*(\cc)$, we call the \emph{$k$-th Betti number} $\beta_k$ the rank of $H_k(\cc)$,
	\[
	\beta_k=\operatorname*{rank} H_k(\cc).
	\]
\end{definition}

\subsection{Hodge Laplacians and the relation to geometry}
While \emph{topology} and in particular \emph{homology} deals with global information on cell complexes, \emph{geometry} allows us to relate the structure to local properties on individual cells.
As we can view cell complexes in geometric applications as discretisations of manifolds, many geometrical notions naturally carry over to cell complexes.
The manifold Laplacian $\Delta$ of a function $f$ encodes for every point $x$ the difference of the value of $f$ on $x$ and on a small neighbourhood of $x$.
Analogously, the $0$-Laplacian $L_0$ of a cell complex $\cc$ takes a function $f\in C_0$ on $0$-cells, i.e., in the $0$-th signal/chain space of $\cc$ and returns, for every $0$-cell $c^i_0$, the difference of $f$ on $c^i_0$ and its neighbours:
\begin{align}
L_0(f)(c_0^i)&=\sum_{c_1^h\in \cc_1:\partial (c_1^h) =\{c_0^i,c_0^j\}}\left(f(c_0^i)-f(c_0^j)\right)\\
&= (B_1B_1^\top f)_i.
\label{eq:GraphLaplacian}
\end{align}
As a matrix, we can thus write the $0$-Laplacian in terms of the boundary operators $L_0= B_1B_1^\top$.
The $0$-Laplacian corresponds to the ordinary graph Laplacian $L=D-A$ on the underlying graph of $\cc$.
The associated quadratic form 
\[C_0\to\R_{\ge 0}\quad f\mapsto f^\top Lf=\langle B_1^\top f,B_1^\top f\rangle_1=\lVert B_1^\top f\rVert^2,
\]
where $\langle-,-\rangle_k$ denotes the inner product on the space on the $k$-th signal space $C_k$,
can be interpreted as the (squared) \emph{total variation} of the node signal $f$ or the so-called Dirichlet energy.

\begin{figure}[t]
    \includegraphics[width=\linewidth]{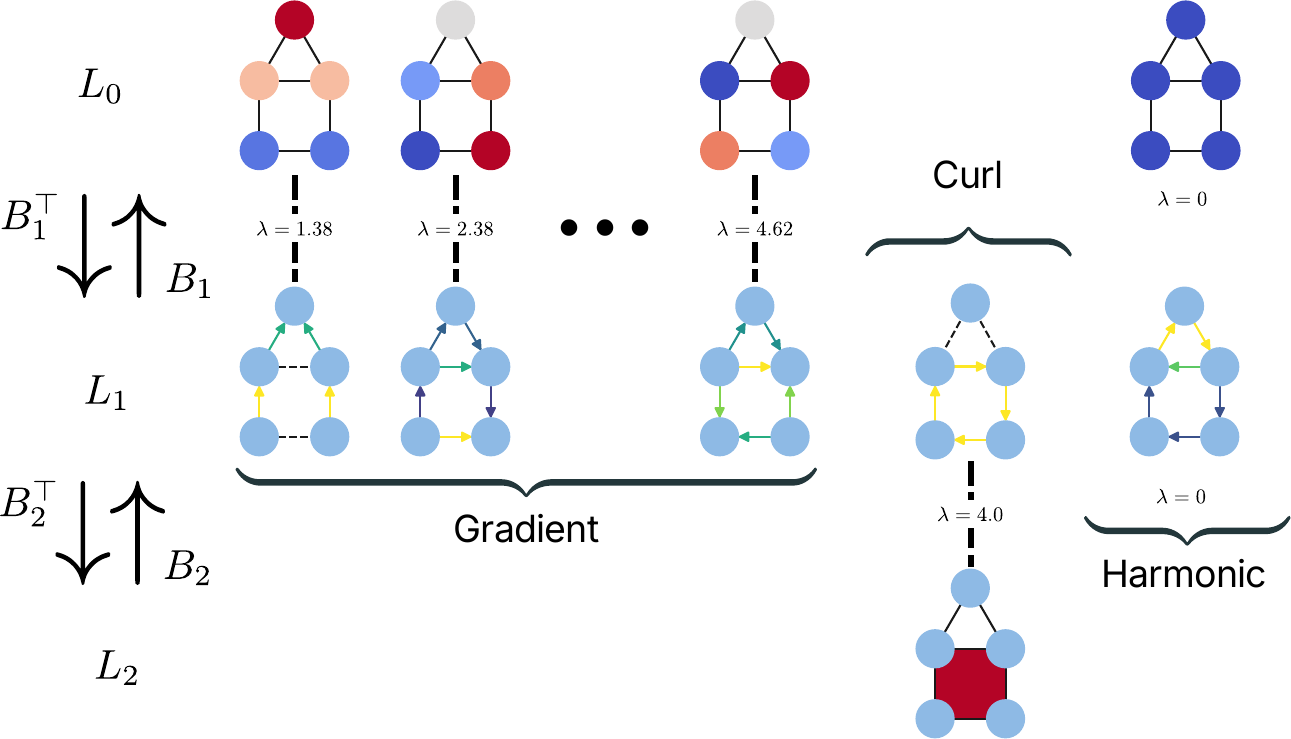}
    \caption{The eigendecompositions of the 0-, 1-, and 2-dimensional Hodge Laplacians on the \CC{} from \cref{fig:toy-homology} (that has the square, but not the triangle 2-cell). The gradient eigenvectors of $L_1$ correspond to eigenvectors of $L_0$ with the same non-zero eigenvalues; the curl eigenvectors of $L_1$ correspond to eigenvectors of $L_2$ with the same non-zero eigenvalues. The harmonic eigenvectors of $L_0$, $L_1$, and $L_2$ are unrelated; the harmonic space of $L_2$ is trivial in this case.
    }
    \label{fig:eigendecomp}
\end{figure}

While the $0$-Laplacian operates on $0$-cell signals, we now look for a generalisation $L_k\colon C_k\to C_k$ sending signals on $k$-cells to signals on $k$-cells.
The naive generalisation of \cref{eq:GraphLaplacian} would be what we call the Up-$k$-Laplacian,
\[
L_k^\text{up}= B_{k+1}B_{k+1}^\top.
\]
The Up-Laplacian, together with the \emph{Down}-Laplacian forms the Hodge Laplacian or combinatorial Laplacian:
\begin{definition}[Hodge Laplacian on Cell Complexes]
	\label{def:hodgelaplacian}
Given a cell complex $\cc$ with boundary matrices $B_i$, the \emph{$k$-th Hodge Laplacian} is the matrix given by
\[
L_k =  \underbrace{B_{k+1}B_{k+1}^\top}_{L_k^\text{up}}+\underbrace{B_{k}^\top B_{k}}_{L_k^\text{down}}.
\]
\end{definition}
Again, the associated quadratic form is
\[
f_k\mapsto f_k^\top L_k f_k = \langle B_{k+1}^\top f_k,B_{k+1}^\top f_k\rangle_{k+1}+ \langle B_kf_k,B_kf_k\rangle_{k-1}=
\lVert B_{k+1}^\top f_k\rVert^2+\lVert B_k f_k\rVert^2
\]
measuring some notion of variation or smoothness of the signal on $k$-cells.
\paragraph{The spectrum of the Hodge Laplacians}
The eigenspectra of the Hodge Laplacian and their corresponding eigenvectors have a meaningful underlying structure and are useful for various tasks.
Let us first assume that $v$ is an eigenvector for eigenvalue $\lambda>0$ of the up-Laplacian.
We then have that 
\[
\lambda v=L_k^\text{up}v = B_{k+1}(B_{k+1}^\top v)
\]
and hence $v$ is in $\Ima B_{k+1}$.
We remember that we constructed the boundary matrices of cell complexes in such a way that the composition of two boundary matrices returns $0$, i.e.\ $B_{k}B_{k+1}=0$ for all valid $k$.
When we now evaluate the down-Laplacian on $v$, we obtain
\[
L^\text{down}_kv=B_k^\top B_kv=\frac{1}{\lambda}B_k^\top\underbrace{B_kB_{k+1}}_{=0}B_{k+1}^\top v=0
\]
and any non-zero eigenvector of $L_k^\text{up}$ is in the kernel of $L^\text{down}_k$ and thus an eigenvector for the same eigenvalue of the Hodge Laplacian $L_k$.
Furthermore, there is a correspondence between the non-zero eigenpairs $(v,\lambda)$ of $L_k^\text{up}$ and non-zero eigenpairs $(B_{k+1}^\top v,\lambda)$ of $L_{k+1}^\text{down}$.
We assume that $v$ is an eigenvector with eigenvalue $\lambda$ of $L_k^\text{up}$ and now check that the vector $B_{k+1}^\top v$ is an eigenvector with the same eigenvalue $\lambda$ of the down-Laplacian $L_{k+1}^\text{down}$ one dimension above:
\[
L_{k+1}^\text{down}B_{k+1}^\top v=B_{k+1}^\top \underbrace{B_{k+1} B_{k+1}^\top}_{L_k^\text{up}} v=\lambda B_{k+1}^\top v.
\]
All of the above holds analogously for the down-Laplacian of dimension $k$ and the Up-Laplacian of dimension $k-1$.
Given two eigenvectors $v$ and $v'$ with eigenvalues $\lambda,\lambda'>0$ of $L_k^\text{down}$ and  $L^\text{up}_k$ respectively, we can see that they are orthogonal by computing their inner product
\[
\langle v,v'\rangle=\frac{1}{\lambda\lambda'}v^\top B_k^\top \underbrace{B_kB_{k+1}}_{=0}B_{k+1}^\top v'=0.
\]
Combining the above observations gives rise to the following structure theorem of the signal spaces (\cite{Eckmann:1944}).
\begin{theorem}[Hodge decomposition for cellular complexes]
	\label{thm:hodgedecomposition}
Given a cell complex $\cc$ with Hodge Laplacians $L_k$, signal/chain spaces $C_k$ and boundary matrices $B_k$, there is an orthogonal decomposition
\[
\R^{|\cc_k|}\cong C_k=\underbrace{\Ima B_{k+1}}_\text{curl/coexact}\oplus\underbrace{\Ima B_k^\top }_\text{gradient/exact}\oplus \underbrace{\ker L_k}_\text{harmonic}
\]
where $\ker L_k=\ker B_k\cap\ker B_{k+1}^\top$, and there exists a basis of $C_k$ of eigenvectors of $L_k$ respecting this decomposition.
The terms gradient and curl are most literal for $k=1$, while exact and coexact describe the corresponding summands in arbitrary dimension.
\end{theorem}
For an illustration of this, see \cref{fig:eigendecomp}.

\paragraph{The total chain space and the Dirac operator}
Given an $n$-dimensional cell complex $\cc$, we will call the direct sum of the individual chain spaces of $\cc$ the \emph{total chain space} or \emph{topological spinor} of $\cc$
\[
\mathbf{C}=C_0\oplus C_1\oplus \dots \oplus C_n
\]
The total chain space encodes all possible combinations of signals on the nodes, edges, $2$-cells and so on.
We can view every element $x$ of $\mathbf{C}$ as a vector $(x_0,x_1,\dots,x_n)^\top$ where $x_0\in C_0$ is a node signal, $x_1\in C_1$ an edge signal, and so on.
We can define the total boundary operator $\mathbf{B}\colon \mathbf{C}\to \mathbf{C}$ acting on the total chain spaces:
\[
\mathbf{B}=
\begin{pmatrix*}
	0 &B_1 &0 &0&0\\
	0 &0&B_2 &0&0\\
	0&0&0&\ddots&0\\
	0& 0 &\ddots& 0 &B_n&\\
	0 & 0 & 0& 0&0
\end{pmatrix*}.
\]
This leads us to the construction of the total Hodge Laplacian $\mathbf{L}\colon \mathbf{C}\to \mathbf{C}$ as the direct sum of the individual Hodge Laplacians, or simply as $\mathbf{B}^\top\mathbf{B}+\mathbf{B}\mathbf{B}^\top$:
\[
\mathbf{L} = L_0\oplus L_1 \oplus \dots \oplus L_n = \begin{pmatrix*}
L_0 &0 &0 &0\\
0 &L_1 &0 &0\\
0&0 &\ddots&0\\
0& 0 &0&L_n
\end{pmatrix*}
\]
Analogously, the definition of the Dirac operator \cite{calmon2023dirac} as the sum of the total boundary and its transpose is now immediate:
\begin{align*}
\mathbf{D}&=\mathbf{B}+\mathbf{B}^\top \\
&=\begin{pmatrix*}
	0 &B_1 &0 &0&0\\
	B_1^\top &0&B_2 &0&0\\
	0&B_2^\top &0&\ddots&0\\
	0& 0 &\ddots& 0 &B_n&\\
	0 & 0 & 0& B_n^\top &0
\end{pmatrix*}
\end{align*}
We have that $\mathbf{B}^2=0$ because of the individual relations $B_{k-1}B_k=0$.
Thus, we can rewrite the total Hodge Laplacian as follows:
\[
\mathbf{L} = \mathbf{B}^\top\mathbf{B}+\mathbf{B}\mathbf{B}^\top =\left(\mathbf{B}^\top+\mathbf{B}\right)^2=\mathbf{D}^2.
\]
We see that the Dirac operator thus acts as some form of generalised \enquote{square root} of the total Hodge Laplacian.
Such a relation is only possible by working in the total chain spaces.
The total Dirac operator and its $k$-dimensional counterparts $D_k=\left(\begin{smallmatrix}0&B_k\\B_k^\top&0\end{smallmatrix}\right)$ can be used for processing of joint signals on $k$-cells and $(k-1)$-cells, analogously to how the Hodge Laplacian is used in topological signal processing \cite{calmon2023dirac}, cf.\ \cref{par:TSP}.

The spectrum of $\mathbf{L}$ is the combination of the individual spectra of the Hodge Laplacians, where most eigenvalues appear twice (once for $L_k^\text{up}$ and once for $L_{k+1}^\text{down}$).
In turn, the eigenvalues of the Dirac operator are the square roots of the spectrum of $\mathbf{L}$, but the eigenvectors are linear combinations of the eigenvectors of $\mathbf{L}$: 
Let $v$ be an eigenvector with eigenvalue $\lambda>0$ of $L_k^\text{down}$ and hence an eigenvector of the complete $L_k$ with eigenvalue $\lambda$ and we have that $B_{k+1}^\top v=0$.
We denote by $\tilde{v}$ the zero-padded embedding of $v$ into the full chain space $\mathbf{C}$.
We now claim that $\sqrt{\lambda}\tilde{v}\pm \mathbf{B} \tilde{v}\in\mathbf{C}$  is an eigenvector of $\mathbf{D}$ with eigenvalue $\pm \sqrt{\lambda}$:
\begin{align*}
\mathbf{D}(\sqrt{\lambda}\tilde{v}\pm \mathbf{B} \tilde{v})&=\sqrt{\lambda}(\mathbf{B}\tilde{v}+\underbrace{\mathbf{B}^\top \tilde{v}}_{=0})\pm\underbrace{\mathbf{B}^2}_{=0} \tilde{v}\pm\underbrace{\mathbf{B}^\top\mathbf{B}}_{\mathbf{L}^\text{down}} \tilde{v}\\
&=\sqrt{\lambda} (\mathbf{B} \tilde{v}\pm \sqrt{\lambda}\tilde{v})=\pm \sqrt{\lambda}(\sqrt{\lambda}\tilde{v}\pm \mathbf{B}\tilde{v}).
\end{align*}

\section{Weights and/or Non-Oriented Data}
\label{sec:weights}

So far, we have assumed all cells in the same dimension to be equal.
However, in many cases, cells are not uniform and should not contribute equally; this is modeled by assigning \emph{weights}.
Depending on the application and the dimension, these weights can represent different quantities (for example, distance- or similarity-based edge strengths on graphs).

In this work, weights determine the relative scaling of cells in operators and energies. Their precise effect depends on the cell dimension and on whether they enter an up- or down-Laplacian term.

While weight vectors would have the same shape as signal vectors, they are not conceptually elements of the chain space as weights do not have an orientation\,---\,instead, they are part of the structure of the cell complex itself.
To avoid confusion, we represent weights as diagonal weight matrices $W_k \in \mathbb R^{|\cc_k|\times |\cc_k|}$ with positive diagonal entries.

There are a multitude of different, and sometimes conflicting, notions of what weights on cell complexes should represent and how and where they should interact with the computations.
In the next subsections, we first discuss the formal definition based on modifying the inner products on the chain spaces in \cref{sec:weightsandinnerproducts}.
Then, we discuss an intuition from resistor nets (\cref{sec:weightsintuitionelectricalsystems}) and show an example from a geometric complex (\cref{sec:weightsgeometricccs}) that both mirror this formal definition.
We generalize the normalized random walk Hodge Laplacian from \cite{schaub2020random} to cell complexes by defining weights in \cref{sec:normalisedrandomwalk}.
Finally, we discuss some results from the literature on recovering continuous Laplace operators using weighted simplicial complexes in \cref{sec:weightsrecovery}.

\subsection{Weighted cell complexes and inner products}
\label{sec:weightsandinnerproducts}
Let us consider the space of signals on $k$-cells $C_k$ and recall that $c_k^i\in C_k$ denotes a signal that has value $1$ on the $i$-th $k$-cell and is $0$ on all other cells.
So far, we have implicitly assumed that the $c_k^i$ form an \emph{orthonormal basis} of the signal space, i.e. $\lVert c_k^i\rVert_{C_k}^2=1$ (cells have norm $1$) and $\langle c_k^i, c_k^j \rangle_{C_k}=0$, if $j\neq i$ (orthogonality).

Orthogonality corresponds to our central notion of each cell being independent and being able to carry independent signals, so we will keep this assumption.\footnote{If one uses cell complexes as discretisations of a manifold of dimension $d$, one could for example argue that $1$-cells correspond to different directions in some tangent space of dimension $d$, and thus that $1$-cells pointing in similar directions should not be orthogonal. However, we will not adopt this perspective.}

Setting $\lVert c_k^i\rVert_{C_k}^2 = W^{-1}_{k,ii}$ and \emph{changing the norm of our selected basis} is the natural way to incorporate weights into our mathematical framework. Then, we can write the inner product $\langle x,y\rangle_{C_k}$ of $C_k$ as $x^\top W^{-1}_k y$.
The fact that $W_k$ is diagonal reflects our assumption that the signals on single cells are orthogonal.

The boundary matrices are unchanged by this: $B_1$ still sends a flow on $1$-cells to the net-inflow on $0$-cells, $B_1 x=0$ still  indicates a flow without source and sink and $B_{k-1}B_{k}=0$ still holds. However, the coboundary matrices $B^k$, i.e. the adjoints of $B_k$, change:
We recall that in \cref{rmk:coboundary} they were given by an equality between inner products
\begin{align*}
\langle B_k s_k,s_{k-1}\rangle_{C_{k-1}}&=\langle s_k,B_k^*s_{k-1}\rangle_{C_{k}}\\
s_{k}^\top B_k^\top W^{-1}_{k-1}s_{k-1} &= s_k^\top W^{-1}_{k}B_{k}^*s_{k-1}
\end{align*}
for all signals $s_k\in C_k$ and $s_{k-1}\in C_{k-1}$, which gives $B_k^*=W_{k}B_k^\top W^{-1}_{k-1}$.
Changing the inner product to take into account the weights will thus change the adjoints.
Plugging in the updated adjoints $B_k^*$ into \cref{def:hodgelaplacian} then gives rise to a Hodge Laplacian on weighted cells
\begin{align*}
L^{W,u}_k &= B_k^*B_k+B_{k+1}B_{k+1}^*\\
&=W_{k}B_k^\top W^{-1}_{k-1}B_k+B_{k+1}W_{k+1}B_{k+1}^\top W^{-1}_{k}
\end{align*}
We can then continue to use $L^{W,u}_k$ as in prior applications and recover the unweighted Laplacian for $W_i=I$.
However, there are three caveats with this definition:
\begin{enumerate}
	\item Our basis elements have different norms, making the basis choice unintuitive
	\item The corresponding Hodge Laplacians $L^{W,u}_k$ are not symmetric anymore, leading to different left- and right eigenvectors
	\item The weighted coboundary is not the ordinary transpose of the boundary in the original basis: $B_k^*=W_kB_k^\top W_{k-1}^{-1}\neq B_k^\top$ in general.
\end{enumerate}
We can solve the above issues by choosing an orthonormal basis of $C_k$ by rescaling the original basis entries:
In the new basis, the $i$-th basis vector is $\tilde{c}_k^i=W^{1/2}_{k,ii}c_k^i$.
Equivalently, if $\tilde{x}_k$ denotes coordinates in this orthonormal basis and $x_k$ coordinates in the original basis, then $x_k=W_k^{1/2}\tilde{x}_k$.
Changing into this basis means conjugating $L_k^{W,u}$ with $W_k^{-1/2}$ thus obtaining a symmetric weighted Laplacian:
\begin{align*}
L_k^W&=W_k^{-1/2}L_k^{W,u}W_k^{1/2}\\
& = W_k^{-1/2}\left(W_{k}B_k^\top W^{-1}_{k-1}B_k+B_{k+1}W_{k+1}B_{k+1}^\top W^{-1}_{k}\right)W_k^{1/2}\\
& = W_{k}^{1/2}B_k^\top W^{-1}_{k-1}B_kW_{k}^{1/2}+W_{k}^{-1/2}B_{k+1}W_{k+1}B_{k+1}^\top W_{k}^{-1/2}
\end{align*}
This now leads to the following definitions:

\begin{definition}[Weighted Boundary Matrix]\label{def:weighted_boundary_matrix}
	Given a weighted cell complex $\cc$ with boundary matrices $B_i$ and weight matrices $W_i$, the $k$-th \emph{weighted boundary matrix} $B_k^W$ is given by
	\[B_k^W = W_{k-1}^{-\frac{1}{2}}B_kW_k^{\frac{1}{2}}.\]
\end{definition}

\begin{definition}[Symmetric Weighted Hodge Laplacian]\label{def:weighted_hodge_laplacian}
	Given a weighted cell complex $\cc$ with weighted boundary matrices $B_k^W$, the $k$-th \emph{weighted Hodge Laplacian} is given by
	\[
	L_k^W =  \underbrace{B_{k+1}^W[B_{k+1}^{W}]^{\top}}_{{L_k^W}^\text{up}}+\underbrace{[B_{k}^{W}]^\top B_{k}^W}_{{L_k^W}^\text{down}}.
	\]
\end{definition}
While this rectifies the earlier identified issues, the symmetric weighted boundary matrices don't allow for an interpretation of $B_k^W$ simply as a net-inflow, or $[B_k^W]^\top$ as summing flow/signal along the boundary of a high-dimensional cell (a basis that fulfills the last requirement can be found by applying another scaling step with $W_k^{1/2}$.).
However, we still have that the boundary of a boundary is $0$, i.e., $B_k^W B_{k+1}^W=0$.

We will discuss the intuition behind weights and signals with concrete examples in the upcoming paragraphs.
\Cref{tab:weightsinfluence} shows the influence of changing individual weights on  gradient, curl, and harmonic eigenvectors and eigenvalues.

\begin{table}[h]
	\centering
	\begin{tabular}{lcccccc}
		\toprule
		& $\lambda_\text{grad}$ & $\lambda_\text{harm}$ & $\lambda_\text{curl}$ & $e_\text{grad}$ & $e_\text{harm}$ & $e_\text{curl}$\\
		\midrule
		$W_{k-1}\uparrow$ &  \textcolor{red}{$\downarrow$}& \textcolor{gray}{$\rightarrow$} & \textcolor{gray}{$\rightarrow$}& \textcolor{blue}{$\uparrow\downarrow$} & \textcolor{gray}{$\rightarrow$} & \textcolor{gray}{$\rightarrow$}\\
		$W_{k-1}\downarrow$ &\textcolor{green}{$\uparrow$} &\textcolor{gray}{$\rightarrow$}& \textcolor{gray}{$\rightarrow$}&  \textcolor{blue}{$\uparrow\downarrow$}& \textcolor{gray}{$\rightarrow$} & \textcolor{gray}{$\rightarrow$} \\
		$W_k\uparrow$ & \textcolor{green}{$\uparrow$}& \textcolor{gray}{$\rightarrow$} & \textcolor{red}{$\downarrow$}&\textcolor{blue}{$\uparrow\downarrow$} & \textcolor{blue}{$\uparrow\downarrow$}& \textcolor{blue}{$\uparrow\downarrow$} \\
		$W_k\downarrow$ & \textcolor{red}{$\downarrow$} & \textcolor{gray}{$\rightarrow$}& \textcolor{green}{$\uparrow$} & \textcolor{blue}{$\uparrow\downarrow$}& \textcolor{blue}{$\uparrow\downarrow$}& \textcolor{blue}{$\uparrow\downarrow$}\\
		$W_{k+1}\uparrow$ & \textcolor{gray}{$\rightarrow$}&  \textcolor{gray}{$\rightarrow$}& \textcolor{green}{$\uparrow$}& \textcolor{gray}{$\rightarrow$} & \textcolor{gray}{$\rightarrow$} & \textcolor{blue}{$\uparrow\downarrow$}\\
		$W_{k+1}\downarrow$ & \textcolor{gray}{$\rightarrow$}&  \textcolor{gray}{$\rightarrow$} &\textcolor{red}{$\downarrow$}& \textcolor{gray}{$\rightarrow$}& \textcolor{gray}{$\rightarrow$} &\textcolor{blue}{$\uparrow\downarrow$}\\
		\bottomrule
	\end{tabular}
	\caption{\textbf{Influence of changing individual weights on eigenvalues and eigenvector entries of both the symmetric and unsymmetric Hodge Laplacians.} The arrows indicate an increase \textcolor{green}{$\uparrow$}, decrease \textcolor{red}{$\downarrow$}, no change \textcolor{gray}{$\rightarrow$}, or both changes possible \textcolor{blue}{$\uparrow\downarrow$}.}
	\label{tab:weightsinfluence}
\end{table}

\subsection{Intuition from electrical systems behind the weighted Hodge Laplacian}
\label{sec:weightsintuitionelectricalsystems}
We consider the intuition of electrical systems given by graphs, where edges correspond to resistors.
We first assume that the signal space on $0$-cells corresponds to electric potentials, that $W_1$ denotes the conductances of the $1$-cells, and that $W_0=I$.
Given a set of potentials $\phi\in C_0$ on the nodes, applying the adjoint of the boundary matrix $B_1^\top \phi=V$ retrieves the voltages on the edges.
Multiplying this by the conductance weights $W_1$, we obtain the edge currents $I=W_1B_1^\top \phi$.
If one wants to encode resistances instead, the conductance matrix $W_1$ has to be replaced by the inverse resistance matrix.
The total power of the system is now given by applying the $0$-Hodge Laplacian
\[
P=V^\top I=\phi^\top \underbrace{B_1 W_1B_1^\top }_{=L_0^W}\phi.
\]
This reveals one important detail: Depending on our interpretation of signals on edges either as current or as voltages we need to factor in the weights $W_1$ at different positions.

We can define an inner product/norm on the $1$-chains that returns the power of the given constellation.
If the $C_1$ express currents, we get the inner product
\[
P=I^\top W_1^{-1}I=\langle I,I\rangle_{W_1^{-1}}
\]
whereas for voltages we get
\[
P=V^\top W_1V=\langle V,V\rangle_{W_1}.
\]
If we were to choose the orthonormal basis leading to the symmetric Hodge Laplacians of \cref{sec:weightsandinnerproducts}, the signals on the $1$-chains would encode $\sqrt{V\cdot I}$, something inbetween voltage and current.
This underlines the earlier claim that the symmetric Hodge Laplacian enjoys desirable mathematical properties, but restricts physical intuition.
If we stick to the current interpretation of $C_1$, $B_1I$ measure the net inflow of current of every node and $B_1 I=0$ would formalise Kirchhoff's junction law.
In the same interpretation, $B_2^\top I$ would measure the circular flow around $2$-cells, which is relevant when we assume to have a $2$-cellular complex representing a $2$-dimensional conductor where a current is magnetically induced.
When identifying $C_1$ with voltages across edges, $B_2^\top V=0$ would indicate that integrating voltages around returns $0$, mirroring another Kirchhoff's law.
Multiplying the above terms by weights on the $0$-cells, $1$-cells, or $2$-cells will change the underlying physical concepts.
This means that there is not one general way of dealing with weights, but rather that the choice of weights should be guided by the individual applications.

\subsection{Weights in geometric cell complexes}
\label{sec:weightsgeometricccs}
\begin{figure}[t]
	\centering
	\begin{minipage}{.32\linewidth}
		\centering
		\includegraphics[width=\linewidth]{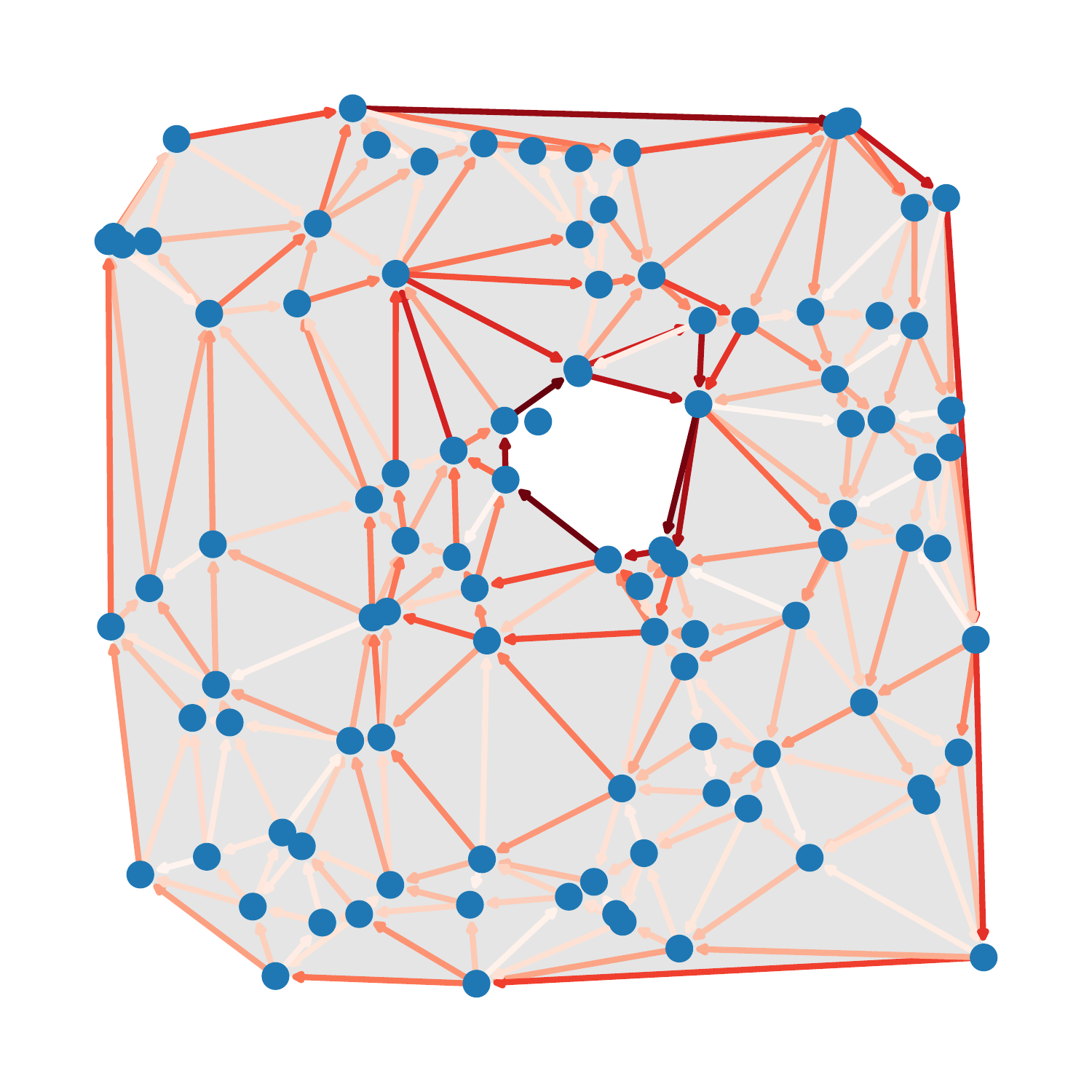}
		(a)
	\end{minipage}
	\begin{minipage}{.32\linewidth}
		\centering
		\includegraphics[width=\linewidth]{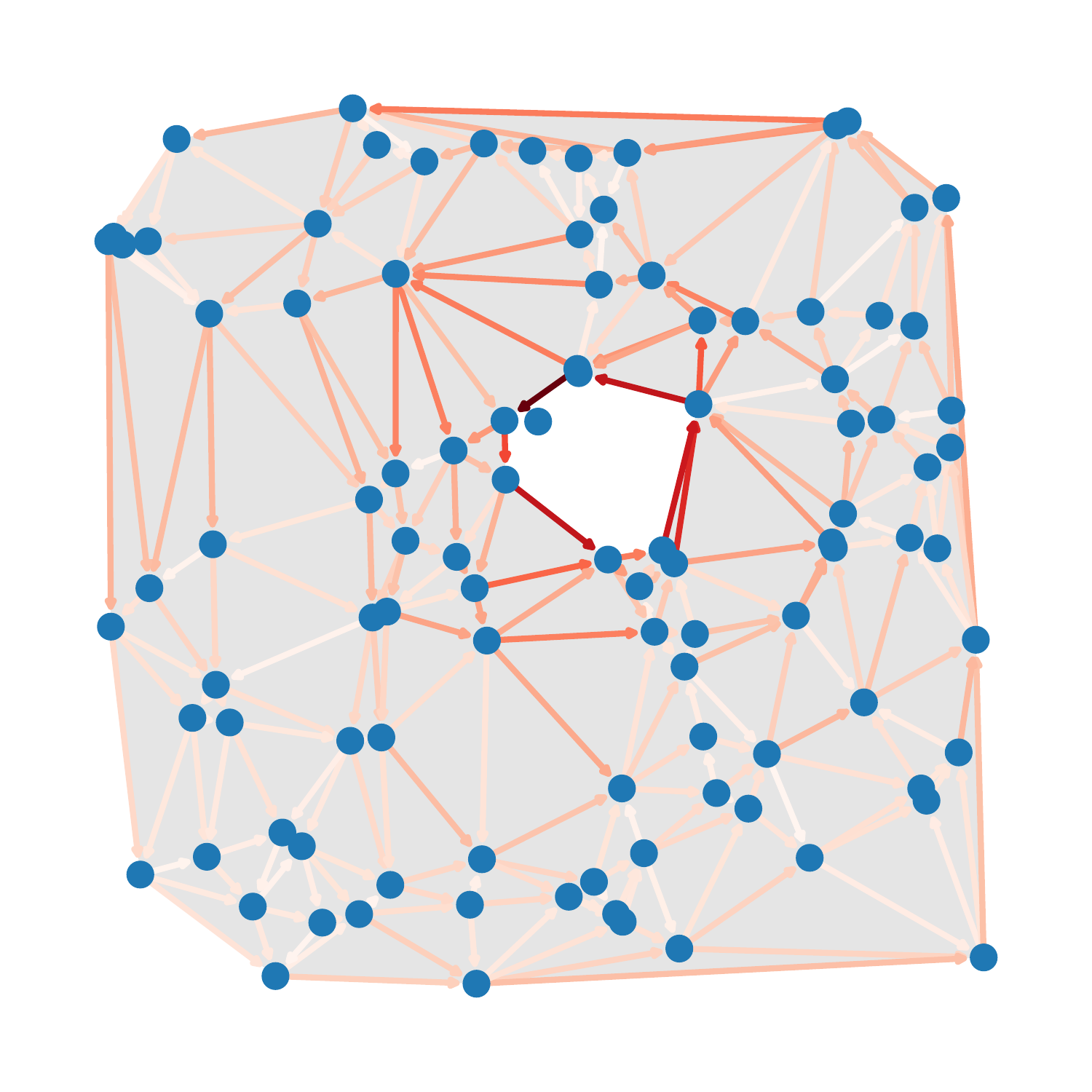}
		(b)
	\end{minipage}
	\begin{minipage}{.32\linewidth}
		\centering
		\includegraphics[width=\linewidth]{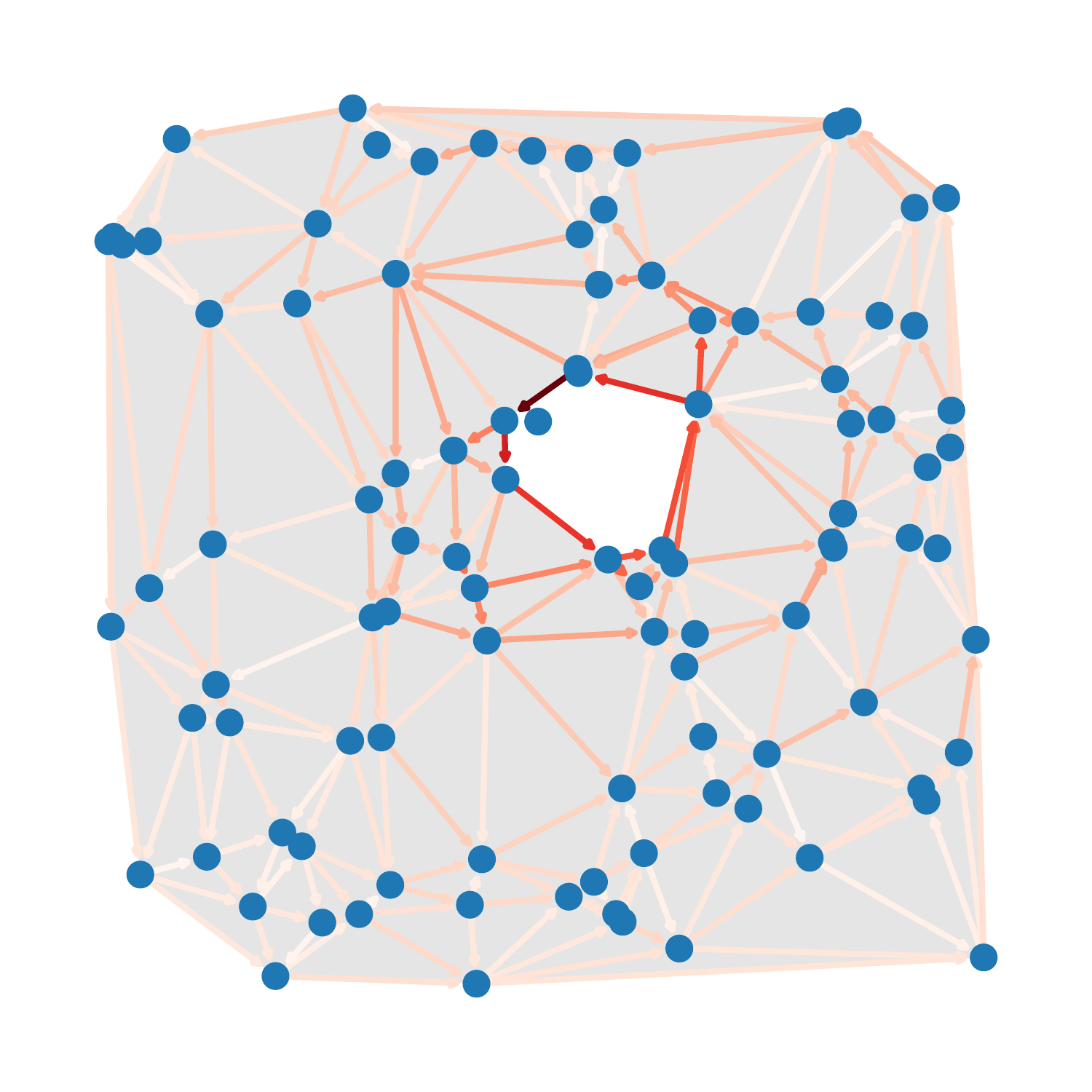}
		(c)
	\end{minipage}
	\caption{$2$-cell complex based on a Delaunay triangulation of a point cloud, with artificial hole in the center created by deleting simplices. The edge color shows the entries of the single harmonic eigenvector of the (a) unweighted and (b) symmetric weighted Hodge Laplacian. In (b), the weights are the inverse of the edge lengths. (c) depicts the re-scaled values according to \cref{eq:non-sym}.}\label{fig:weights}
\end{figure}

In geometric graphs, the size of cells (lengths, areas, volumes) constitutes a relevant property.
In particular, a cell complex and a subdivision of the same cell complex into more and smaller cells represent identical geometric objects and thus should be \enquote{similar} in terms of eigenvalues and eigenvectors of the associated Hodge Laplacians.
For the unweighted Hodge Laplacian, this is not the case, as \cref{fig:weights} (a) shows the harmonic eigenvector of the unweighted Hodge Laplacian.
There, we witness that longer edges will get assigned comparably larger eigenvector entries than shorter edges at similar positions.

This is correct in the sense that these edges contribute more to the harmonic flow.
However, if we want to interpret $1$-chains as the \emph{speed or flow rate of the flow}, the length should not matter.

Naively, we can try to account for this by setting the weights to the inverse of the edge lengths and then compute the eigenvectors of the associated symmetric weighted Hodge Laplacian.
This is shown in \cref{fig:weights} (b), but longer edges still have a larger flow value.
This behavior is consistent with the theoretical considerations, which tell us that the entries are the square root of the product of the total flow and flow rate per edge.

Let us instead focus on the two conditions which we want to be fulfilled for our interpretation a harmonic flow:
\begin{enumerate}
	\item \emph{Flow preservation/Divergence-free flow}, the net-inflow at every $0$-cell needs to be $0$, i.e.  $B_1c_H=0$.
	\item \emph{Curl-free flow}: Integrating the flow around the boundary of any $2$-cell needs to be $0$. Because the integral is proportional to the length of edge and $W_1$ encodes the inverse of the edge length, this means in mathematical terms that $B_2^\top W_1^{-1}c_H=0$.
\end{enumerate}

This matches the condition of being in the kernel of the unsymmetric Hodge Laplacian:
\begin{equation}\label{eq:non-sym}
	{L'}_1^W = W_1B_1^\top W_0^{-1} B_1 + B_2 W_2 B_2^\top W_1^{-1}.
\end{equation}
Because $W_0$ and $W_2$ are assumed positive definite, they do not change the nullspace conditions $B_1c_H=0$ and $B_2^\top W_1^{-1}c_H=0$; they affect scaling and nonzero spectral information, but not this harmonic-flow subspace.

\Cref{fig:weights} (c) shows the harmonic eigenvector of ${L'}_1^W$, where the values now correspond to the flow speed.

Thus, the eigenvectors of the weighted symmetric Hodge Laplacian (and thus the $k$-chains) are, again, an intermediate between the lower-order interpretation of flow speed and the higher-order interpretation of flow mass.
A higher cell weight will make the cell have more influence in computing the down-Laplacian and the associated down-energy and less importance in computing the up-Laplacian.
This matches the intuition that potential differences along short edges (and thus close nodes) and round flow along long edges will result in a high Hodge Laplacian energy.

\subsection{The normalized random walk Hodge Laplacian}
\label{sec:normalisedrandomwalk}
Analogously to the normalized graph Laplacian \cite{chung1997spectral} we can define a normalized random walk Hodge Laplacian \cite{schaub2020random} for different dimensions.
The original paper \cite{schaub2020random} defines the normalized random walk Hodge Laplacian in another non-symmetric way, but we will use the symmetric version as introduced in \cref{def:weighted_hodge_laplacian} for simplicity.
The two variants have the same eigenspectra; corresponding eigenvectors can be obtained via a transformation that is similar to the one used for the non-symmetric Hodge Laplacian above.

The normalization uses a different weight matrix in each dimension.
In \cite{schaub2020random}, every triangular $2$-cell has weight $1/3$, the reciprocal of its number of boundary $1$-cells.
We extend this convention by assigning each $2$-cell the reciprocal of its boundary size.
The resulting weight matrices are

\begin{align}
	W_2 &= \text{diag}\left(\left|B_2\right|^\top\mathbbm{1}\right)^{-1}\\
	W_1 &= \max\left(\text{diag}\left(\left|B_2\right| \mathbbm{1}\right), I\right)\\
	W_0 &= 2 \text{diag}(|B_1|W_1\mathbbm{1}).
\end{align}
Here, the maximum in the definition of $W_1$ is taken entrywise.

\subsection{Weighted Hodge Laplacians recover manifold Laplacians}
\label{sec:weightsrecovery}
For computational purposes, it is often practical to use a discrete simplicial or cell complex to approximate a continuous manifold.
Usually, we sample a set of points $X$ from a manifold $\mathcal{M}$.
As the number of sampled points increases, our intuition demands that the discretization approximates the behavior of the original manifold $\mathcal{M}$ with increasing accuracy.

There are multiple ways of constructing graphs, simplicial or cell complexes with vertices $X$ to recover the geometry and topology of the manifold $\mathcal{M}$.
Its non-negative Laplace--Beltrami operator $\Delta$ sends a smooth function $f\colon\mathcal{M}\to\R$ to
\[
\Delta f=-\divg\nabla f.
\]
The Laplace--Beltrami operator determines the Riemannian metric on $\mathcal{M}$, and thus encodes the geometry and topology of $\mathcal{M}$.
In the widely known diffusion maps paper \cite{coifman2006diffusion}, the authors prove a convergence guarantee of a weighted discrete $0$-Laplacian to the Laplace--Beltrami operator on the manifold.
Let
\[
K(x,y)=\begin{cases}
	\exp \left( -\frac{\lVert x-y\rVert^2}{\sigma^2}\right) & \text{if }\lVert x-y\rVert^2<\varepsilon\\
	0& \text{else}
\end{cases}
\]
be an exponential kernel with radius $\varepsilon$ and bandwidth $\sigma$ and let $K$ be the diagonal matrix whose entry $K_{ii}$ is the kernel evaluated on the two vertices of the $i$-th $1$-cell $c_1^i$.
Then we define the vertex degree vector $d=\lvert B_1\rvert K\mathbbm{1}$ and the diagonal vertex degree matrix $D=\diag(d)$.
Writing
\[
D_-=\diag\left((-B_1^{<0})^\top d\right),
\qquad
D_+=\diag\left((B_1^{>0})^\top d\right),
\]
we define
\[
\tilde{K}\coloneqq D_-^{-1}KD_+^{-1},
\]
which is the density-normalised version of $K$ where the value on every $1$-cell is divided by the product of the local densities at its two endpoints, where $B_1^{>0}$ and $B_1^{<0}$ denote the boundary matrices where only the non-zero entries larger or smaller than $0$ remain.
Let
\[
\tilde{D}\coloneqq\diag\left(\lvert B_1\rvert\tilde{K}\mathbbm{1}\right).
\]
We then set $W_0=\tilde{D}$ and $W_1=\tilde{K}$.
With the convention of \cref{def:weighted_hodge_laplacian}, the resulting symmetric diffusion Laplacian is
\[
L^\text{sym}_\text{dif}
=\tilde{D}^{-1/2}B_1\tilde{K}B_1^\top\tilde{D}^{-1/2}.
\]
Under the usual sampling and regularity assumptions, a suitably bandwidth-rescaled version of this operator converges to the non-negative Laplace--Beltrami operator as the sample size increases and the bandwidth tends to zero \cite{coifman2006diffusion}.
 
Even though the $0$-Laplacian (also graph Laplacian) is the most important, there are several works studying the convergence behaviour of higher-order Hodge Laplacians.
Chen et al.~\cite{chen2021helmholtzian} prove spectral consistency of the $1$-down-Laplacian and pointwise convergence of the $1$-up-Laplacian for Vietoris--Rips simplicial complexes with radius $\varepsilon$, an exponential gaussian edge kernel $K$ like above with bandwidth $\sigma$.
Then, the weight of a $2$-simplex $(x,y,z)$ is given by the product of the three edge kernels
\[
W_2(x,y,z)\coloneqq K(x,y)K(y,z)K(x,z),
\]
the weight on $1$-simplices given by generalised degrees
$
W_1=\diag(\lvert B_2\rvert W_2\mathbbm{1})
$
and analogously the weight on $0$-simplices by $W_0=\diag(\lvert B_1\rvert W_1\mathbbm{1})$.
The bandwidth $\sigma$ and maximum radius $\varepsilon$ should be chosen such that $\sigma \sim \varepsilon^{2/3}$.
The authors only show the convergence separately for the up- and down-Laplacian, which means that the scaling between the two parts of the Hodge Laplacian needs to be determined empirically.
Furthermore, pointwise convergence of the up-Laplacian is less strong than full spectral convergence.

The authors of \cite{lerch2025empirical} again study the behaviour of the discretised Hodge Laplacian on simplicial complexes and its approximation of the manifold Laplace--Beltrami operator, although with slightly different notation and a different setting.
They propose the following weighting of the $k$-simplex $(x_0,\dots,x_k)$ on point cloud $X$
\[
W(x_0,\dots,x_k)=\frac{1}{\binom{|X|}{k+1}}\frac{k!}{(2\sigma)^k}
\left(
\frac{1}{k+1}\sum_{i=0}^k\prod_{\substack{j=0\\j\ne i}}^k K_\sigma(x_i,x_j)
\right).
\]
It is immediately clear that these weights differ from the weights introduced in \cite{chen2021helmholtzian}.

Combining the above insights, it seems weights with the correct convergence behaviour need to
\begin{enumerate}
	\item Converge to $0$ fast enough for simplices with vertices that are too far apart.
	\item Have some form of consistency between different dimensions.
	\item Correctly account for sampling heterogeneity.
	\end{enumerate}
In summary, choosing the correct weights for empirically good recovery, even before passing to the limit, is still an open research question.
This is true for simplicial complexes, but it is even more true for cell complexes.
Despite all the open questions, weighting cell complexes is \emph{crucial} in virtually all applications and can significantly improve computational results.
However, as of now, there is not a single one-fits-all recipe for choosing the correct weights.
In contrast, the ramifications of the weights on the problem at hand and at the computational steps and matrices involved need to be carefully considered.
 
\section{Applications and Tools}\label{sec:applications}
This section aims to provide an introduction to methods involving abstract cell complexes.
Also note that due to their similarity, many methods that utilize the boundary structure of simplicial complexes can be easily adapted to cell complexes.

\subsection{Topological signal processing}
\label{par:TSP}
Graph Signal Processing \cite{dong2020graph,ortega2018graph} enables signal processing over signals that are defined on the nodes of a graph.
For this, the Graph Laplacian $L = D - A$ (where $D$ is the (diagonal) degree matrix of $G$) is used as a differential operator.
For simplicial complexes and \CC{}s, Topological Signal Processing \cite{roddenberry2023signal,roddenberry2022signal,sardellitti2024topological,schaub2020random,schaub2021signal} uses the boundary matrices and the Hodge Laplacian to enable signal processing over simplices or cells of any dimension.
See also \cite{isufi2024topological} for an overview of recent advances and \cite{petruso2026topological} for a more in-depth application-oriented tutorial mostly focused on simplicial complexes.
Depending on the exact goal, different variants of the Hodge Laplacian (see \cref{sec:topology,sec:weights}) can be used.
For simplicity, we will only mention the standard Hodge Laplacian $L_k$ in the following; however, the other variants can simply be substituted.

The boundary matrices already act as operators of signals (chains) on $k$-cells.
For an edge signal $s_1 \in C_1$, we can calculate the divergence (net in- or outflow per node) as $B_1 s_1$.
Analogously, the \emph{curl} (net flow around the boundaries of $2$-cells) can be calculated as $B_2^\top s_1$.
While simple, these operations extract information that has direct interpretations in many systems, for example for processing physical flows in water distribution systems \cite{cattai2025leak}.

The $k$-th Hodge Laplacian induces the $k$-th Hodge decomposition, partitioning the signal space $C_k$ into signals induced by $k-1$-dimensional structures ($\Ima B_{k}^\top$), $k+1$-dimensional structures ($\Ima B_{k+1}$), and $k$-dimensional holes or homology ($\ker L_k$), see \cref{thm:hodgedecomposition}.
\Cref{fig:eigendecomp} shows the Hodge decomposition of the $1$-chains of a cell complex, which are the most commonly used kinds of signals in TSP.
We can see that the curl space (induced by higher-dimensional structure) of $L_1$ corresponds to the space induced by lower-dimensional structure in $L_2$, while the harmonic spaces correspond to homology in different dimensions and are not related by the same non-zero up/down spectral correspondence.
This decomposition has been used, for example, to denoise and sparsely represent edge flows, to design filters acting differently on gradient, curl, and harmonic subspaces, and to analyze biological signals such as RNA-velocity fields and biomolecular structures \cite{schaub2018flow,hoppe2023representing,barbarossa2020topological,yang2022simplicial,liu2024unrolling,yang2024hodge,su2024hodge,wei2022hodge}.
See \cref{sec:trajectory-classification} for its use in trajectory classification.

Topological Signal Processing can also utilize the entire Eigenspectrum of the Hodge Laplacian via the eigendecomposition:
\begin{equation}
	L_k = U_k \Lambda_k U_k^\top
\end{equation}
\noindent
where $U_k$ is the matrix of eigenvectors and $\Lambda_k$ is the diagonal matrix of eigenvalues.
It is also possible to filter gradient, curl, and harmonic eigenvectors separately.

The Eigenbasis $U_k$ is analogous to the Fourier basis in classical signal processing.
Consequently, we can define filters based on the eigenvalues that are defined by a filter function $f\colon \mathbb{R} \to \mathbb{R}$.
In the filter, the eigenspace of eigenvalue $\lambda_i$ is amplified by a factor of $f(\lambda_i)$.
Given a signal $x \in C_k$, we can construct a coefficient matrix $f(\Lambda_k)=\diag(f(\lambda_0),f(\lambda_1),\dots)$ and apply the filter as a linear operation:
\begin{equation}
	\hat x = U_k f(\Lambda_k) U_k^\top x
\end{equation}

In many cases, the exact eigendecomposition is difficult to compute.
For large networks, it may be necessary to operate on a truncated selection of eigenvectors.
However, common filtering operations can be performed more efficiently.
For example, after normalising the spectrum to lie in $[0,1]$, the low-pass filter $f(\lambda)=1-\lambda$ can be written as $\hat x=(I-L_k)x$.

Apart from classical pre-defined filtering operations, the function $f$ can also be learned.
For example, \cite{canbolat2025cellular} learns a polynomial function of the eigenvalue to fit an autoregressive model.
This reduces the number of parameters in the model, making training much simpler.
In their examples, the authors use this methodology to predict weather parameters in the ocean and movements in a football game.

\subsection{Nonlinear topological signal processing across $2$ ranks}
\paragraph{Motivating nonlinearities in TSP}
Due to the strict structure of cell complexes, applying the boundary operators twice in a row will always return $0$.
This represents the fact that the boundary of a boundary is always empty, i.e., $B_{k-1}B_k=0$.
Thus, trying to send signals across two dimensions using concatenations of boundary operators will always fail, and canonical boundary operations can only process signals using adjacent dimensions of cells and their structural relations.
In particular, trying to define an upper $2$-rank-hop Laplacian
\[
L_{k,\mathrm{up}}^{(2)}
\coloneqq (B_{k+1}B_{k+2})(B_{k+1}B_{k+2})^\top
=B_{k+1}B_{k+2}B_{k+2}^\top B_{k+1}^\top
\]
will always return $0$.
Hence, in the classical linear processing of node signals, $2$-cells do not have any influence.
On the flipside, using only node signals, it is impossible to infer any $2$-cells through pure boundary-operator based methods like Laplacians. 

One way to augment this is to add another processing step between applying the two adjacent boundary operators:
The most straightforward way to do this is to apply an elementwise nonlinearity $\sigma$ to the intermediate result.
Specifically nonlinear TSP applies nonlinearities before or after (co-) boundaries, e.g., from the signals on nodes $c_0 \in C_0$ to the signals on polygons $c_2 \in C_2$:
\begin{align}
	c_2 = B_2^\top \sigma(B_1^\top c_0)
\end{align}
Note that this is in line with how artificial neural networks apply nonlinearities to intermediate results between matrix operations to increase the expressivity of the model.

However, if we want preserve the interpretation of the signals on the $k$-cells in terms of their orientation, we have to impose some conditions on the elementwise nonlinearity $\sigma$~\cite{roddenberry2021principled}.
Namely, since the orientation of cells is arbitrary and a change of orientation changes the sign of the corresponding signal, the function must be odd to ensure orientation equivariance of the overall processing pipeline:
\begin{align}
	\sigma(-x) = -\sigma(x).
\end{align}
Note that many common nonlinearities (e.g. ReLU and its variants) do not fulfill this.

\begin{figure}
	\includegraphics[width=\linewidth]{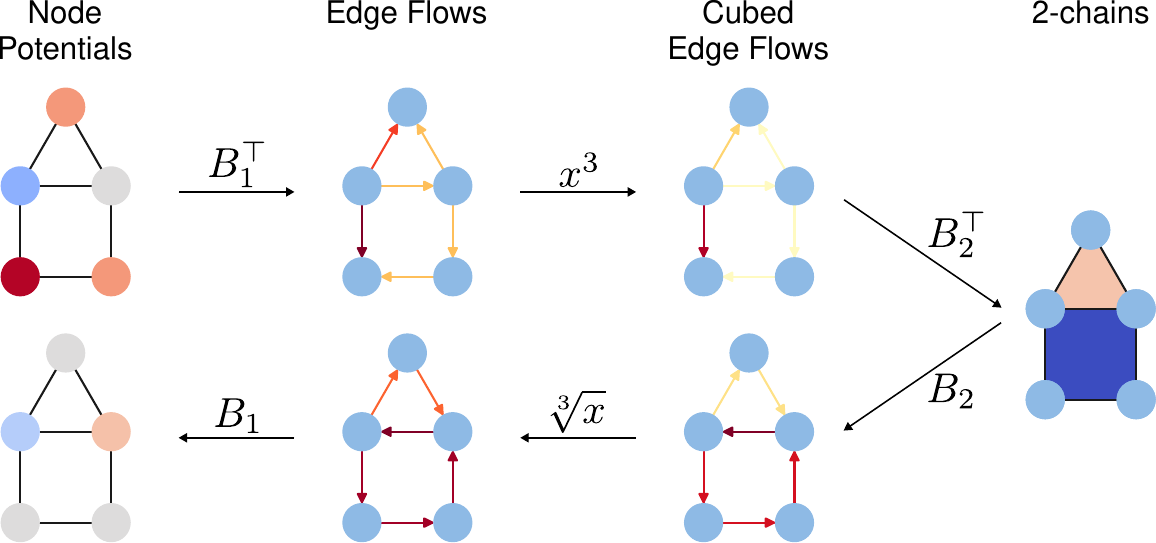}
	\caption{Minimal example for nonlinear TSP. By applying a nonlinearity before projecting the gradient signal up onto the $2$-cells, the resulting $2$-chain is not zero, increasing with increasing heterogeneity of clockwise and counterclockwise gradients. When projecting this down to the space of flows, we get the oriented sum of these heterogeneities. Taking the cubic root then tapers these sums, effectively introducing a divergence in opposite direction to the strongest flows.}\label{fig:nonlinear}
\end{figure}

\paragraph{Discussion of different nonlinearities}
First, we will consider some nonlinear functions $\sigma\colon \R\to\R$ that can be applied element-wise and fulfill this property and discuss the implications of using them.
With that intuition, we will consider the different functions in more detail to see their potential advantages and shortcomings.

The space of functions fulfilling the necessary conditions is very large, so we consider three different kinds of functions.
In machine learning, nonlinear functions that map monotonically from $\R$ to $[-1,1]$, such as the hyperbolic tangent or softsign, are commonly used and have the antisymmetric property.
The sine function is periodic and antisymmetric.
Odd power functions ($x^1$, $x^3$, $\dots$) and their inverses ($x^{1/1}$, $x^{1/3}$, $\dots$) are antisymmetric. Multiplication by non-zero constants preserves both antisymmetry and invertibility. More generally, odd polynomials are also antisymmetric but not always invertible.

Nonlinearities that map from $\mathbb{R}$ to $[-1,1]$ monotonically, such as $\tanh$, have to be tuned to the magnitude of the signal.
For signals with very small magnitude, the functions are approximately linear (often the identity) and thus will not bring any nonlinear effect.
For signals with very large magnitude, the functions are approximately the sign function, almost making even big differences between signals indistinguishable.

The periodicity of the sine function has to match the underlying data:
Santoro et al.\ \cite{santoro2025nodes} used $\sigma(x)\coloneqq\sin(x)$ for node-level fMRI (brain imaging) data.
They first extracted a phase from node-wise time series data, motivated by analogies from the Kuramoto model.
By processing the data to phases, the sine function is a natural nonlinearity that is compatible with the data.
In this instance, the $2$-chains ultimately measure the asynchronicity of signals on the nodes of a triangle.

Odd power functions and invertible odd polynomials have the advantage that they have a similar behavior at different scales.
Furthermore, by applying an invertible nonlinear function $\sigma$, performing an action (e.g. measuring the energy of the resulting chain) and applying the inverse $\sigma^{-1}$, we can revert the result to the original scale.

All three families of functions can benefit from adaptive scaling of the input vector to adapt to the scale of the input data or to fit some desired output data.
For example, we can make the hyperbolic tangent adaptable by scaling the input by a learnable weight $w$, i.e., $\sigma(x) \coloneqq \text{tanh}(wx)$.
The reader will recognize this as similar to the methodology of artificial neural networks.
In fact by applying multiple layers of boundary matrices and nonlinearities with learnable scale, we start designing a simple cell complex neural network.
See \cref{sec:tdl} for more information on this avenue.

\paragraph{Laplacian-like operators across $2$ ranks}
We already discussed that the invertibility of odd power functions makes it possible to design nonlinear methods that revert to the original scale.
To understand the effects of this better, we now consider a modification to basic methods in TSP by introducing nonlinearities to bridge the gap between $0$-chains and $2$-chains.
In linear \textsc{tsp}, the $0$th Hodge Laplacian (i.e., graph Laplacian) $L_0=B_1B_1^\top$ is associated with the Dirichlet form
\[
\mathcal E_0(x,y)
= \left\langle B_{1}^\top x, B_{1}^\top y\right\rangle_{C_1}
= \left\langle x, L_0 y \right\rangle_{C_0}.
\]
On the diagonal, $\mathcal E_0(x,x)=\|B_1^\top x\|_{C_1}^2$ is the sum of the squared differences of $x$ across the edges and therefore measures how strongly the node signal varies over the graph.

For the nonlinear construction, we define the node-to-polygon map
\[
\Phi_\sigma\colon C_0\to C_2,
\qquad
\Phi_\sigma(x)=B_2^\top\sigma\left(B_1^\top x\right).
\]
The quantity
\[
K_\sigma(x,y)
=\left\langle \Phi_\sigma(x),\Phi_\sigma(y)\right\rangle_{C_2}
\]
compares node signals through the nonlinear projection to $2$-cells.
It is a positive-semidefinite kernel on the node signals.
The corresponding scalar quantity is more naturally described as the nonlinear energy
\[
\mathcal E_\sigma(x)=\frac{1}{2}\left\|\Phi_\sigma(x)\right\|_{C_2}^2.
\]
In the case of $\sigma(t)=t^3$, the nonlinearity amplifies larger edge gradients, and $\Phi_\sigma(x)$ records the oriented sum of the cubed gradients around each $2$-cell.
Thus, it measures a nonlinear imbalance among the boundary gradients rather than an ordinary curl, which would vanish because $B_2^\top B_1^\top=0$.
\Cref{fig:nonlinear} \emph{Top} shows the effects of the individual steps.

The linear graph Laplacian can be recovered as the gradient of its energy,
\[
L_0x=\nabla_x\left(\frac{1}{2}\left\|B_1^\top x\right\|_{C_1}^2\right).
\]
This suggests defining the $2$-ranks Laplacian-like operator, or \enquote{nonlinear cross-rank energy operator}, associated with the nonlinear energy in the same way:
\[
\mathcal L_\sigma(x)
\coloneqq \nabla_x\mathcal E_\sigma(x)
=\mathrm D\Phi_\sigma(x)^\top\Phi_\sigma(x).
\]
If $\sigma$ is differentiable,
then $\mathrm D\Phi_\sigma(x)=B_2^\top \diag\left(\sigma'\left(B_1^\top x\right)\right)B_1^\top$ and hence we can write the nonlinear $2$-ranks Laplacian as
\[
\mathcal L_\sigma(x)
=B_1\diag\left(\sigma'\left(B_1^\top x\right)\right)B_2B_2^\top\sigma\left(B_1^\top x\right).
\]
Thus, we have constructed a $2$-ranks Laplacian, processing node signals across two dimensions by allowing a nonlinearity to act on the intermediate edge signals.
This then means that this Laplacian-like operator is no longer linear and hence cannot be represented by a matrix.
A separate version of the nonlinear Hodge Laplacian, which only incorporates up- and down-ward projections across a single dimension, has been considered in \cite{deville2021consensus}.

When we instead try to construct a Laplacian-like $2$-ranks operator that is equivariant to a change of scale, or that at least tries to limit the influence of the nonlinearity, we will end up with a different approach.
Let us assume that $\sigma$ is invertible and define the reverse face-to-node map and the corresponding $2$-ranks Laplacian-like operator, or \enquote{scale-equivariant node--face feature-transport operator}, as
\begin{align*}
\Phi_{\sigma}^{\text{rev}}(y) &\coloneqq B_1 \sigma^{-1}\left(B_2 y\right)\\
\mathcal{L}_{0,\sigma}(x) &\coloneqq \Phi_{\sigma}^{\text{rev}}(\Phi_{\sigma}(x)).
\end{align*}
For $\sigma(t) = c\cdot t^n$ with odd $n$ and $c\neq 0$, ensuring global invertibility over $\mathbb R$, this operator is equivariant to scaling of the input $x$ by a constant factor $a$, i.e., $\mathcal{L}_{0,\sigma}(ax) = a \mathcal{L}_{0,\sigma}(x)$.
However, this construction is less interpretable in practice and only has limited divergent flows.
In conclusion, while nonlinear node-to-polygon maps such as $\Phi_\sigma$ have been shown to work in multiple instances, designing reverse face-to-node maps with similarly useful behavior remains an open problem.

Due to the limitations of element-wise nonlinearities applied between boundaries, it is useful to consider more complex nonlinear relationships:
One such avenue of research is the use of stochastic partial differential equations that can model a latent space which informs dynamics of signals on the nodes \cite{liu2026topologicalkalman}.
By fitting this model for time series prediction, one can then infer higher-order structure informed by these dynamics.
The authors apply this method to water distribution systems, sensor networks, and traffic between bikesharing stations \cite{liu2026topologicalkalman}.

In conclusion, introducing nonlinearities into \textsc{TSP} is a promising avenue of research with broad potential applications.
As we have seen, while some ideas work well, there are many open problems that may be addressed by future research.

\subsection{Topological trajectory classification}\label{sec:trajectory-classification}
Many observations in the real world can be represented as trajectories, e.g.\ the movement of taxis \cite{liu2012understanding}, ocean drifters \cite{sykulski2016lagrangian}, and actions of robots \cite{hawasly2014multiscale}.
Trajectory Classification \cite{da2019survey} is concerned with classifying trajectories, either in a supervised or unsupervised setting.
Current topological trajectory classification methods embed trajectories into a simplicial complex and use the homology or harmonic projection \cite{hawasly2014multiscale,frantzen2021outlier} for classification. 
By diffusing the trajectory using the Hodge Laplacian, it is possible to increase the robustness of the classification \cite{grande2024topological}.
Further recent advances include the detection of optimal holes in a given complex \cite{grande2024topological}.

For spaces without much structure, such as the ocean, the triangulation works well.
However, there are many cases where cell complexes are a more natural representation than simplicial complexes.
For example, in a city, the road network can be represented by intersections (nodes), edges (roads), and blocks (polygons).
This principle transfers well to other geographic networks such as internet cables, power grids, or water pipes; and to regional and national scales.

Except for some special cases such as tunnels, the blocks (polygons) are simply the windows in the geographic embedding of the network.
If these cases do not represent a hole, they are usually small and can be spanned by properly configured cells.
In a simplicial complex, the blocks have to be triangulated, adding spurious edges.

\subsection{Topological data analysis}
\begin{figure}
	\centering
	\includegraphics[width=\textwidth]{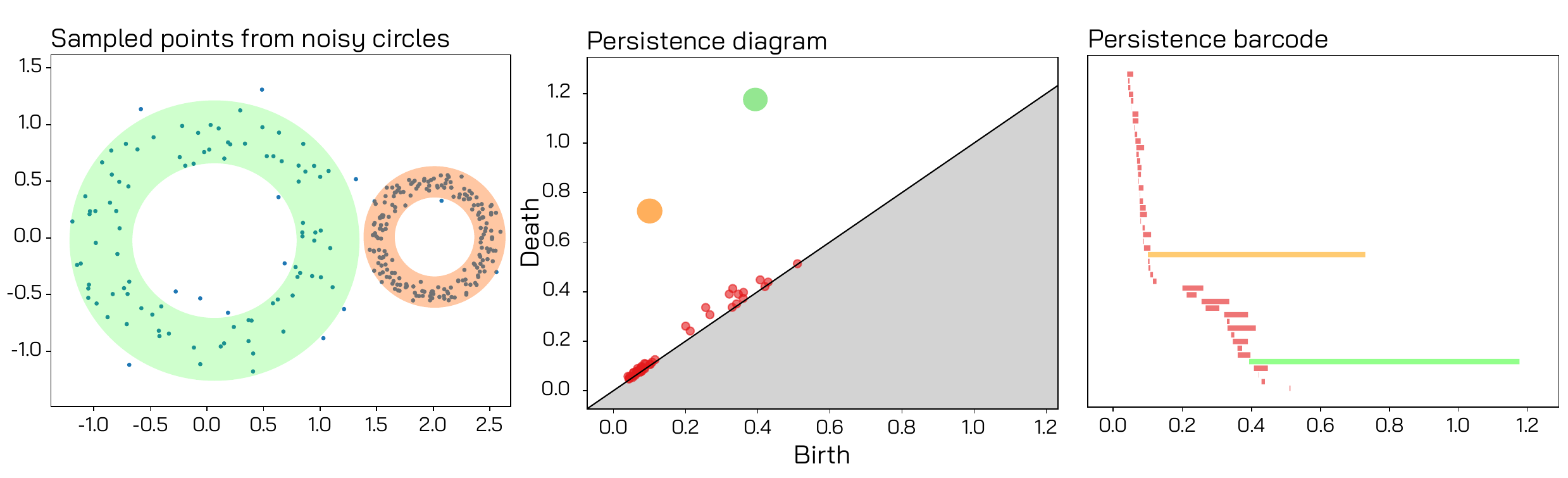}
	\caption{\textbf{Persistent homology diagram and barcodes.}
	\textbf{Left:} Original 2d point cloud with two most relevant loops highlighted.
	\textbf{Centre:} Associated persistent homology diagram. Topological features corresponding to the two most relevant loops are highlighted. We note that there are no points below the diagonal, as this would correspond to features dying before they were born.
	\textbf{Right:} Persistence barcodes with two most relevant bars highlighted.
	Computations were done with gudhi \cite{maria2014gudhi}.
	}
	\label{fig:PersistentHomology}
\end{figure}
The idea of topological data analysis (\textsc{tda}) is to uncover meaningful topological patterns in data like point clouds or images.
\emph{Topological pattern} here refers to loops, holes and voids across different scales, regions of low density surrounded by high-density regions, and complex connectivity patterns.
The most common idea is to associate a topological space, usually a simplicial complex for point clouds and a cubical complex for images (both special cases of cell complexes), to the data and then compute its homology; see \cref{def:homology}.
To capture homology across different scales, one can build a nested sequence (called \emph{filtration}) of complexes indexed by a scale parameter $\varepsilon$ (\cref{def:VRcomplex}).
By tracking the homology of the complex across the sequence, we end up with something called the \emph{persistent homology} of the data set.

From a technical point of view, assume we are given a filtration of cell complexes $\cc_\varepsilon$ for $\varepsilon \in \R_{\ge 0}$ with $\cc_\varepsilon\subseteq \cc_{\varepsilon'}$ for $\varepsilon\leq\varepsilon'$.
We call the inclusion of cell complexes $\iota_{\varepsilon,\varepsilon'}\colon \cc_\varepsilon \to \cc_{\varepsilon'}$.
The functoriality of homology gives us now an associated map between the homology groups,
\[
H_*(\iota_{\varepsilon,\varepsilon'})\colon H_*(\cc_\varepsilon) \to H_*(\cc_{\varepsilon'}).
\]
We now call the image of this map the persistent homology between $\varepsilon$ and $\varepsilon'$,
\[
P_{\varepsilon,\varepsilon'}(\cc_*)=\Ima H_*(\iota_{\varepsilon,\varepsilon'}).
\]
The collection of vector spaces $H_*(\cc_\varepsilon)$ and maps $H_*(\iota_{\varepsilon,\varepsilon'})$ is called the \emph{persistence module} of the filtration.
The beautiful insight about this theory is that the above persistence module decomposes into a direct sum of persistence bars, i.e.\ of homological features being born at a birth step $b$ and dying at a death step $d$ of the filtration.

This allows us to represent persistent homology in a persistence barcode, which is just a summary of the individual bars, or in a persistence diagram, where the $x$- and $y$-coordinate of every feature corresponds to its birth and death time.
For an example of this, see \cref{fig:PersistentHomology}.

\textsc{tda} has been used to analyse cancer progression (e.g. in \cite {lawson2019persistent}) or  protein structures \cite{benjamin2023homology}.
While most techniques in \textsc{tda} produce a single global description of the entire dataset, \cite{grande25topf} construct point-level features based on the most representative holes/homology classes, linking global topology to locally defined vectors.
Furthermore, \cite{deSilva2011persistent} extract circular (i.e.\ angular) coordinates defined on points for every single selected cohomology class.
Detailed introductions to \textsc{tda} include \cite{chazal2021introduction}, \cite{carlsson2021topological} or \cite{munch2017user}.

\begin{figure}
	\centering
	\includegraphics[width=.5\linewidth]{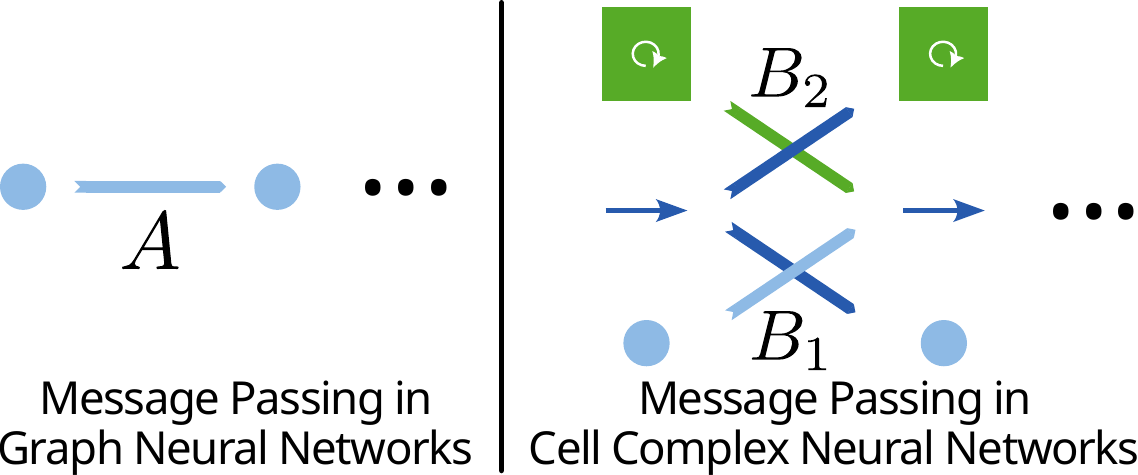}
	\caption{Message passing in graph neural networks compared to cell complex neural networks. Graph neural networks generally apply message passing via the adjacency matrix (i.e., along the edges). Cell complex neural networks generally propagate messages via the boundary matrices (i.e., to the boundary and coboundary). However, there is a variety of propagation methods that are possible, replacing or supplementing message passing along all boundaries. This includes propagating messages to upper- or lower-adjacent cells, only propagating along some boundaries or in certain directions.}\label{fig:ccnn}
\end{figure}

\subsection{Topological deep learning and cell complex neural networks}\label{sec:tdl}
Graph neural networks \cite{wu2020comprehensive} apply artificial neural networks to graphs (compare \cref{fig:ccnn}).
While many different architectures exist, most of them are based on multiple rounds of message passing.
Node $n$ is initialized with feature vector $x_n^{(0)}$.
In each round $i$, features from neighbors of a node are weighted with weights $W^{(i)}$ and aggregated using an aggregation function $\sum$.
Finally, a nonlinear activation function $\sigma$ is applied to obtain the updated feature vector $x_n^{(i+1)}$:

\begin{equation}
	x_n^{(i + 1)} = \sigma\left(\sum_{m \in \mathcal N(n)} W^{(i)} x_m^{(i)} + W'^{(i)} x_n^{(i)} + b^{(i)}\right).
\end{equation}

This represents a basic message passing architecture.
Depending on the variant, this can be modified with skip connections, two-hop neighbors, additional processing of messages using edge features or attention, etc.

Cell complex neural networks \cite{bodnar2021weisfeiler,hajij2022topological} generalize this idea to cell complexes by passing messages along the boundaries and coboundaries, and to upper- and lower-adjacent cells.
The additional structure improves the expressiveness compared to \textsc{gnn}s, increasing the prediction accuracy where the given $2$-cells capture meaningful information.
Let $c\in\cc_k$ be a $k$-cell, let $\mathcal N_k(c)$ denote its same-dimensional adjacency set, let $\partial c$ denote its boundary cells, and let $\operatorname{co}\partial(c)$ denote its cofaces.
A simplified cell-complex message passing update can then be written as
\begin{equation}
\begin{aligned}
x_c^{(i+1),k} = \sigma\Big(&\sum_{c'\in \mathcal N_k(c)} W^{(i)}_{k,k}x_{c'}^{(i),k}
{}+ \sum_{c'\in \partial c} W^{(i)}_{k,k-1}x_{c'}^{(i),k-1}\\
&{}+ \sum_{c'\in \operatorname{co}\partial(c)} W^{(i)}_{k,k+1}x_{c'}^{(i),k+1}
{}+ W_k'^{(i)}x_c^{(i),k} + b^{(i)}\Big).
\end{aligned}
\end{equation}

This is a unified and simplified definition of message passing in \textsc{ccnn}s.
For example, \cite{bodnar2021weisfeiler} introduces message passing between upper-adjacent cells that also uses the features of the shared coboundary.
Similar to \textsc{gnn}s, the full space of possible configurations is too vast to cover in detail in this overview.
Related architectures include simplicial neural networks \cite{ebli2020simplicial,roddenberry2021principled}, message-passing simplicial networks \cite{bodnar2021weisfeilerTopological}, attention-based simplicial networks \cite{giusti2022simplicialAttention,goh2022simplicialAttention,battiloro2023generalizedSimplicial}, and block Hodge-Laplacian based simplicial models such as \textsc{bscnet}s \cite{chen2022bscnets}.

\section{Outlook and Challenges}\label{sec:outlook}
In this section, we will identify five key challenges for the future of research on computational methods on cell complexes:
\begin{enumerate}
	\item Bringing data in cell complex form.
	\item Creating a generative cell complex model zoo like we have for graphs.
	\item Controlling the computational complexity and overhead with respect to graphs.
	\item Finding meaning in the higher-dimensions of cell complexes.
	\item Providing benchmark data sets and a standard data format.
\end{enumerate}
We will now discuss the above identified points one-by-one:
\paragraph{Obtaining cell complexes from data}
Cell complexes are powerful models that can abstract many real-world relationships.
However, data in the wild often does not come in the form of cell complexes.
While there are many methods to turn data sets into graphs, simplicial complexes or even hypergraphs, similar techniques for cell complexes are underdeveloped.
This presents a significant challenge for practitioners.
Thus, a very important problem remains how to meaningfully abstract data to take the form of cell complexes.
The challenge in obtaining cell complexes lies in the sophisticated structure presented by the boundary maps.
While there is a single possible hyperedge connecting all $n$ nodes of a fully connected graph, there are $(n-1)!/2$ different possibilities for $n>2$ for attaching a single $2$-cell connecting all nodes.
\paragraph{Generative models for cell complexes}
There is a wide variety of generative models for graphs that capture different aspects, such as community structure, degree distribution, or local neighborhoods.
For abstract cell complexes, available generative models currently control only a limited set of basic properties: graph liftings can prescribe the expected number of cells of each boundary size \cite{hoppe2024random} (analogously to the Erdős--Rényi model), while geometric approaches typically derive cells from Delaunay triangulations of random point clouds.

Part of the challenge to create a meaningful generative model for cell complexes is identifying the \emph{right properties to capture}.
As \CC s are more complex than graphs, there are more possible properties to choose from.
Locally, we can look at the number of faces and cofaces of cells, the number of lower- and upper-adjacent cells, and the size of the boundary of $2$-cells.
If possible, preserving local neighborhood structure would be very interesting as a more nuanced null model.
Globally, we have topological properties such as the Betti numbers, and (upper) connectedness in every dimension, and geometric properties like curvature.

Some straightforward properties on graphs have multiple possible generalizations to cell complexes: For example, a block structure could be modeled via cells spanning an individual block vs. multiple blocks, but the combinatorial complexity means the number of cells that possibly span multiple blocks is, in general, much larger than the number of cells that span a single block.

Thus, a key challenge for future research is to create a zoo of generative cell-complex models like we have for graphs. This will make it easier and more transparent to benchmark current and future computational cell complex methods on flexible and meaningful synthetic data.

\paragraph{Conquering computational complexity}

A major challenge for computational methods on cell complexes is their complexity.
While many methods behave nicely on graphs, increasing the dimension to higher-order cell complexes potentially introduces an exceedingly large number of possible cells and possible combinations to consider.

This challenge has to be addressed at two separate steps:
First, focus has to be put on extracting meaningful \emph{sparse} higher-order cell complexes.
This is often not easy, as methods either produce very few, or far too many higher-order cells:
Whereas the possible number of edges on a graph is quadratic in the number of nodes, the number of possible $2$-cells (simple cycles) is exponential.
While there already exist some methods for $2$-cell complexes,  \cite{battiloro2024latent, bodnar2021weisfeiler, hoppe2023representing}, extracting sparse cell complexes of dimension $3$ and above is still virtually uncharted territory.

Second, once a cell complex is constructed, computational methods have to take into account the possible large size of the cell sets and thus make limited use of expensive computation steps polynomial in the number of cells. 

\paragraph{Finding meaning in higher dimensions}

Cell complexes are incredibly versatile tools that can model relationships in arbitrary dimensions.
However, it is not always clear what these relations would mean on real-world datasets.
For $0$-cells, there are clear interpretations as single data points, locations, or points of interest.
The $1$-cells represent semantic, local, or geometric connections between $0$-cells.
In some applications dealing with flows, $2$-cells can keep track of the flow around a certain object or area.
It is a challenge for future work to unlock the full potential of the cell complex model by finding meaningful interpretations of the \emph{true} higher-order nature of cell complexes beyond dimension $2$.
This would mean being able to leverage the multi-dimensional interactions on data sets through the powerful structure in the form of boundary operators.

\paragraph{Benchmark data sets and standard data formats}

Standard benchmark data sets are common in areas like image processing (ImageNet, \cite{russakovsky2015imagenet}), point cloud learning (ShapeNet, \cite{chang2015shapenet}) or graph learning (TUDataset \cite{morris2020tudataset}).
Having benchmark datasets is important because a) they provide a way to meaningfully assess the strength of a newly proposed method and b) they stimulate competition and encourage development of novel and innovative methods in the relevant fields.
Despite all of this, there are no benchmark data sets yet for computational methods on cell complexes.
Recent tools such as TopoBench and TopoX already provide support for cell complexes, either through benchmarking pipelines with graph-to-cell liftings or through software infrastructure for constructing and learning on cell complexes \cite{telyatnikov2025topobench,hajij2024topox}.
However, they are not purpose-built benchmark collections for native cell-complex data.
Even for the comparatively popular simplicial complexes, general benchmark sets are lacking and most of the time just taken from graph learning.
The introduction of benchmark sets for cell complex data would thus be an important next step for the computational cell complex community and provide the cornerstone for promising future research.

Closely related is the adoption of a common data structure for cell complexes.
While packages like TopoX \cite{hajij2024topox} or cell-flower \cite{hoppe2023representing} provide formats for cell complexes, there is no commonly adopted computational representation, further hindering availability of standard cell complex data sets.
Thus, providing specifications of a universal cell complex data format is another task to be solved by a future comprehensive computational cell complex benchmark set.

\section{Conclusion}

In this paper we introduced abstract regular cell complexes as a computational framework for higher-order network modelling.
Our definition is self-contained and directly expressed in terms of boundary matrices and associated chain-space operators.
For dimensions up to $2$, we showed that this is equivalent to regular \CW{} complexes; in higher dimensions, our formulation yields a tractable algebraic model that captures the structures used in signal processing and network-science applications.

On this basis, we provided an overview of core methods and use cases, geared towards application-oriented readers without a background in algebraic topology.
This includes practical routes for constructing cell complexes from data, the role of homology and Hodge Laplacians for topological and geometric analysis, and extensions based on weighted and nonlinear operators.
Where appropriate, we translated methods from the setting of simplicial complexes to the more general setting of cell-complexes.

In closing we want to highlight once more that the use of cell complexes for data analysis task outside topological data analysis is still a relatively nascent area, and several issues remain open for future work.
As discussed in the previous section these issues include the design of constructions of complexes from data, the computational scalability of cell-complex based algorithms in higher dimensions, better benchmarks, and software toolkits.
We believe that progress on these fronts could substantially improve the adoption of cell-complex based methods in practice.
We hope that this work, by having exposed the mathematical structure of cell complexes in a more accessible form, will encourage readers to engage in these research areas.

\vspace{3cm}
\begin{center}
\large\textsc{Don't be Afraid of Cell Complexes!}
\end{center}
\newpage

\section{Declarations}

\subsection{Availability of Data and Materials}

The code and Jupyter notebooks accompanying this article are available at \url{https://github.com/josefhoppe/cell-complex-playground}.
No new datasets were generated or analysed as part of this work.

\subsection{Competing interests}

The authors declare no competing financial interests. A preprint version of this manuscript is available on arXiv, \href{https://doi.org/10.48550/arXiv.2506.09726}{doi:10.48550/\allowbreak arXiv.2506.09726}.

\subsection{Funding}

Funded by the European Union (\textsc{erc}, \textsc{high-hop}e\textsc{s}, 101039827) and by the German Research Council (\textsc{dfg}) within Research Training Group 2236. Views and opinions expressed are however those of the author(s) only and do not necessarily reflect those of the European Union or the European Research Council Executive Agency. Neither the European Union nor the granting authority can be held responsible for them.

\subsection{Authors' contributions}

All authors conceived of the scope and content of the manuscript. VPG and JH wrote the manuscript. VPG and JH created the figures and examples. All authors revised, read and approved the final manuscript.
\subsection{Acknowledgements}

Not applicable.

\subsection{Use of generative AI}
During the preparation and revision of this manuscript, the authors used generative AI tools to assist with drafting and revising passages, language editing, grammar checking, and identifying potential errors in the exposition and mathematical arguments. All outputs and suggestions from these tools were critically reviewed, independently verified, and edited by the authors, who take full responsibility for the final content of the manuscript.
	
	\appendix
	\crefalias{section}{appendix}

\section{Historical notions of cell complexes}
\label{sec:historical-definitions}
In the early days of algebraic topology in the first half of the 20th century, topologists were considering a wide range of possible notions for cell complexes.

The Mayer complexes, introduced by Mayer in \cite{Mayer1929AbstrakteTopologie,Mayer1929AbstrakteTopologieII}, are in modern language chain complexes of free $\mathbb{Z}$-modules, i.e., they are given by a sequence of free abelian groups $C_k$ together with boundary maps $\partial_k\colon C_k\to C_{k-1}$ such that $\partial_{k-1}\circ \partial_k=0$ for all $k$ \cite{EilenbergMacLane1945NaturalEquivalences}.
We note that modern cell complexes and abstract regular cell complexes are special cases of Mayer complexes (after forgetting the additional structure), where the $C_k$ are generated by the $k$-cells of the complex and the boundary maps are given by the boundary matrices.
However, Mayer complexes neither specify nor identify the underlying \emph{cells}, nor do they encode the attaching and regularity data from which the boundary maps of \CW{} and abstract regular cell complexes arise.

Tucker complexes, introduced by Tucker in \cite{Tucker1933AbstractApproach}, deal with the first above issue: They are essentially Mayer complexes with a fixed basis for each $C_k$ \cite{EilenbergMacLane1945NaturalEquivalences}.
The basis elements can then be viewed as the $k$-cells of the complex, and the coefficients of the boundary maps with respect to these bases play the role of incidence numbers.

In \cite{Lefschetz1941AbstractComplexes,Lefschetz1942AlgebraicTopology}, Lefschetz complexes are essentially Tucker complexes, but allow for negative dimensions and non-finiteness.

Finally, the complexes introduced by Newman in \cite{Newman1926FoundationsCombinatoryAnalysisSitusI,Newman1926FoundationsCombinatoryAnalysisSitusII} are essentially simplicial complexes whose simplices are given with an ordering of their vertices, which determines an orientation of each simplex.

\section{Regular \textsc{cw} Complexes and Combinatorial Complexes}
\label{sec:topdefinition}
We will now introduce and discuss the two most commonplace definitions of \CW{} complexes and discuss their drawbacks for computational topology.
In particular, we propose that a definition of regular \CW{} complexes for computational purposes must fulfil the following three requirements:
\begin{enumerate}
	\item The algebraic structure of the \CW{} complex in terms of its boundary matrices and associated Hodge Laplacians must be easily obtainable from the definition.
	\item Checking whether a given object $X$ satisfies the definition of \CW{} must be computationally feasible.
	\item The definition must align with the underlying geometric intuition of cells representing real-world lines, areas, volumes, etc.
\end{enumerate}
For the first point, all computational methods introduced or discussed in this paper rely on the boundary matrices or the Hodge Laplacians in some form or another.
Hence a definition which does not give us access to these matrices cannot be used in practice.
For the second point, it is very useful to be able to check whether a constructed object actually satisfies the given definition.
If this is not the case, we cannot give any theoretical guarantees for the correctness of the applied methods.
The standard modern definition uses the concept of pushout squares from category theory and builds the cell complexes iteratively over each dimension:
\begin{definition}[Finite regular cell complex I]
	\label{def:topCWcomplexI}
	A \emph{finite regular cell complex} $(X_*,\alpha_*)$ is a sequence of topological spaces $X_0\subset X_1\subset \dots \subset X_N=X$, where $X_0$ is a finite discrete set of $0$-cells, together with a series of maps $\alpha_k^1\dots \alpha_k^{n_k}$ with $\alpha^i_k\colon  S^{k-1}\to X_{k-1}$ for each $k>0$ such that each $\alpha_k^i$ is a homeomorphism onto its image and the following square
	\[
	\begin{tikzcd}
		\coprod_i S^{k-1}\ar[d, hook, "\iota"] \ar[r,"\coprod_i\alpha_i"] & X_{k-1}\ar[d]\\
		\coprod_i D^k\ar[r]&X_{k}
	\end{tikzcd}
	\]
	where $\iota \colon \coprod_i S^{k-1}\to \coprod_i D^k$ denotes the coproduct of the canonical inclusions of the $(k-1)$-sphere $S^{k-1}=\{x\in \R^k:\lVert x\rVert =1\}$ into the $k$-disk $D^k=\{x\in \R^k:\lVert x\rVert \leq 1\}$, is a pushout square.
\end{definition}
The requirement of the $\alpha_k^i$ being homeomorphisms on their images is the regularity condition.
Equivalently, one may formulate regularity in terms of characteristic maps $\varphi_k^i:D^k\to X_k$: their restrictions to $\mathring D^k$ are homeomorphisms onto open $k$-cells, and in the regular case the maps $\varphi_k^i$ are homeomorphisms from $D^k$ onto the corresponding closed cells.
Dropping this, we would obtain the definition of finite \CW{} complexes.
While the powerful language of category theory is beautiful in its own right, this level of abstraction is not required for the computational purpose of this paper.
In particular, checking the universality condition of the pushout square is not always feasible in practice.
The same applies for the homeomorphism condition.
The other definition, adapting \cite{hansen2019toward}, is closer examining the structure of $X$ being made up of individual cells:
\begin{definition}[Finite regular cell complex II]
	A \emph{finite regular cell complex} $X$ is a topological space with a partition into finitely many subspaces $X_1, \dots, X_N$, called \emph{open cells}, satisfying the following conditions
	\begin{enumerate}
	\item For all subspaces $X_i$ and $X_j$, the intersection of $X_j$ and the closure of $X_i$ is nonempty, $\bar{X_i}\cap X_j\not = \emptyset$ only if $X_j$ is a subset of $\bar{X_i}$, $X_j\subseteq\bar{X_i}$
		\item For every subspace $X_i$, there is a homeomorphism from $\bar{X_i}$ to a closed disk $D^{n_i}=\{x\in \R^{n_i}:\lVert x\rVert \leq 1\}$ restricting to a homeomorphism between $X_i$ and the open disk $\mathring{D}^{n_i}$.
	\end{enumerate}
\end{definition}
While this definition is very concise, it provides us with little insight on how to \emph{construct} the cell complex.
In particular, using the first condition together with the dimensions $n_i$ can help us piece together the face relations of the cell complex.
To compute the boundary maps, we would further need to extract some kind of orientations, which is made even more difficult by a direct application of the definition.

In summary, while the above definitions work well from a theoretical perspective of algebraic topology or category theory, they are not geared towards computational applications.

Finite regular \CW{} complexes also admit a face poset recording closure incidences among cells.
Under additional conditions, such as the \CW{}-poset conditions studied in the poset literature, a poset can reconstruct a regular \CW{} complex \cite{bjorner1984posets,hersh2014cwposets}.
This is closely related to the present boundary-matrix viewpoint, but encodes incidence order rather than oriented cellular boundary maps.
However, checking these conditions is not a purely computationally feasible task in arbitrary dimensions, for the same sphere-recognition obstructions discussed in \cref{rmk:twoways}.

\paragraph{Combinatorial complexes}
Hajij et al.~\cite{hajij2022topological} introduce the notion of combinatorial complexes:
\begin{definition}[Combinatorial Complex]
A \emph{combinatorial complex} is a triple  $(S,\mathcal{X},\text{rk})$ consisting of a set $S$ called \emph{vertices}, a set $\mathcal{X}$ of non-empty subsets of $S$ called \emph{cells} and a function $\text{rk}\colon \mathcal{X}\to \Z_{\geq 0}$ called \emph{rank function} with the following properties:
\begin{enumerate}
	\item For all vertices $s\in S$, $\{s\}$ is a cell in $\mathcal{X}$
	\item For two cells $x,y\in \mathcal{X}$ with $x\subseteq y$, we have that $\text{rk}(x)\leq \text{rk}(y)$.
	\end{enumerate}
\end{definition}

While the above definition is very easy to verify, it does not allow for a geometric intuition and does not have the structural properties needed for computational applications requiring the boundary matrices.

\section{Checking the Abstract Regular Cell Complex Conditions}
\label{sec:checking-arcc}

We will now describe how the conditions of \cref{def:abstractcellcomplex} can be checked from the finite boundary-matrix data.
This makes concrete our claim that deciding whether given finite data define an abstract regular cell complex is computationally feasible.
We note that in many examples of interest, this will not be necessary, as the cell complex construction procedure will already guarantee that it outputs an abstract regular cell complex.
We will manually verify the conditions for the toy complex in \cref{fig:1} in \cref{rmk:checking-toy-complex}, and then give a small example showing the difference between homological checks and spherical attaching boundaries in dimension $3$.

\begin{algorithm}[t]
	\caption{Checking the abstract regular cell complex conditions}
	\label{alg:check-arcc}
	\footnotesize
	\begin{tabbing}
		\qquad\=\qquad\=\qquad\=\kill
		\textbf{Input:} ordered cell sets $\cc_0,\dots,\cc_n$ and integer matrices $B_1,\dots,B_n$.\\
		\textbf{Output:} \texttt{true} if the data define an abstract regular cell complex, otherwise \texttt{false}.\\[2pt]
		\textbf{if} some $B_k$, $1\le k\le n$, does not have size $|\cc_{k-1}|\times |\cc_k|$ \textbf{return} \texttt{false}.\\
		\textbf{if} some entry of some $B_k$, $1\le k\le n$, is not in $\{0,\pm1\}$ \textbf{return} \texttt{false}.\\
		\textbf{if} some column of $B_1$ does not have exactly one $1$ and one $-1$ \textbf{return} \texttt{false}.\\
		\textbf{for} $k=0,\dots,n$ \textbf{do}\\
		\> \textbf{for} each cell $c_k^i\in\cc_k$ \textbf{do}\\
		\>\> Set $\hat{\cc}_k=\{c_k^i\}$.\\
		\>\> \textbf{for} $r=k,k-1,\dots,1$ \textbf{do}\\
		\>\>\> Set $\hat{\cc}_{r-1}=\partial \hat{\cc}_r$, using the non-zero entries of $B_r$.\\
		\>\> \textbf{for} $r=1,\dots,k$ \textbf{do}\\
		\>\>\> Form $\hat{B}_r\colon \Z^{|\hat{\cc}_r|}\to \Z^{|\hat{\cc}_{r-1}|}$ with the induced orders.\\
		\>\> \textbf{if} $\hat{B}_{r-1}\hat{B}_r\ne0$ for some $2\le r\le k$ \textbf{return} \texttt{false}.\\
		\>\> Compute the integral homology groups $H_q(\hat{\cc})$ for $0\le q\le k$ as below.\\
		\>\> \textbf{if} $H_0(\hat{\cc})\not\cong\Z$ or some $H_q(\hat{\cc})\ne0$ for $q>0$ \textbf{return} \texttt{false}.\\
		\textbf{return} \texttt{true}.
	\end{tabbing}
\end{algorithm}

\paragraph{Collecting iterated boundary cells}
For a fixed cell $c_k^i$, the first step is to form the smallest subcomplex containing $c_k^i$ respecting the boundary relation.
Starting with $\hat{\cc}_k=\{c_k^i\}$, one repeatedly takes boundaries:
\[
	\hat{\cc}_{r-1}=\partial \hat{\cc}_{r}.
\]
Here $\partial$ is the set-valued boundary from \cref{def:abstractcellcomplex}: a cell lies in $\partial \hat{\cc}_{r}$ if it appears with non-zero coefficient in the corresponding column of $B_r$ for some cell in $\hat{\cc}_r$.
Thus this step only uses the support pattern of the boundary matrices.

\paragraph{Building restricted matrices}
After the cells of the generated subcomplex have been collected, the matrices $\hat{B}_r$ are obtained by restricting $B_r$ to the rows indexed by $\hat{\cc}_{r-1}$ and the columns indexed by $\hat{\cc}_r$, with the induced orders.
The signs are not recomputed; they are inherited from the original boundary matrices.
In particular, the products $\hat{B}_{r-1}\hat{B}_r$ must vanish.
If they do not, then the boundary of a boundary is non-zero already inside the generated subcomplex, and the data fail the definition.

\paragraph{Checking homology}
The remaining test is the local homology condition.
For the generated subcomplex $\hat{\cc}$, form the chain complex
\[
	\Z^{|\hat{\cc}_0|}
	\longleftarrow
	\Z^{|\hat{\cc}_1|}
	\longleftarrow
	\dots
	\longleftarrow
	\Z^{|\hat{\cc}_k|}.
\]
The definition asks that this complex have the homology of a point:
\[
	H_0(\hat{\cc})\cong \Z,\qquad H_q(\hat{\cc})=0\quad\text{for }q>0.
\]
Let $m_q=|\hat{\cc}_q|$.
For integer matrices, ranks below are taken over $\mathbb{Q}$, equivalently as the number of non-zero Smith invariant factors.
The degree-$0$ condition is checked from the cokernel of $\hat{B}_1$.
If $k=0$, it is automatic, since the generated subcomplex consists of one $0$-cell.
Otherwise, compute a Smith normal form
\[
	U_1\hat{B}_1V_1=D_1,
\]
where $U_1$ and $V_1$ are unimodular and $D_1$ is diagonal.
Since
\[
	H_0(\hat{\cc})=\Z^{m_0}/\Ima \hat{B}_1
\]
we have $H_0(\hat{\cc})\cong \Z$ precisely when $D_1$ has exactly $m_0-1$ non-zero diagonal entries, all equal to $1$.

For $1\le q<k$, one has to check that $\Ima \hat{B}_{q+1}$ is the whole kernel of $\hat{B}_q$, as a lattice over $\Z$.
Compute
\[
	U_q\hat{B}_qV_q=D_q
\]
and let $K_q$ be the matrix formed by the last $s_q=m_q-\operatorname{rank}\hat{B}_q$ columns of $V_q$.
These columns form a $\Z$-basis of $\ker \hat{B}_q$.
Because $\hat{B}_q\hat{B}_{q+1}=0$, the columns of $\hat{B}_{q+1}$ can be written in this basis, giving an integer matrix $M_q$ with
\[
	K_qM_q=\hat{B}_{q+1}.
\]
Then the homology group is
\[
	H_q(\hat{\cc})\cong \Z^{s_q}/\Ima M_q .
\]
Thus $H_q(\hat{\cc})=0$ precisely when the Smith normal form of $M_q$ has $s_q$ non-zero diagonal entries, all equal to $1$.
If $k>0$, the top-degree condition is simply $\ker \hat{B}_k=0$, or equivalently $\operatorname{rank}\hat{B}_k=m_k$.
If one only needs the Betti numbers over a field $\mathbb F$, then rank-nullity checks suffice:
\[
	\dim_{\mathbb F}\ker \hat{B}_q^{\mathbb F}
	=
	\operatorname{rank}_{\mathbb F}\hat{B}_{q+1}^{\mathbb F}
	\quad(q>0),
	\qquad
	|\hat{\cc}_0|-\operatorname{rank}_{\mathbb F}\hat{B}_1^{\mathbb F}=1.
\]
Here $\hat{B}_{k+1}$ is interpreted as the zero map.
These field computations are often enough for real-valued Hodge-theoretic applications, but they do not by themselves exclude integral torsion.
For the literal definition over $\Z$, Smith normal form is the appropriate exact check.

\paragraph{Complexity and low-dimensional simplifications}
The boundary-collection step is linear in the number of non-zero entries inspected, for each generated subcomplex.
The expensive part is the homology computation.
Dense Smith normal form algorithms are polynomial in the matrix size and bit length, but can be costly because of coefficient growth; sparse and modular variants are preferable in implementations.
In many applications, however, the complexes are low-dimensional and the generated subcomplex around a single cell is small.
In dimension $2$, one usually does not need a full Smith normal form computation.
For a candidate $2$-cell with boundary column $b$, one checks that $B_1b=0$, that the selected edges and vertices form a connected graph, and that the resulting cycle space is generated by $b$.
With coefficients in $\{0,\pm1\}$, this is the graph-theoretic test that the boundary is a single simple oriented cycle, as in \cref{def:2dCellComplexes}.

\begin{remark}[Checking the toy complex]
	\label{rmk:checking-toy-complex}
	Consider the toy complex in \cref{fig:1}.
	The first boundary condition is immediate from the displayed matrix $B_1$: every edge column has exactly one $-1$ and one $1$.
	The condition for $0$-cells is automatic, and for each $1$-cell the generated subcomplex consists of one edge and its two endpoints, hence has the homology of a point.
	For the orange $2$-cell, the generated subcomplex consists of the vertices $\encircle{0},\encircle{3},\encircle{4}$, the edges $0\to3$, $0\to4$, and $3\to4$, and the orange $2$-cell itself.
	The restricted matrices are
	\[
		\hat{B}_1=
		\bordermatrix{ & 0\hspace{-1pt}\rightarrow\hspace{-1pt}3 & 0\hspace{-1pt}\rightarrow\hspace{-1pt}4 & 3\hspace{-1pt}\rightarrow\hspace{-1pt}4\cr
			\encircle{0} & -1 & -1 & 0\cr
			\encircle{3} & 1 & 0 & -1\cr
			\encircle{4} & 0 & 1 & 1},
		\qquad
		\hat{B}_2=
		\bordermatrix{ & \celltri\cr
			0\hspace{-1pt}\rightarrow\hspace{-1pt}3 & 1\cr
			0\hspace{-1pt}\rightarrow\hspace{-1pt}4 & -1\cr
			3\hspace{-1pt}\rightarrow\hspace{-1pt}4 & 1}.
	\]
	We have $\hat{B}_1\hat{B}_2=0$.
	The graph on the three selected vertices is connected, so $H_0\cong\Z$.
	Moreover, $\operatorname{rank}\hat{B}_1=2$, so $\ker \hat{B}_1$ has rank $1$, and it is generated by the primitive vector $\hat{B}_2\celltri$.
	Thus $H_1=0$.
	Finally, $\hat{B}_2$ is a non-zero map from $\Z$ to $\Z^3$, so $\ker \hat{B}_2=0$ and hence $H_2=0$.
	The green $2$-cell is checked in the same way.
\end{remark}

\subsection{A 3-dimensional ARCC that is not a CW{} complex}
\label{subsec:nonmanifold-3cell-boundary}

The definition of ARCCs checks for homological conditions, but as already mentioned in \cref{rmk:twoways}, this is not enough to guarantee that the attaching maps of the cells are homeomorphisms onto their images.
Already in dimension $3$, a candidate boundary can have the homology of $S^2$ without being a surface:

\begin{example}[A non-manifold boundary passing the homological check]
	Let $\cc_0=\{0,1,2,3\}$ and let $\cc_1$ consist of the six oriented edges
	\[
		e_{01},e_{02},e_{03},e_{12},e_{13},e_{23},
	\]
	oriented from the smaller to the larger vertex.
	Let $\cc_2=\{f_1,f_2,f_3,f_4\}$ with
	\[
	\begin{aligned}
		B_2f_1&=e_{02}-e_{03}-e_{12}+e_{13},\\
		B_2f_2&=e_{12}-e_{13}+e_{23},\\
		B_2f_3&=e_{01}-e_{03}+e_{13},\\
		B_2f_4&=e_{01}-e_{02}+e_{13}-e_{23}.
	\end{aligned}
	\]
	Finally, add one $3$-cell $g$ and set
	\[
		B_3g=f_1+f_2-f_3+f_4.
	\]
	With the boundary matrices induced by these formulas, one checks that
	\[
		B_1B_2=0,\qquad B_2B_3=0.
	\]
	Each displayed $2$-cell boundary is a simple oriented cycle, and adding the corresponding $2$-cell kills that cycle.
	Hence every generated subcomplex below dimension $3$ has the homology of a point.
	Moreover, the column of $B_3$ is a primitive generator of $\ker B_2$.
	Thus adding $g$ kills the only second homology class and, since $B_3g\ne0$, creates no third homology.
	The subcomplex generated by $g$ therefore has the homology of a point, so the boundary-matrix data pass the homological condition in \cref{cond:individualcells}.
	Together with the preceding lower-dimensional checks, these data satisfy the conditions of \cref{def:abstractcellcomplex}.

	Nevertheless, $K$ is not a cell decomposition of $S^2$.
	The edge $e_{13}$ occurs in the boundary of all four $2$-cells $f_1,f_2,f_3,f_4$.
	Thus an interior point of $e_{13}$ has four local sectors around it.
	In a surface, only two local sectors meet along such an edge; equivalently, the transverse link is $S^0$, i.e.\ two points.
	Hence the boundary of $g$ has the homology of $S^2$, but is not a surface and therefore is not homeomorphic to $S^2$.

	This is the first obstruction from \cref{rmk:twoways} in concrete form.
	The algebraic condition checks that the generated chain complex is homologically disk-like.
	A regular \CW{} $3$-cell requires more: its boundary must be a genuine $2$-sphere, and this includes local manifoldness.
\end{example}

\section{Abstract regular cell complexes are closed under taking products}
\label{sec:products-preserve-arcc}

In \cref{def:prodcellcomplexes}, we introduced the product of two cell complexes by taking pairs of cells and using the usual tensor-product sign convention for the boundary maps.
We now verify that this construction is compatible with \cref{def:abstractcellcomplex}.

\begin{theorem}[Products of abstract regular cell complexes]
\label{thm:products-preserve-arcc}
	Let $\cc=(\cc_*,B^\cc_*)$ and $\cc'=(\cc'_*,B^{\cc'}_*)$ be abstract regular cell complexes of dimensions $m$ and $n$.
	Choose any ordering of the product cells
	\[
	(\cc\times \cc')_{\hat{k}}=\mathop{\dot\bigcup}_{k+k'=\hat{k}}\cc_k\times\cc'_{k'}
	\]
	and define the product boundary matrices as in \cref{def:prodcellcomplexes}, i.e. on basis elements by
	\[
	B^{\cc\times\cc'}_{k+k'}(c_k\times c'_{k'})
	=
	(B^\cc_kc_k)\times c'_{k'}
	+
	(-1)^k c_k\times (B^{\cc'}_{k'}c'_{k'}),
	\]
	where missing terms for $k=0$ or $k'=0$ are interpreted as zero.
	Then $\cc\times\cc'$ is an abstract regular cell complex of dimension $m+n$.
	\end{theorem}
	
	Before proving the theorem, we recall the relevant parts of the definitions.
	An abstract regular cell complex consists of finite ordered sets of cells and boundary matrices with entries in $\{0,\pm1\}$, whose $1$-cell columns have one positive and one negative entry and whose subcomplex generated by any single cell has the homology of a point.
	The product construction of \cref{def:prodcellcomplexes} takes pairs of cells, orders them by total dimension, and equips them with the tensor-product boundary formula displayed above.
	Thus the proof has to check that these product boundary matrices still have coefficients in $\{0,\pm1\}$, satisfy Condition~\ref{cond:B1columns}, and satisfy the local homology requirement in Condition~\ref{cond:individualcells}.
	
	\begin{proof}
		We first check the incidence coefficients.
	Let $c_k\times c'_{k'}$ be a product cell.
	The two summands in the product boundary lie in the two different bidegrees $(k-1,k')$ and $(k,k'-1)$.
	Thus no cell can occur in both summands.
	Since the boundary coefficients of $\cc$ and $\cc'$ are in $\{0,\pm 1\}$, the same is true for the boundary coefficients of $\cc\times\cc'$.
	Moreover, the top-dimensional product cells are exactly the pairs $c_m\times c'_n$ with $c_m\in\cc_m$ and $c'_n\in\cc'_n$, so the dimension is $m+n$.
	
		We now verify Condition~\ref{cond:B1columns} of \cref{def:abstractcellcomplex}.
	Every $1$-cell of the product is either of the form $c_1\times c'_0$ or of the form $c_0\times c'_1$.
	In the first case,
	\[
	B^{\cc\times\cc'}_1(c_1\times c'_0)=(B^\cc_1c_1)\times c'_0.
	\]
	Since $\cc$ is an abstract regular cell complex, the column $B^\cc_1c_1$ has exactly one positive and one negative entry.
	The same is therefore true for the column corresponding to $c_1\times c'_0$.
	The case $c_0\times c'_1$ is identical, using $B^{\cc'}_1c'_1$ instead.
	Hence every column of $B^{\cc\times\cc'}_1$ has exactly one positive and one negative entry.
	
		It remains to verify Condition~\ref{cond:individualcells} of \cref{def:abstractcellcomplex}.
		Fix a product cell $x=c_p\times c'_q$ of dimension $p+q$.
		Let $\hat{\cc}^{\,c}$ be the subcomplex of $\cc$ generated by $c_p$, with $i$-cells $\hat{\cc}^{\,c}_i$, and let $\hat{\cc}'^{\,c'}$ be the subcomplex of $\cc'$ generated by $c'_q$, with $j$-cells $\hat{\cc}'^{\,c'}_j$.
		Furthermore, let $\hat{\mathcal E}$ denote the subcomplex of $\cc\times\cc'$ generated by $x$.
	We claim that its cells are precisely
	\begin{equation}
	\label{eq:product-cell-closure}
	\hat{\mathcal E}_r=
	\mathop{\dot\bigcup}_{i+j=r}\hat{\cc}^{\,c}_i\times \hat{\cc}'^{\,c'}_j.
	\end{equation}
		Indeed, one set-valued boundary step in the product lowers exactly one of the two factors by one dimension.
		Thus, after iterating the boundary relation, a cell in dimension $r$ must be a pair consisting of an $i$-cell in $\hat{\cc}^{\,c}$ and a $j$-cell in $\hat{\cc}'^{\,c'}$ with $i+j=r$.
	This gives one inclusion in \eqref{eq:product-cell-closure}.
	Conversely, any pair on the right-hand side is obtained by applying the boundary relation $p-i$ times in the first factor and $q-j$ times in the second factor, in any order.
	This proves the claim.
	
	Write
	\[
	A_i=\Z^{|\hat{\cc}^{\,c}_i|},\qquad A'_j=\Z^{|\hat{\cc}'^{\,c'}_j|}
	\]
	for the integral chain groups of the two generated subcomplexes.
	By \eqref{eq:product-cell-closure}, the integral chain group of $\hat{\mathcal E}$ in degree $r$ is canonically
	\[
	E_r=\Z^{|\hat{\mathcal E}_r|}
	\cong
	\bigoplus_{i+j=r}A_i\otimes_\Z A'_j.
	\]
	Under this identification, the restricted product boundary is the total tensor-product boundary
	\[
	\partial(a_i\otimes a'_j)
	=
	\hat{B}^{\cc}_i a_i\otimes a'_j
	+
	(-1)^i a_i\otimes \hat{B}^{\cc'}_j a'_j.
	\]
	Since $\cc$ and $\cc'$ are abstract regular cell complexes, the generated complexes $A_*$ and $A'_*$ have the homology of a point:
	\[
	H_0(A_*)\cong \Z,\qquad H_l(A_*)=0\text{ for }l>0,
	\]
		and analogously for $A'_*$. For cells of positive dimension, this is exactly Condition~\ref{cond:individualcells}; for $0$-cells, it is immediate.
	The K\"unneth formula for chain complexes of free abelian groups \cite[Section~3.B]{hatcher2002algebraic} gives
	\[
	H_r(E_*)
	\cong
	\bigoplus_{i+j=r}H_i(A_*)\otimes_\Z H_j(A'_*)
	\oplus
	\bigoplus_{i+j=r-1}\operatorname{Tor}^{\Z}_1(H_i(A_*),H_j(A'_*)).
	\]
	All higher homology groups of $A_*$ and $A'_*$ vanish, and the only non-zero degree-$0$ homology group is $\Z$.
	Moreover, the relevant Tor terms vanish because $\operatorname{Tor}^{\Z}_1(\Z,\Z)=0$.
	Hence
	\[
	H_0(E_*)\cong \Z,\qquad H_r(E_*)=0\text{ for all }r>0.
	\]
	Translating this back into the notation of \cref{def:abstractcellcomplex}, we obtain
	\[
	\ker \hat{B}^{\cc\times\cc'}_{p+q}=0,
	\qquad
	\ker \hat{B}^{\cc\times\cc'}_{l-1}
	=
	\Ima \hat{B}^{\cc\times\cc'}_l
	\quad \text{for }2\le l\le p+q,
	\]
	and
	\[
	\Z^{|\hat{\mathcal E}_0|}/\Ima \hat{B}^{\cc\times\cc'}_1\cong \Z.
	\]
		Thus Condition~\ref{cond:individualcells} holds for the product cell $x$.
		Since $x$ was arbitrary, Condition~\ref{cond:individualcells} holds for all product cells, and we conclude that $\cc\times\cc'$ is an abstract regular cell complex.
\end{proof}

\begin{remark}[Products of more than two factors]
	By induction, the same argument shows that every finite product of abstract regular cell complexes is again an abstract regular cell complex.
	In particular, the cubical complexes introduced in \cref{subsec:cc-wild-product-structure} are abstract regular cell complexes because they are finite products of path graphs.
\end{remark}

\end{document}